\documentclass[12pt,preprint]{aastex}

\slugcomment{draft version \today, }
\shorttitle{A new sub-nuclear density equation of state for core-collapse}
\shortauthors{Furusawa et al.}

\begin{document}

\title{A new baryonic equation of state at sub-nuclear densities for core-collapse simulations}

\author{Shun Furusawa\altaffilmark{1}, Shoichi Yamada\altaffilmark{1,2}, Kohsuke Sumiyoshi\altaffilmark{3} and Hideyuki Suzuki\altaffilmark{4} }

\altaffiltext{1}{Department of Science and Engineering,
 Waseda University, 3-4-1 Okubo, Shinjuku, Tokyo 169-8555, Japan}
\altaffiltext{2}{Advanced Research Institute for Science and Engineering, Waseda University, 3-4-1
Okubo, Shinjuku, Tokyo 169-8555, Japan}
\altaffiltext{3}{Numazu College of Technology, Ooka 3600, Numazu, Shizuoka 410-8501, Japan}
\altaffiltext{4}{Faculty of Science and Technology, Tokyo University of Science, Yamazaki 2641, Noda, Chiba 278-8510, Japan}
\email{furusawa@heap.phys.waseda.ac.jp}

\begin{abstract}
We calculate a new equation of state for baryons at sub-nuclear densities meant for the use in core-collapse simulations of massive stars. 
The abundance of various nuclei is obtained together with the thermodynamic quantities. 
The formulation is the NSE description and the liquid drop approximation of nuclei. 
The model free energy to minimize is calculated by relativistic mean field theory for nucleons and the mass formula for nuclei with the atomic number up to $\sim 1000$. 
We have also taken into account the pasta phase, thanks to which the transition to uniform nuclear matter in our EOS occurs in the conventional manner: nuclei are not dissociated to nucleons but survive right up to the transition to uniform nuclear matter. 
We find that the free energy and other thermodynamical quantities are not very different from those given in the H. Shen's EOS, one of the standard EOS's that adopt the single nucleus approximation. 
The average mass is systematically different, on the other hand, which may have an important ramification to the rates of electron captures and coherent neutrino scatterings on nuclei in supernova cores. 
It is also interesting that the root mean square of the mass number is not very different from the average mass number, 
since the former is important for the evaluation of coherent scattering rates on nuclei but has been unavailable so far. 
The EOS table is currently under construction, which will include the weak interaction rates. 
\end{abstract}

\section{Introduction}
Equations of state (EOS's) of hot and dense matter play an important role in the dynamics of core collapse supernovae both at sub- and 
supra-nuclear densities (see e.g. \citet{Janka2007}). Not only thermodynamical quantities such as pressure, internal energy, entropy,
sound velocity but also information on matter composition is provided by EOS. At sub-nuclear densities, the latter affects the rates of 
electron captures and coherent neutrino scatterings on nuclei, both of which in turn determine the electron fraction of 
collapsing cores, one of the most critical ingredients for the core dynamics. This is exactly the issue we would like to address in this 
paper.

The EOS employed for the core collapse simulation must cover a wide range of density ($10^5 \lesssim \rho_B \lesssim 10^{15} \rm{g/cm^3}$) and 
temperature ($10^9 \lesssim T \lesssim 10^{12}$K), including both neutron-rich and proton-rich matter. One of the difficulties in constructing
the EOS originates from the fact that depending on the density, temperature and proton fraction, the matter consists of either dilute free 
nucleons or a mixture of an ensemble of nuclei and free nucleons or strongly interacting dense nucleons. Another complication is the
existence of the so-called nuclear pasta phase, in which nuclear shapes change from droplet to rod to slab to anti-rod and 
bubble(anti-droplet) as the density increases toward the nuclear saturation density, at which uniform nuclear matter is realized 
\citep{Ravenhall1983, Hashimoto1984,Nakazato2009}. At high temperatures ($T \gtrsim 5\times 10^9$K), chemical equilibrium is
achieved for all strong and electromagnetic reactions, which is referred to as nuclear statistical equilibrium, or NSE, and the nuclear 
composition in the matter is determined as a function of density, temperature and proton fraction.\citep{Blinnikov2009} At lower temperatures, the matter 
composition is an outcome of preceding nuclear burnings and cannot be obtained by statistical mechanics. In this paper we are concerned 
with the high temperature regime, in which the nuclear composition is a part of EOS.

At present, there are only two EOS's in wide use for the core-collapse simulation. The Lattimer-Swesty's EOS~\citep{Lattimer1991} is 
based on Skyrme-type nuclear interactions and the so-called compressible liquid drop model for nuclei surrounded by dripped nucleons. 
The EOS by H. Shen, Toki, Oyamatsu and Sumiyoshi~\citep{Shen1998} employs a relativistic mean field theory (RMF) to describe nuclear
matter and the Thomas-Fermi approximation for finite nuclei with dripped nucleons. It should be emphasized here that both EOS's 
adopt the so-called single nucleus approximation (SNA), in which only a single representative nucleus is included. In other words, 
the distribution of nuclei is ignored. \citet{Burrows1984} demonstrated that SNA is not a bad approximation for thermodynamical quantities 
such as pressure. It will not be the case, however, for the weak interaction rates, since the electron capture rates are sensitive to
nuclear shell structures and the greatest contributor is not the most abundant nuclei that the single representative nuclei in SNA are
supposed to approximate \citep{Langanke2003}. It is also noted that the coherent scattering of neutrinos on nuclei has a cross section 
proportional to $A^{2}$ with $A$ being the nuclear mass number. In SNA we have no information on the average of mass number squared 
$\overline{A^2 }$ and replace it by $A_{rep}^2$ with $A_{rep}$ being the mass number of the representative nucleus. This is certainly an 
approximation that needs justification. 

Recently \citet{Hempel2010} published a new EOS at sub-nuclear densities meant for the use in supernova simulations. Assuming 
NSE, they calculated the abundances of nuclei as a function of density, temperature and proton fraction, minimizing the model 
free energy. They employed RMF for the nucleon contribution and adopted masses of nuclei from two mass tables in public domain. 
One of them is the experimental atomic mass table 2003 \citep{Audi2003} and the other is a tabulated theoretical prediction for 
nuclei with no available experimental data, which is based on RMF \citep{Geng2005} with the TMA parameter set \citep{Toki1995}. 
The translational and Coulomb energies of nuclei are designed so that the free energy could match the one given by 
RMF for uniform nuclear matter at the saturation density. It should be noted that the solution of the Saha equation for the mass data, 
which are obtained in laboratory \citep{Audi2003} or calculated for isolated nuclei \citep{Geng2005},
is not appropriate at high densities ($\rho_B \gtrsim 10^{12}$g/cm$^3$), since nuclei are not isolated and affected by 
surrounding other nuclei, nucleons and electrons, resulting in the modification of masses of nuclei. 

Our formulation is quite similar to the one employed by \citet{Hempel2010}. The major difference, however, is that we calculate very 
heavy nuclei ($Z \gtrsim 100$ with $Z$ being the atomic number) with a mass formula 
whereas \citet{Hempel2010} ignored these nuclei owing to the limitation of the mass tables. 
It is stressed, however, that these massive nuclei do appear in the H. Shen's EOS  \citep{Shen1998} and, as shown later, they are 
indeed abundant at high densities ($\rho_B \gtrsim 10^{13} {\rm g / cm^3}$) in our EOS also. Another important improvement in this paper is 
the treatment of the nuclear pasta phase, which was entirely ignored in the previous paper \citep{Hempel2010}. It turns out that 
this is important to reproduce the conventional manner of transition to uniform nuclear matter, that is, 
nuclei do not melt down to nucleons but survive up to the nuclear saturation density. 

The aim of our project is in a sense to merge all the previous EOS's to provide all the information needed for core collapse simulations: 
thermodynamical quantities such as pressure, internal energy, entropy, sound velocity and some other important derivatives as well as  
information on the abundance of nuclei required for the calculation of the weak interaction rates; the description of uniform
nuclear matter is based on RMF in H. Shen's EOS; the approximate treatment of the pasta phase is similar to the one employed in 
Lattimer-Swesty's EOS; we solve a Saha-like equation to obtain the abundance of nuclei. In this paper, we describe the formulation in
detail and show some features of our EOS, which are particularly different from those of the previous EOS's. More systematic and
thorough presentation of the EOS will be given in the sequel after we have constructed the EOS table that includes somehow the electron 
capture rates. 

This article is organized as follows. In section 2 we explain in detail the model free energy to be minimized. 
Then the results of minimization are shown in section 3, with an emphasis being put on the comparison with the previous EOS's. 
The paper is wrapped up with a summary and some discussions in section~4.

\section{Model Free Energy}
The strategy to obtain the multi-component EOS is to construct a model free energy and minimize it with respect to the 
parameters involved in the expression of the free energy. The matter in the supernova core at sub-nuclear densities consists of 
nucleons, nuclei together with electrons and photons. The latter two are not treated in this paper although the inclusion
of them as ideal Fermi and Bose gases respectively are quite simple and now a routine. Neutrinos are not always in thermal or 
chemical equilibrium with the matter and cannot be included in the EOS. Their non-equilibrium distributions should be computed 
by the transport equations. 

The free energy is thus a sum of the contributions from the individual nucleus and those from the nucleons that are not contained 
in the nuclei. We employ the RMF free energy for the nucleons outside the nuclei. In constructing the model free energy for the nuclei, 
the following points are appropriately taken into account: 
\begin{enumerate}
\item
Not only the nuclei that are known to exist in laboratory but exotic ones such as very neutron-rich and/or very heavy nuclei should be
also included somehow even if experimental data are unavailable. They are indeed obtained in supernova simulations with EOS's 
that employ SNA. \label{req1}
\item 
At low densities, experimental mass data should be used whenever available. This is important to guarantee that the most stable 
iron group nuclei are populated with correct abundances at low temperatures. \label{req2}
\item
At higher densities, strong and electromagnetic interactions among free nucleons, nuclei and electrons cannot be ignored and 
should be properly reflected in the surface and Coulomb energies of nuclei. \label{req3}
\item
The bulk energy of nuclei should approach the one for uniform nuclear matter as the density increases to the nuclear saturation density. 
This should be taken into consideration to obtain a continuous transition to uniform nuclear matter. \label{req4}
\item
Near the nuclear saturation density, the nuclear pasta phase should be treated somehow. This is also a necessary ingredient
to realize a seamless transition to uniform nuclear matter, in which nuclei are not dissociated to nucleons but continue to exist
right up to the transition. \label{req5}
\end{enumerate}

The model free energy for nuclei described in detail in the following subsections is based on the mass formula. As mentioned above in 
point~\ref{req2}, at low densities, experimental mass data are utilized, whenever available, to obtain the bulk energy of nuclei and 
RMF is employed otherwise.  At high densities, however, even for the nuclei, for which experimental data exist, the bulk energy is 
modified so that points~\ref{req3} and~\ref{req4} are satisfied. As for point~\ref{req5}, only the bubble phase is explicitly considered, 
following the phenomenological prescription adopted in  \citep{Lattimer1991}. Although it is easy to change it, we suppose in this paper 
that H. Shen's EOS is used for uniform nuclear matter at supra-nuclear densities. The continuous transition from the EOS at sub-nuclear densities 
to the one at supra-nuclear densities is guaranteed by the following prescriptions: the free energy of the nucleons outside nuclei is 
based on the same RMF model as for uniform nuclear matter; the bulk energy of nuclei is modified to approach that of uniform nuclear 
matter as the density increases; the surface, Coulomb and translational energies of nuclei go to zero as the transition density is approached. 

The details of these prescriptions will be given in the following subsections. We briefly explain the RMF model we employ for 
the nucleons in section~\ref{secnucleon}. Then we proceed to the description of the model free energy for nuclei in section~\ref{secnuclei}, 
in which the Coulomb and surface energies are treated in subsections~\ref{subsecc} and~\ref{subsecs}, respectively, the bulk energy is 
presented in subsection~\ref{subsecb} and the translational energy is given in subsection~\ref{subsect}. The method to minimize the model free 
energy is explained in section~\ref{secmin} and the evaluation of thermodynamical quantities from the free energy is mentioned in 
section~\ref{secth}. 

\subsection{Free Energy for Nucleons outside Nuclei \label{secnucleon}}
The free energy of the nucleons that are not contained in nuclei is base on RMF in this paper~ \citep{Shen1998}. 
The choice is not essential, however, and one can replace it by any other theory or model indeed. The important thing 
to ensure the continuous transition 
to uniform nuclear matter is to use the same theory or model for the nucleons outside nuclei at sub-nuclear densities and uniform nuclear 
matter at supra-nuclear densities. The details of the RMF model employed in this paper can be found in the original paper~ \citep{Shen1998}. 
In the following, we summarize only the essential ingredients.

In RMF nuclear interactions are described by the exchange of mesons. The employed Lagrangian is the following:
\begin{eqnarray}
{\cal L}_{RMF} & = & \bar{\psi}\left[i\gamma_{\mu}\partial^{\mu} -M
-g_{\sigma}\sigma-g_{\omega}\gamma_{\mu}\omega^{\mu}
-g_{\rho}\gamma_{\mu}\tau_a\rho^{a\mu}
\right]\psi  \\ \nonumber
 && +\frac{1}{2}\partial_{\mu}\sigma\partial^{\mu}\sigma
-\frac{1}{2}m^2_{\sigma}\sigma^2-\frac{1}{3}g_{2}\sigma^{3}
-\frac{1}{4}g_{3}\sigma^{4} \\ \nonumber
 && -\frac{1}{4}W_{\mu\nu}W^{\mu\nu}
+\frac{1}{2}m^2_{\omega}\omega_{\mu}\omega^{\mu}
+\frac{1}{4}c_{3}\left(\omega_{\mu}\omega^{\mu}\right)^2   \\ \nonumber
 && -\frac{1}{4}R^a_{\mu\nu}R^{a\mu\nu}
+\frac{1}{2}m^2_{\rho}\rho^a_{\mu}\rho^{a\mu}. 
\end{eqnarray}
In the above expression, the notation is the same as in \citet{Shen1998}: $\psi$, $\sigma$, $\omega$ and $\rho$ denote nucleons 
(proton and neutron), scalar-isoscalar meson, vector-isoscalar meson and vector-isovector meson, respectively, and 
$W_{\mu\nu} =\partial^{\mu}\omega^{\nu}- \partial^{\nu}\omega^{\mu} $ and 
$R^{a}_{\mu\nu}= \partial^{\mu}\rho^{a\nu}- \partial^{\nu}\rho^{a\mu} +g_{\rho}\epsilon^{abc}\rho^{b\mu}\rho^{c\nu} $. 
The nucleon-meson interactions are given by the Yukawa couplings and the isoscalar mesons ($\sigma$ and 
$\omega$) interact with themselves, which are expressed as the cubic and quartic terms. In the mean field theory, the mesons are assumed 
to be classical and replaced by their ensemble averages whereas the Dirac equation for nucleons is quantized and the free energy is 
evaluated based on the energy spectrum of nucleons obtained this way. $M$ is the mass of nucleons and assumed to be $938$MeV. 
We use the TM1 parameter set as in \citet{Shen1998}, in which the masses of mesons, $m_{\sigma}$, $m_{\omega}$, $m_{\rho}$, and 
the coupling constants, $g_{\sigma}$, $g_{\omega}$, $g_{\rho}$, $g_2$, $g_3$, $c_3$, are determined so that not only the saturation of 
uniform nuclear matter but also the properties of finite nuclei could be best reproduced \citep{Sugahara1994}. 

It is interesting to study how the resulting EOS is affected by the choice of RMF parameter or, more generally, by the choice of EOS 
for uniform nuclear matter. If the preceding paper~\citep{Hempel2010}, in which several RMF parametrizations were compared, is any guide, 
differences in the RMF parametrization will be reflected in thermodynamic quantities near the saturation density. Since RMF is employed 
also for the bulk energy of nuclei in our EOS as explained later in subsection~\ref{subsecb}, differences in symmetry energy originated 
from different parametrizations may be reflected in the neutron-richness of nuclei, for which we have no available experimental mass 
data and the empirical mass formula is employed. The abundance of nuclei with experimental mass data, on the other hand, will not be 
changed very much. It should be also noted that RMF in general tends to overestimate the neutron matter energy at sub-nuclear 
densities~\citep{Sumiyoshi1995,Akmal1998, Steiner2005,Gandolfi2010,Hebeler2010}. It is stressed that one of the 
features of our EOS is that it is easy to adopt different EOS's at supra-nuclear densities as long as they are available. 
The above issues will be discussed more in detail in the forthcoming paper. 

In this paper, the free energy density of uniform nuclear matter, $f^{RMF}(n_B,T,Y_p)$ with $n_B$, $T$ and $Y_p$ being the baryon number 
density, temperature and proton fraction, respectively, is used mainly at sub-nuclear densities. 
In so doing, one important thing to be taken into account is the fact that the free nucleons cannot exist in the volume occupied by nuclei. 
The volume, $V'$, that can accommodate the nucleons is hence given by 
\begin{equation}
 V' =V-\sum_{i} N_i V_i^N, 
\end{equation}
where index $i$ specifies a nucleus, $V$ is the total volume, $N_i$ is the particle number of nucleus $i$ 
and  $V_i^N$ is the volume of nucleus $i$, which is expressed by the mass number, $A_{i}$, and 
number density, $n_i$, of the same nucleus as well as the nuclear saturation density, $n_{si}$, as follows: 
\begin{equation}
V_i^N= \frac{A_i}{n_{si}}, 
\end{equation}
where we assume that the nuclei are homogeneous, having the nuclear saturation density, $n_{si}$. The reason why $n_{si}$ has the subscript 
$i$ will be made clear later (see section~\ref{secnuclei}). Then the densities of free protons
and neutrons in the volume $V'$, which are denoted by $n'_p$ and $n'_n$, respectively, are given by 
\begin{equation}
n'_{p/n}=\frac{N_{p/n}}{{V'}}= \ \frac{n_{p/n}}{\eta}, 
\end{equation}
where $N_{p/n}$ and $n_{p/n}$ are the particle numbers and the number densities in the total volume $V$ of nucleons and $\eta$ is expressed as $\eta=V'/V=1-\sum_{i} A_i n_i/n_{si}$. 
The free energy density for the nucleons in $V'$ should be evaluated at $Y_p= n'_p/(n'_p+n'_n)$ and $n_B=(n'_p+n'_n)$ and 
is given for the entire volume as 
\begin{equation}
f_{p,n}  =  \eta \, f'_{p,n}
\end{equation}
with $f'_{p,n}  =  f^{RMF} (n'_p,n'_n,T)$. 

\subsection{Free Energy for Nuclei \label{secnuclei}}
This is the most crucial part of our model free energy and is described in detail in the following. The free energy of nuclei consists of 
two parts, one of which concerns a translational energy whereas the other part is roughly a nuclear mass. The modeling of the latter is 
based on a semi-empirical mass formula and takes into account the Coulomb, surface and bulk energies. We explain each contribution to 
the mass formula one by one and then move on to the treatment of the translational energy. 

Before proceeding, however, we state here the definition of the nuclear saturation density, $n_{si}(Z_i/A_i,T)$, more precisely, since it will 
appear frequently in the following. It is in principle the baryon number density at which 
the free energy per baryon, $F^{RMF}(T,n_B, Y_p)$ with $Y_p=Z_i/A_i$, by RMF 
takes its minimum value or, put another way, the uniform nuclear matter {\it with the same proton fraction as the nucleus} has the lowest energy 
in RMF. Note that the proton fraction of each nuclei is different from the one for the whole system and that our nuclear saturation
density is actually different from nucleus to nucleus and hence has the subscript that indicates which nucleus is referred to. There are
two exceptions to this definition, though. The first one is the case, in which the saturation density given above is smaller than the 
number density of the whole system, $n_B$. This happens if $n_B$ reaches the saturation density of some nucleus, $n_{si}(Z_i/A_i,T)$,
before the whole system becomes uniform. This is indeed possible because the saturation density tends to be smaller for more neutron-rich 
nuclei. In such a case the nucleus will be overcompressed and we replace the original saturation density 
with the number density of the whole system, $n_B$. The second case occurs at high temperatures, for which the free energy, 
$F^{RMF}(T,n_B, Z_i / A_i )$, has no minimum because the entropy contribution, the $TS$ term with $S$ being entropy, overwhelms the internal
energy. In that case the nuclear saturation density, $n_{si}$, is set to be the density where the matter becomes uniform in the H. Shen's EOS. 
\subsubsection{Coulomb energy\label{subsecc}}
To calculate the Coulomb energy density in the free energy for nuclei we introduce a Wigner-Seitz cell (W-S cell) for each species of 
nuclei and impose charge neutrality in it. It is assumed that a nucleus is centered in the W-S cell, the volume of which is denoted as 
$V_i$ and that the cell contains also nucleons as a vapor outside the nucleus as well as electrons, which are assumed to be uniform 
in the entire cell. The charge neutrality in the cell can be expressed as 
\begin{equation}
V_i n_e =Z_i +(V_i-V_i^N)n'_p, 
\end{equation}
which can be solved for $V_i$ and gives 
\begin{equation}
V_i = \frac{Z_i - n'_p V_i^N}{n_e-n'_p}, 
\end{equation}
where $V_i^N$ is the nuclear volume in the cell and $n_e$ denotes the number density of electrons. 
The vapor volume and nuclear volume fraction for each nucleus are given by $ V_i^B = V_i-V_i^N  $ and $u_i =  V_i^N / V_i$, respectively.

As the density approaches the nuclear saturation density, the nuclear pasta phase is encountered, in which various shapes of nuclei
exist. As mentioned already, this phase has to be somehow treated in order to obtain the smooth transition to uniform nuclear matter.
It is much beyond the scope of our project to treat all known pastas accurately, however. We hence resort to a rather crude approximation, 
in which the bubble phase is explicitly considered. It is known that bubble nuclei, in which the vapor nucleons are surrounded by the 
nucleus, appear at the end of the pasta phase. We assume there that a spherical vapor is located at the center of W-S cell surrounded
by a shell-shaped nucleus. In this paper we assume further that each nucleus enters the nuclear pasta phase individually when 
the volume fraction, $u_i$, reaches $0.3$ and that the bubble shape is realized when it exceeds $0.7$ \citep{watanabe2005}. 
The intermediate states ($ 0.3<u_i<0.7 $) are simply interpolated from the normal and bubble states. 
This criterion is admittedly rather arbitrary. We have tried another choice, $0.4<u_i<0.6$, and confirmed that the results are 
hardly affected, though. We hence believe that our EOS is not very sensitive to the criterion. 

The evaluation of the Coulomb energy in the W-S cell is straightforward and given by the following expression:
\begin{equation} 
E_i^C= \int_{\rm cell} \frac{q(r) \, dq(r)}{r},
\end{equation}
where the cell is assumed to be spherically symmetric and $q(r)$ is the sum of electron and proton charges up to the radius $r$. 
Assuming further that the protons are uniformly distributed in the nucleus and the vapor separately, we obtain the Coulomb energies for 
the normal and bubble phases from the above integration as follows:
\begin{eqnarray}
E_i^C=\left\{ \begin{array}{ll}
\displaystyle{\frac{3}{5}\left(\frac{3}{4 \pi}\right)^{-1/3}  \frac{e^2}{n_{si}^2} \left(\frac{Z_i - n'_p V_i^N}{A_i}\right)^2 {V_i^N}^{5/3} D(u_i)}    & (u_i\leq 0.3), \\
\displaystyle{\frac{3}{5}\left(\frac{3}{4 \pi}\right)^{-1/3}   \frac{e^2}{n_{si}^2} \left(\frac{Z_i - n'_p V_i^N}{A_i}\right)^2 {V_i^B}^{5/3} D(1-u_i)}   & (u_i\geq 0.7),
\end{array} \right.
\label{eqclen}
\end{eqnarray} 
with $D(u_i)=1-\frac{3}{2}u_i^{1/3}+\frac{1}{2}u_i$, where $e$ is the elementary charge. Note that the two energies coincide with each other at $u_i=0.5$,
since the nuclear volume, $V_i^N$, is equal to the vapor volume, $V_i^B$, there although the value is not used and the interpolation is 
done from the values at $u_i=0.3$ and $u_i=0.7$ to take crudely into account the existence of other pastas.
We employ cubic polynomials of $u_i$ for interpolation. The four coefficients of the polynomial are determined by the condition that the 
Coulomb energy is continuous and smooth as a function of $u_i$ at $u_i=0.3$ and $u_i=0.7$. 

\subsubsection{surface energy \label{subsecs}}
In the semi-empirical mass formula we employ in this paper, the surface energy of nuclei is given by the product of the nuclear surface area
and the surface tension, which is approximated by the following analytic expression:
\begin{equation}
\sigma_i=\sigma_0  - \frac {A_i^{2/3} } {4 \pi r_{N_i}^2} [S_s(1- 2(Z_i/A_i)^2 ],
\end{equation}
where  $r_{N_i} = ( 3/4 \pi V_i^N )^{1/3} $ is the radius of nucleus $i$ and $ \sigma_0$ denotes the surface tension for symmetric nuclei. 
The second term on the right hand side is added to accommodate highly neutron-rich nuclei, for which no experimental data are available 
but a proper consideration of the surface symmetry energy is almost mandatory. 
The values of the constants, $\sigma_0=1.15 \rm{MeV/fm^3}$ and $S_s =45.8 \rm{MeV}$, are adopted from the paper by \citet{Lattimer1991}. 
Note that this expression gives negative values if nuclei become extremely neutron-rich. We hence assume in this paper that only those 
neutron-rich nuclei with a surface tension higher than a prescribed lower limit, $\sigma=0.272 \rm{MeV/fm^3}$, can exist. This value is 
obtained from  the mass data of the most neutron-rich nuclei with $(Z, A) = (1, 7)$~\citep{Audi2003} and the subtraction of the bulk 
and Coulomb energies given by our mass formula.

Noting that the nuclear surface should be replaced by the bubble surface in the bubble phase, we obtain the following expression for the surface energy: 
\begin{eqnarray}
E_i^S=\left\{ \begin{array}{ll}
4 \pi {r^2_{Ni}} \, \sigma_i \left(1-\displaystyle{\frac{n'_p+n'_n}{n_{si}}} \right)^2 =
4\pi \left( \displaystyle{\frac{3}{4 \pi}} V_i^N \right)^{2/3}  \, \sigma_i  \left(1-\displaystyle{\frac{n'_p+n'_n}{n_{si}}} \right)^2   & (u_i\leq 0.3), \\
4 \pi {r^2_{Bi}} \, \sigma_i  \left(1-\displaystyle{\frac{n'_p+n'_n}{n_{si}}} \right)^2=4\pi \left(\displaystyle{\frac{3}{4 \pi}} V_i^B \right)^{2/3} \, \sigma_i 
\left(1-\displaystyle{\frac{n'_p+n'_n}{n_{si}}} \right)^2  & (u_i\geq 0.7), 
\end{array} \right.
\end{eqnarray} 
where $r_{Bi} =( 3/4 \pi V_i^B )^{1/3} $ is the radius of bubble $i$. 
The last factor, $ \left(1-(n'_p+n'_n)/n_{si} \right)^2 $, in these expressions are introduced to take into account the effect that 
the surface energy should be reduced as the density contrast decreases between the nucleus and the nucleon vapor. The specific 
functional form is inspired  by the surface energy employed in the Thomas-Fermi approximation in the H. Shen's EOS, 
$ \left|  \nabla(n_n + n_p) \right|  ^2 $ (Eq.~(35) of \citet{Shen1998} ). As shown later, this prescription is necessary to 
ensure the continuous transition to uniform nuclear matter, since the transition occurs not only by coalescence of nuclei but also 
by the increase in the density of nucleon vapor. For the other pastas ($0.3<u_i<0.7$), we employ the same cubic interpolation as for 
the Coulomb energy (section~\ref{subsecc}).

\subsubsection{bulk energy \label{subsecb}}
The bulk energy of nuclei should be different at high densities, $\rho_B \gtrsim 10^{12}$g/cm$^3$, from the one at lower densities, 
since nuclei cannot be regarded as an isolated object and effects of other nuclei, nucleons and electrons cannot be 
ignored \citep{Vautherin1994}. In this paper, we will take this into account phenomenologically as follows. 

As the density increases, the nuclear bulk energy, $E_i^{B}$, should approach the one for uniform nuclear matter with the same proton 
fraction. We obtain the latter from RMF as 
\begin{equation}
E_i^{B}(T)= A_i F^{RMF}(n_{si},T,Z_i/A_i), 
\label{eq:blkh}
\end{equation}
where $F^{RMF}(n_B,T,Y_p)  $ is the free energy per baryon by RMF. Note that the proton fraction in this expression is not the one for
the whole system but the one for each nucleus; the density is taken to be the saturation density defined earlier (section~\ref{secnuclei}) 
for each nucleus. This implies that nuclei have their own densities that are different from each other. Since the saturation density tends
to be lower for neutron-rich nuclei, the density of the whole system becomes equal to the one for some neutron-rich nucleus one after 
another as it increases. In such a case the density of the nucleus that becomes uniform in its W-S cell is set to be the system density 
as mentioned earlier. These prescriptions ensure the continuous transition to uniform nuclear matter. 

At low densities, on the other hand, experimental mass data are used whenever available~\citep{Audi2003}. This is indeed important 
to reproduce the well-known results of the ordinary NSE. As will be shown later, the nuclear shell and even-odd effects manifest themselves 
in the abundance of nuclei and should be taken into account whenever possible. These experimental mass data are accommodated in our 
mass formula by adjusting the bulk energy, which is expressed as
\begin{equation}
E_i^{B}=M_i^{\rm{data}}-[E_i^{C}]_{vacuum}-[E_i^{S}]_{vacuum},
\label{eq:blkl}
\end{equation}
where subtracted from the experimental mass, $M_i^{\rm{data}}$, are the Coulomb and surface energies $[E_i^{C,S}]_{vacuum}$ of the nucleus  
isolated in vacuum, that is, the energies calculated from our mass formula for $n_e=n'_p=n'_n=0$. This ensures that the experimental 
mass is reproduced exactly in the low density limit. At finite densities, on the other hand, the mass is given by 
\begin{eqnarray}
M_i & = &   E_i^{B} + E_i^{S} + E_i^{C} \nonumber \\
& = & M_i^{\rm{data}}+E_i^{S}(n'_n,n'_p,n_e) + E_i^{C}(n'_p,n_e) -[E_i^{C}]_{vacuum}-[E_i^{S}]_{vacuum}.
\end{eqnarray} 
Note that the bulk energy in our mass formula actually includes other energies such as the bulk symmetry, shell and pair energies. 

For nuclei with no available experimental data, we employ the bulk energy obtained from RMF even at low densities as explained above.
Since these nuclei, which are either very heavy $Z_i \gtrsim 100$ or highly neutron-rich $Z_i/A_i \lesssim 0.35 $, are abundant normally 
at high densities ($\rho_B \gtrsim 10^{12} \rm{g/cm^3}$) alone, the above prescription is justified. The only exception is  
the combination of very low temperatures ($T \lesssim  1$MeV) and low proton fractions ($Y_p \lesssim 0.3$). 
Such cases are not encountered in the simulations of core-collapse supernovae, however, and will not be a problem practically, either. 
These low- and high-density values (Eqs.~(\ref{eq:blkl}) and (\ref{eq:blkh})) are linearly interpolated between  
$\rho_B = 10^{12} \rm{g/cm^3}$ and the nuclear saturation density, $\rho_B \sim 10^{14.2} \rm{g/cm^3}$. 
The linear interpolation makes the free energy not smooth and the pressure discontinuous at the boundaries of the interpolation region. 
In practice, however, the variation of the bulk energy is quite minor compared with those of Coulomb and translational 
energies and the discontinuities of the pressure are negligible. Moreover, nuclei that have experimental mass data and hence need 
the interpolation are not abundant near the high density end, the fact which makes the discontinuities even less important there. 
We hence believe that the above prescription is sufficient. 

\subsubsection{translational energy \label{subsect}}
The translational energy of nucleus $i$ in our model free energy is based on that for the ideal Boltzmann gas and given by 
\begin{equation}
\label{eq:tra}
 E_i^{t}=k_B T \left\{ \log \left(\frac{n_i}{g_i(T) n_{Qi}}\right)- 1 \right\} \left(1-\frac{n_B}{n_s}\right), 
\end{equation}
where $k_B$ is the Boltzmann constant and $n_{Qi} = \left(M_i k_B T/2\pi \hbar ^2   \right)^{3/2}$. 
Since nuclei have internal degrees of freedom, the 
contribution from excited states of nuclei to the free energy cannot be ignored at high temperatures. In the above formula, 
the effect is encapsulated in $g_{i}(T) $, which is approximated by \citet {Fai1982} as
\begin{equation}
\label{eq:ex}
g_{i}(T)=g_{i}^0 +\frac{c_1}{A_i^{5/3}}\int_0^\infty dE e^{-E/T}\exp\left(\sqrt{2 a(A_i) E}\right), 
\end{equation}
where $g_{i}^0$ denotes the contribution of the ground-state (the spin degrees of freedom and normally much smaller than 
$g_{i}(T)$) and is set to be $g_{i}^0=1$ for even nuclei and $g_{i}^0=3$ for odd ones in this paper for simplicity. 
Other coefficients in the contribution from the excited states are given as $a(A_i)=(A_i/8)(1-c_2 A_i^{-1/3})$MeV$^{-1}$, 
$c_1=0.2$MeV$^{-1}$ and $c_2=0.8$. 
In order to estimate the sensitivity of results to $g_{i}(T)$, we multiplied the original 
$g_{i}(T)$ rather arbitrarily by a factor of 0.5 or 2 at $Y_p=0.3$ and $T = 5$MeV and compared the results. We found small 
differences at $\rho \sim 10^{13}$g/cm$^3$. For example, the free energy is decreased or increased by 4\% at most; the entropy 
changes -1\% and +3\% whereas the mass fraction of all nuclei varies -10\% and +8\% from the original values.

The last factor on the right hand side of Eq.~(\ref{eq:tra}) takes account of the excluded-volume effect: each nucleus can move in 
the space that is not occupied by other nuclei and free nucleons. The factor reduces the translational energy at high densities and
is important to ensure the continuous transition to uniform nuclear matter as mentioned earlier. In reality, however, the nuclear translational 
energy is supposed to be suppressed by the formation of Coulomb lattice when the so-called plasma factor, which is defined as 
$\Gamma \equiv  (\bar{Z} e)^2/(\bar{r}  k_B T) $ with the average proton number $\bar{Z}$ and distance between nuclei $\bar{r}$, 
reaches a critical value $\sim 171 $ \citep{Slattery1980} and the introduced factor may be regarded as a very crude phenomenological 
approximation to this situation. The present form of the factor, $(1-n_B/n_s)=(V-V_{baryon})/V$, actually gives a linear suppression 
in terms of the occupied volume $V_{baryon}$. Here we always employ the nuclear saturation density for symmetric nuclei 
$n_s=\left[ n_{si}(Z_i/A_i,T) \right] _{Z_i/A_i=0.5}$ for numerical convenience, which will be justified, provided the approximation 
itself is very crude. Note also that the translational energy is a minor contribution except at low densities, where the factor 
is almost unity.

\subsection{Minimization of Free Energy \label{secmin}}
The abundances of nuclei as a function of $\rho_B$, $T$ and $Y_p$ are obtained by minimizing the model free energy derived so far with 
respect  to the number densities of nuclei and nucleons under the constraints,
\begin{eqnarray}
 n_p+n_n+\sum_i{A_i n_i} & = & n_B=\rho_B/m_B, \nonumber \\
 n_p+\sum_i{Z_i n_i} & = & n_e=Y_p n_B,
\label{eq:cons}
\end{eqnarray}
just as in the ordinary NSE calculations. 
Note that $n_{p/n}$ is the number density of proton/neutron in the total volume $V$ of the system as defined in \ref{secnucleon}.
By introducing Lagrange multipliers, $\alpha$ and $\beta$, for these constraints, the minimization
condition is given by
\begin{equation}
 \frac{\partial}{\partial {n_j}} \{f - \alpha(n_p+n_n+\sum_i{A_i n_i} - n_B   )  - \beta(n_p+\sum_i{Z_i n_i}-n_e) \} =0,
\end{equation}
where$f$ is the free energy density,
\begin{equation}
\ f=\eta f_{p,n}^{RMF}(n'_p,n'_n)+\sum_i{n_i \{E_i^{t} + E_{i}^{B}(T,n_{s}, Z_i / A_i )+E_i^{S}(n'_n,n'_p)+ 
E_i^{C}(n'_p) \}}, 
\end{equation}
and the index $j$ runs over all nuclear species and nucleons.
Taking $j= p, \, n$, we find as usual that the Lagrange multipliers, $\alpha$ and $\beta$, are related  to the chemical potentials of
proton ($\mu_p = \partial f/ \partial n_p$)  and neutron ($\mu_n = \partial f/ \partial n_n$) as 
\begin{eqnarray}
 \alpha & = & \mu_n, \\
 \beta & = & \mu_p -\mu_n.
\end{eqnarray}
The differentiation with respect to the number densities of nuclei, on the other hand, gives the usual relations between the chemical 
potentials of nuclei $\mu _i$ and those of proton and neutron:
\begin{equation}
\label{eq:mun}
 \mu_i=Z_i \mu_p+(A_i-Z_i)\mu_n.
\end{equation}

Differentiating the free energy for nuclei with respective to the number densities of nuclei, $n_i$, and employing Eq.~(\ref{eq:mun}),
we can express $n_i$  as follows:
\begin{equation}
n_i  =  g_i \, n_{Qi} \exp \left( \frac{Z_i \mu_p+(A_i -Z_i) \mu_n - M_i}{k_B T (1-n_B/n_s)} \right). 
\label{eq:nio}
\end{equation} 
Note that the only difference from the counter part in the ordinary NSE calculations,
\begin{equation}
n_i = g_i\, n_{Qi} \exp \left( \frac{Z_i \mu_p+(A_i -Z_i) \mu_n - M_i}{k_B T} \right), 
\label{eq:nin}
\end{equation}
seems to be the factor $(1-n_B/n_s)$ in the denominator in the exponential. 

This is not true, however, and the calculations for our model are much more involved than for the ordinary NSE. In the latter case, 
all we have to do is to solve the two conservation equations, Eq.~(\ref{eq:cons}), for $\mu_p$ and  $\mu_n$, since the number 
densities of all nuclei as well as nucleons are expressed by them alone (e.g. Eq.~(\ref{eq:nin})). In our case, on the other hand,
the masses of nuclei, $M_i$, depend on the number densities of nucleons in the vapor, $n'_p$ and $n'_n$, which are in turn related with
$\mu_p$ and  $\mu_n$ as follows:
\begin{equation}
\mu_{p/n} =\frac{\partial f}{ \partial n_{p/n}}= \mu'^{RMF}_{p/n}(n'_p,n'_n) + \sum_i n_i {\frac{\partial M_i(n'_p,n'_n)}{ \partial n_{p/n}}}, 
\label{eq:chem}
\end{equation}
where  $\mu'^{RMF}_{p/n}$ are the chemical potentials of nucleons in the vapor of volume $V'$, which are obtained from RMF 
(see section~\ref{secnucleon}), and the second term that is originated from the dependence of the mass of nuclei on the nucleon densities
in the vapor and hence is summed over all nuclear species is the source of the complication in the solution. As a result,
we have to solve not only the two conservation equations, Eq.~(\ref{eq:cons}), but also the relations between $\mu_{p/n}$ and 
$n'_{p/n}$, Eq.~(\ref{eq:chem}), for the four variables, $\mu_p$, $\mu_n$, $n'_p$ and $n'_n$, which then give the number density of 
nuclei, $n_i$, by Eq.~(\ref{eq:nio}).

\subsection{Thermodynamical Quantities \label{secth}}
After minimization, we obtain the free energy density together with the abundance of various nuclei 
and free nucleons as a function of $\rho_B$, $T$ and $Y_p$. As mentioned repeatedly, the Coulomb-, surface- 
and translational-energy contributions to the free energy of nuclei are going to zero as the density increases 
and only the nuclear bulk-energy and free-nucleon contributions remain just prior to the transition to uniform 
nuclear matter. Since both of them are entirely based on RMF at this point, the free energy density obtained 
by the sum of them coincides with the one for uniform nuclear matter obtained by RMF. 
At low densities, on the other hand, the excluded-volume effect is negligible and the masses of nuclei agree 
with those of the experimental data. In addition, the RMF free energy for the vapor nucleons reduces to the one for 
the ideal Boltzmann gas of nucleons, since the interactions between nucleons become negligible. As a result, 
the free energy density obtained with the present formulation reproduces the well known free energy of the 
ideal gas composed of nucleons and nuclei, the masses of which are given by the experimental data. 

Once the free energy is obtained, other physical quantities are derived by its partial differentiations.
The baryonic pressure, for example, is obtained by the differentiation with respect to the baryonic density
as follows:  
\begin{eqnarray}
 p_B &=&n_B \left[\partial{f}/\partial {n_B}\right]_{T,Ye}-f  \\
 \label{eq:pr}
   &=&   p_{p,n}^{RMF}+ \sum_i ( p_i^{th}  + p_i^{ex}  + p_i^{bulk} + p_i^{surf} + p_i^{Coul} ),
\end{eqnarray}
where $ p_{p,n}^{RMF}$ is the contribution from the nucleons in the vapor; $p_i^{bulk}$, $p_i^{surf}$ and $ p_i^{Coul} $ 
originate from the bulk, surface and Coulomb energies of nuclei in the free energy, respectively; both 
$p_i^{th} =n_i k_B T (1-n_B/n_s)$  and $p_i^{ex}=n_i k_B T(n_B/n_s) \left(\log (n_i/ n_{Qi}) -1 \right)$ come from the translational energy of nuclei in the free energy, with the former being the ordinary pressure of nuclei as a Boltzmann gas 
and the latter arising from the excluded-volume effect approximated by the factor $(1-n_B/n_s)$. They are minor at high densities. 
Note also that $p_i^{bulk}$ and $p_i^{surf}$ are normally negligible.
The Coulomb-energy contribution, $ p_i^{Coul} $, is negative owing to the attractive electrostatic interactions among 
uniformly-distributed electrons and protons inside nuclei and is particularly important for $\rho_B \gtrsim 10^{11}$g/cm$^3$ and 
$Y_p \gtrsim 0.3$. If nucleus $i$ is in the ordinary droplet phase, it is given as  
\begin{eqnarray}
p_i^{Coul}  &=& n_i n_B \frac{3}{5}\left(\frac{3}{4 \pi}\right)^{-1/3}  \frac{e^2}{n_{si}^2} 
\left(\frac{Z_i - n'_p V_i^N}{A_i}\right)^2 {V_i^N}^{5/3} \nonumber \\
 && \times \frac{1}{2} (- u_i^{1/3}+u_i)   \frac{( \eta n_e -n'_p)(1-u_i)}{(n_e-n'_p) \eta n_B},
\end{eqnarray} 
whereas in the bubble phase it is expressed as 
\begin{eqnarray}
p_i^{Coul}  &=& n_i n_B \frac{3}{5}\left(\frac{3}{4 \pi}\right)^{-1/3}  \frac{e^2}{n_{si}^2} 
\left(\frac{Z_i - n'_p V_i^N}{A_i}\right)^2 \nonumber \\
 && \times \{ {V_i^B}^{5/3}  \frac{1}{2} (- u_i^{1/3}+u_i) \frac{( \eta n_e -n'_p)(1-u_i)}{(n_e-n'_p) \eta n_B} \nonumber \\
 &&+ \frac{5}{3} {V_i^B}^{2/3} D(1-u) \frac{\partial V_i^B}{ \partial n_B} \}.
\end{eqnarray} 

The entropy per baryon is calculated from the following expression:
\begin{eqnarray}
 s& =& -\frac{\left[\partial{f}/\partial {T}\right]_{\rho_B,Ye}}{n_B}   \\
    &=&\eta s^{RMF}_{p,n} +  \sum_i \frac{n_i k_B}{n_B} \Biggl[ \left\{  \frac{5}{2}- \log \left(\frac{n_i}{g_i(T) n_{Qi}} \right)   \right\} (1-n_B/n_s) \nonumber \\ 
\label{eq:ent}
& & +  (1-n_B/n_s) \, T \, g_i'/g_i  -  \partial{M_i}/\partial {T}  \Biggl] ,
\end{eqnarray} 
where $g'_i(T)= \left[\partial{g_i(T)}/\partial {T} \right]_{\rho_B,Ye}$. The first term in Eq.~(\ref{eq:ent}) is the entropy of uniform 
nuclear matter and the second one is the entropy of nuclei as a Boltzmann gas with the excluded-volume effect, which is negligible
at low densities. The third term accounts for the thermal excitations of nuclei with $(1-n_B/n_s)  T g_i'/g_i $ being non-negligible 
at high temperatures as shown later. The last term is originated from the fact that the bulk energy of nuclei is temperature-dependent 
in our formulation and given as follows: 
\begin{equation}
\label{eq:ent2}
 \frac{\partial{M_i}}{\partial {T}} = \frac{\partial{E_i^{B}}}{\partial {T}} = - A_i s_i^{RMF} (T,n_{si}, Z_i / A_i ). 
\end{equation}
The contribution of this term is normally negligible except near the nuclear saturation density.
 
The free energy thus obtained is continuous at any density but not smooth at the saturation density. In fact, the pressure, or 
the density-derivative of the free energy, is not continuous at the saturation density. This is most easily understood from 
Eq.~(\ref{eq:pr}). The contribution from the excluded volume effect in the translational energy, $p_i^{ex}$, in the equation does 
not vanish at the saturation density. The continuity of pressure would have been obtained if we had employed a suppression 
factor (the last term on the right hand side of Eq.~(\ref{eq:tra})) with a higher power. We do not think this is necessary, 
however, since the discontinuity is practically negligible compared with the dominant electron pressure in core-collapse 
simulations. The derivatives of free energy with respect to $T$ and $Y_p$ are continuous at any density as observed in 
Eq.~(\ref{eq:ent}) for entropy. 
\section{Results}
We show the main features of our EOS for selected combinations of densities ($\rho_B > 10^6$g/cm$^{3}$), temperatures ($T=1, 5, 10$MeV) 
and proton fractions ($Y_p = 0.1, 0.3, 0.5$), emphasizing in particular the differences from the H. Shen's \citep{Shen1998} and 
Hempel's \citep{Hempel2010} EOS's. The matter compositions are given first. It is also demonstrated that the transition to uniform 
nuclear matter in our EOS occurs in the conventional way, that is, either by coalescence of nuclei or by the increase in the density 
of nucleon vapors. Then we turn to the thermodynamical quantities such as free energy, pressure and entropy. 

\subsection{Compositions \label{secrecmp}}
\begin{figure}
   \begin{center}
    \begin{tabular}{c}
         \resizebox{130mm}{!}{\plotone{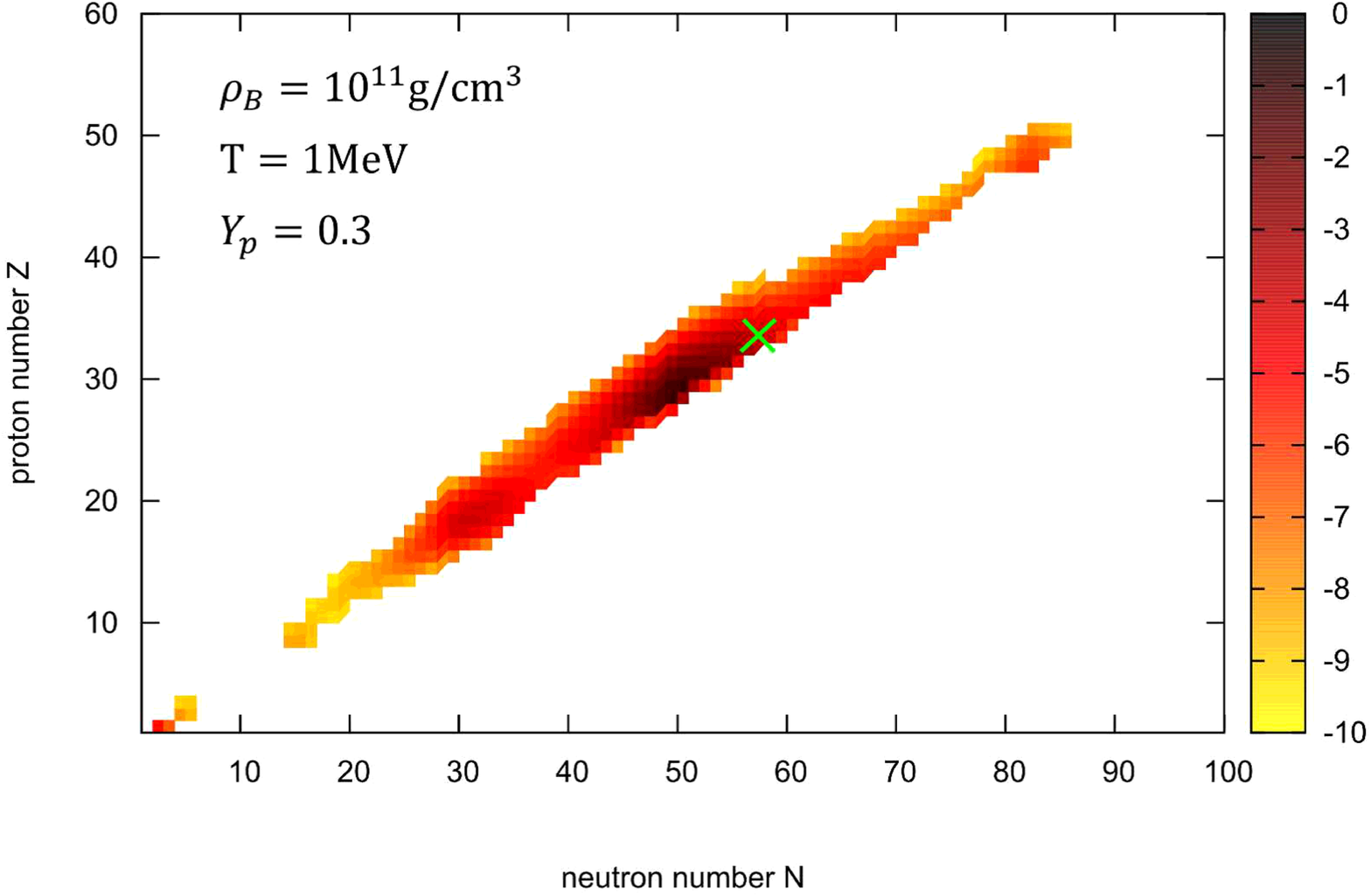}} 
    \end{tabular}
\caption{ The mass fractions of nuclei in the $(N, Z)$ plane for $\rho_B=10^{11}$g/cm$^3$, $T=1$MeV and $Y_p=0.3$. 
The cross indicates the representative nucleus for the H. Shen's EOS under the same condition. }
    \label{dis11}		
        \begin{tabular}{c}
           \resizebox{130mm}{!}{\plotone{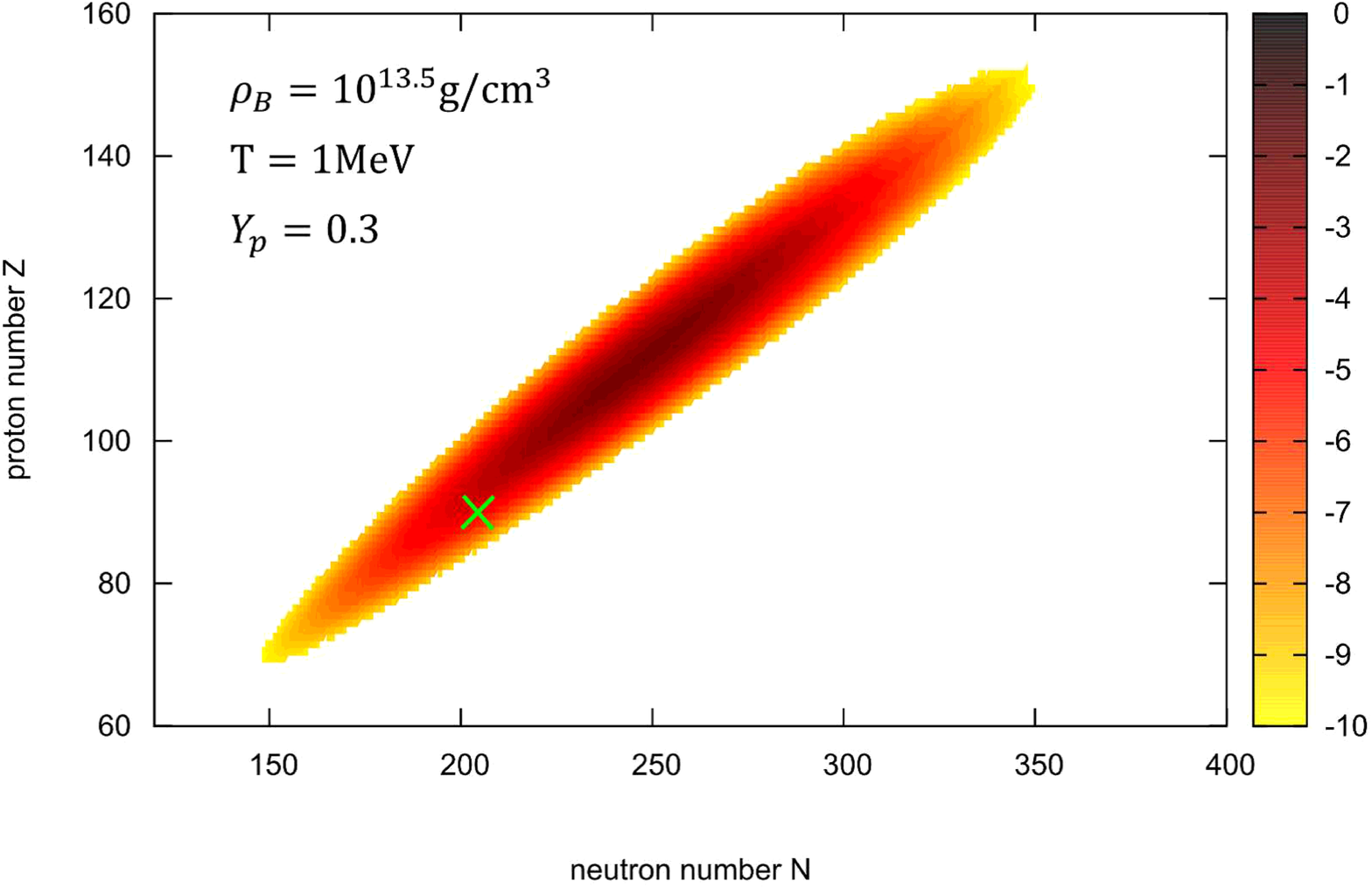}} 
    \end{tabular}
\caption{ The mass fractions of nuclei in the $(N, Z)$ plane for $\rho_B= 10^{13.5} $ g/cm$^3$, $T=1$MeV and $Y_p=0.3$. 
The cross indicates the representative nucleus for the H. Shen's EOS under the same condition. }
    \label{dis13}		
   \end{center}
\end{figure}

The mass fractions of nuclei are shown in the $(N, Z)$ plane with $N$ being the neutron number for $\rho_B = 10^{11}$g/cm$^3$, $T=1$MeV and $Y_p=0.3$ 
in Fig.~\ref{dis11}. It is clear that nuclei are abundant in the vicinities of the neutron magic numbers ($N = 28 $, 50 and 80). 
In fact, the abundance peak arises around $(N, Z)\sim(50,30)$ in our EOS in this case. For the same density, temperature and $Y_p$, 
on the other hand, the H. Shen's EOS predicts that the representative nucleus has $(N, Z)=(57.2, 33.2)$. The difference originates
mainly from the fact that the H. Shen's EOS does not include the shell effect in their Thomas-Fermi approximation. At higher densities, 
heavier nuclei become abundant, which is evident in Fig.~\ref{dis13} that shows the mass fractions of nuclei for 
$\rho_B = 10^{13.5}$g/cm$^3$, $T=1$MeV and $Y_p=0.3$. It should be noted that nuclei with $Z \gtrsim 100$ are quite abundant and,
in fact, are dominant in this case. Hence the neglect of these nuclei in \citet{Hempel2010} cannot be justified at these high 
densities. It should be also mentioned that the nuclear distribution is much smoother in this case than that for the lower density. 
This is because the shell effect is reduced in our EOS at high densities, where the nuclear bulk energy is interpolated from the
one derived experimentally and the one obtained theoretically for uniform nuclear matter, with the latter being dominant at this high
density. 
The H. Shen's EOS predicts that the most abundant nucleus is the one with $(N, Z)=(203.8, 89.4)$, which is still different
from the prediction of our EOS, $(N, Z)=(243,110)$. 
These differences may have an important ramification for the electron capture 
in the collapsing core~\citep{Hix2003}. 

\begin{figure}
   \begin{center}
    \begin{tabular}{c}
               \resizebox{63mm}{!}{\plotone{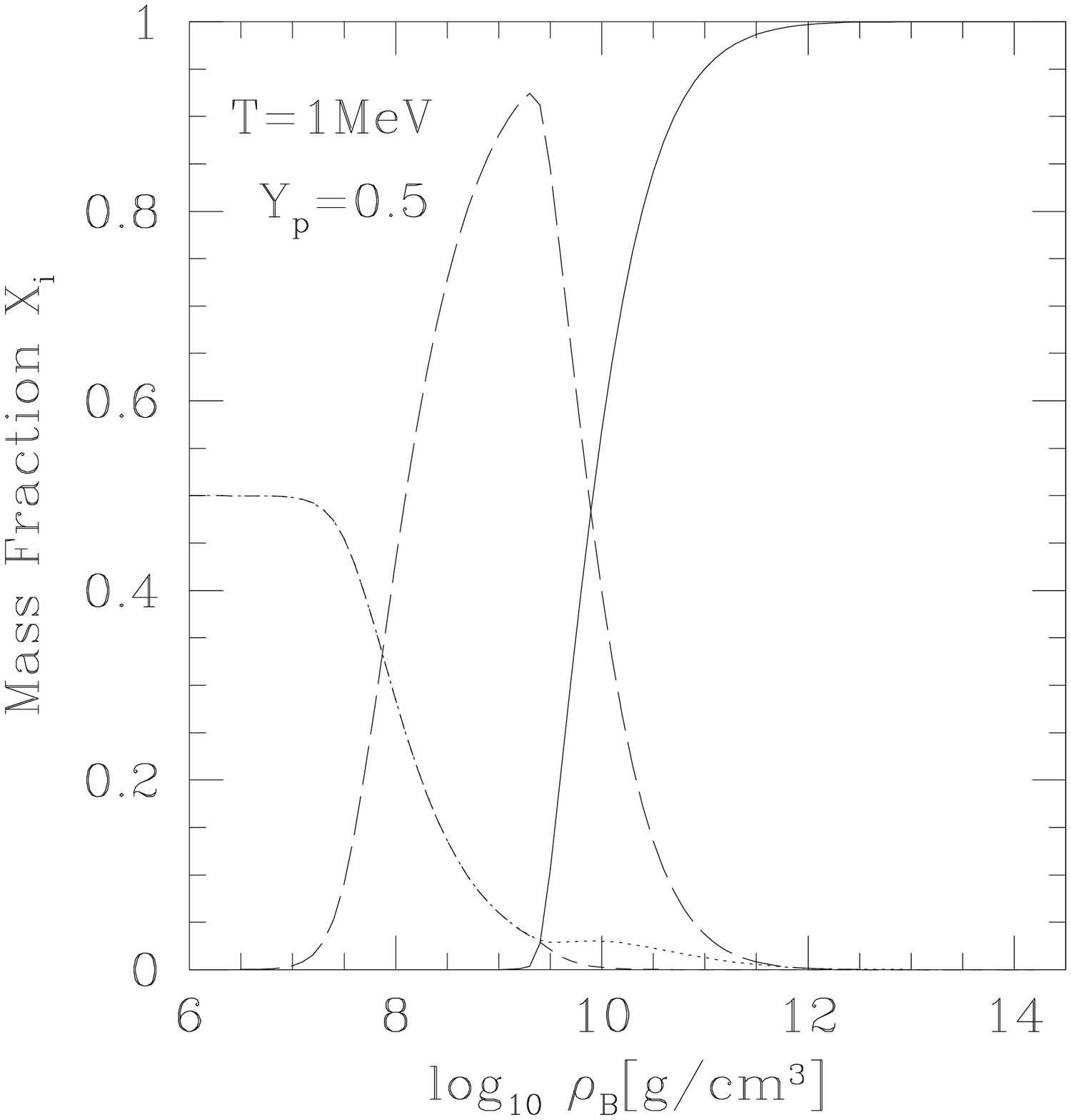}} \\
               \resizebox{63mm}{!}{\plotone{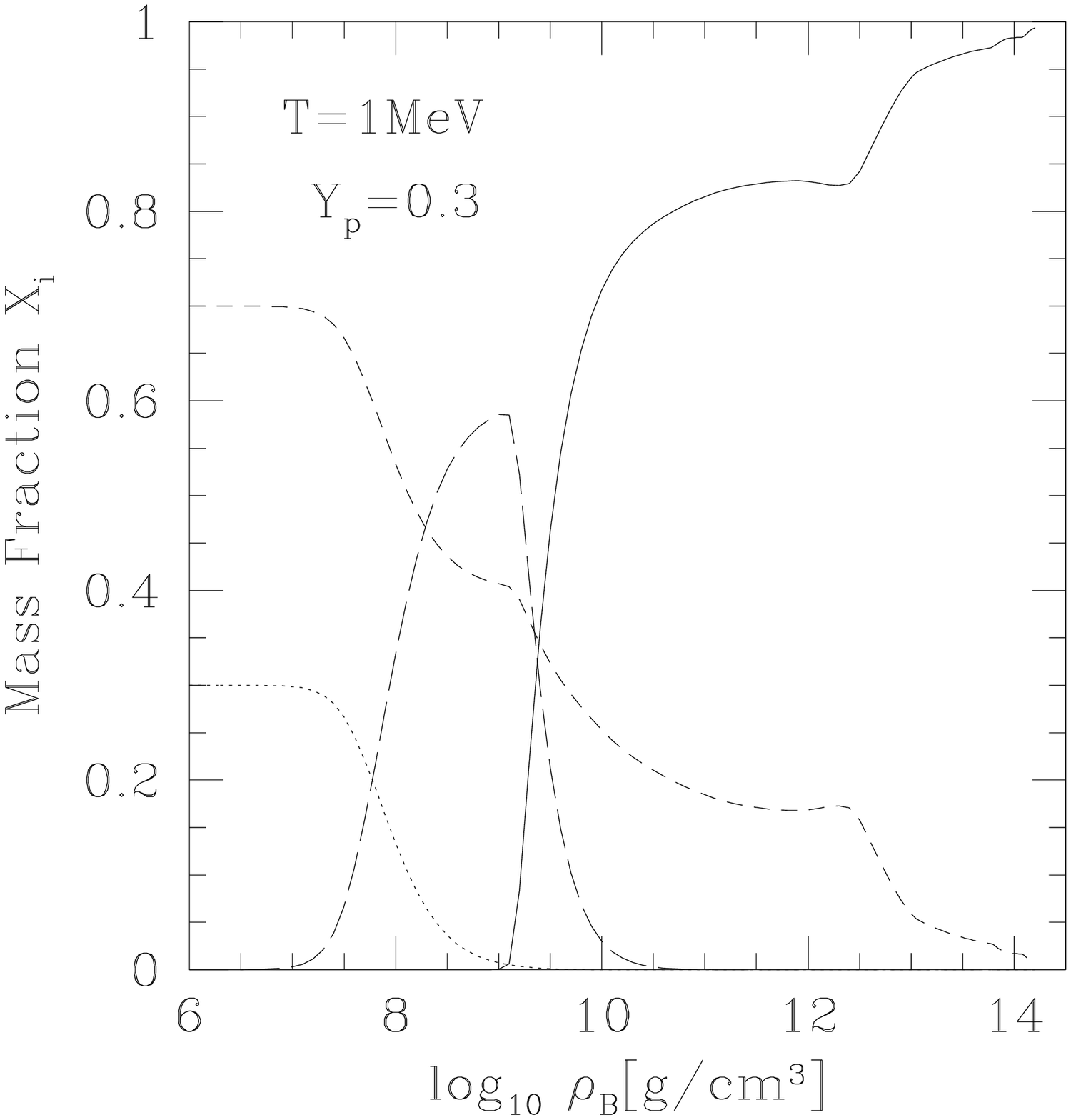}} \\
               \resizebox{63mm}{!}{\plotone{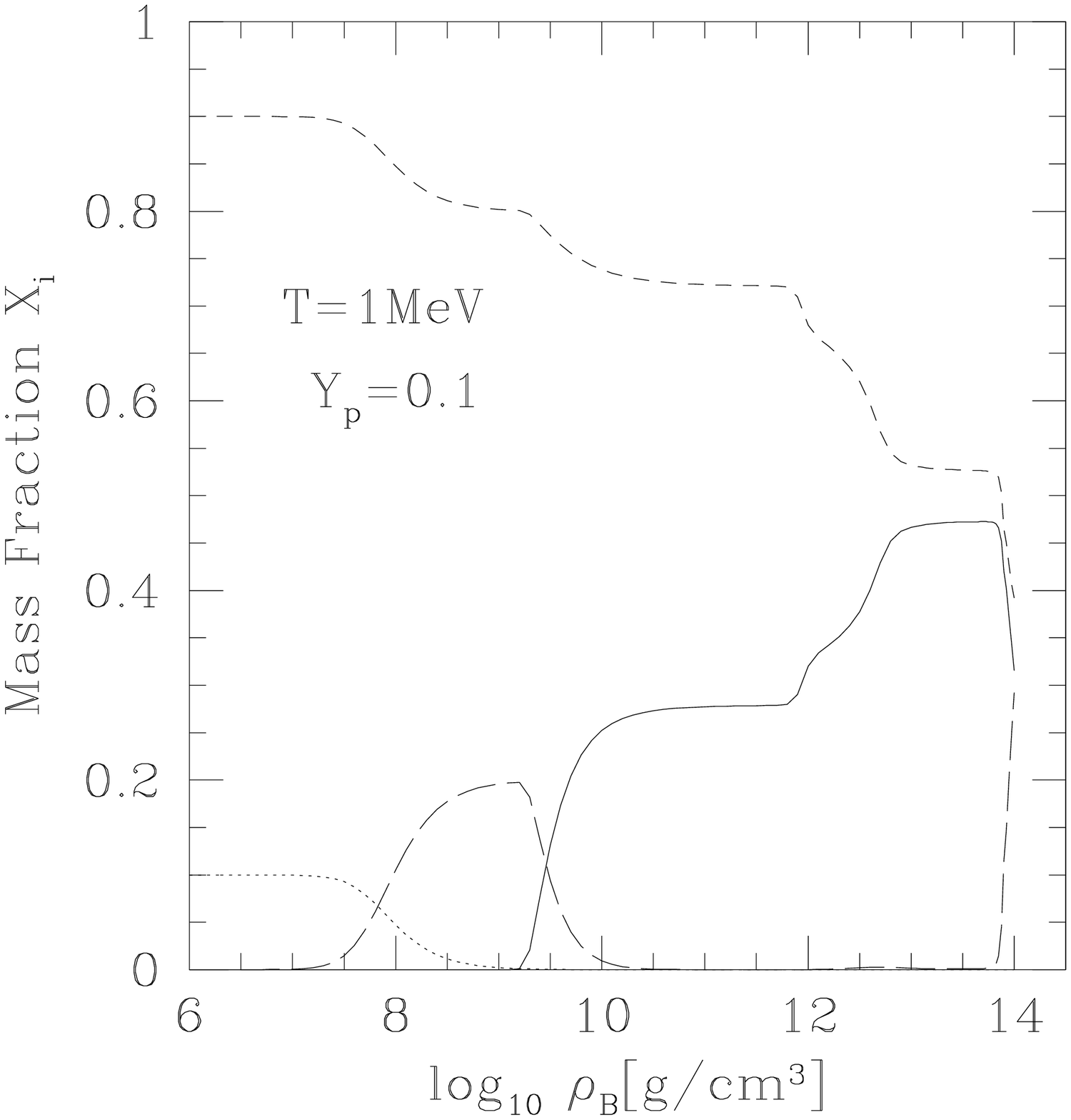}} \\
    \end{tabular}
\caption{The mass fractions of free protons (dotted lines), free neutrons (dashed lines), light nuclei ($ Z\leq 5$, long dashed lines) 
and heavy nuclei ($Z \geq 6$, solid lines) as a function of density for $T=1$MeV and $Y_p=0.5, 0.3, 0.1$ from top to bottom.}
    \label{cmp1}		
   \end{center}
\end{figure}

\begin{figure}
   \begin{center}
    \begin{tabular}{c}
               \resizebox{63mm}{!}{\plotone{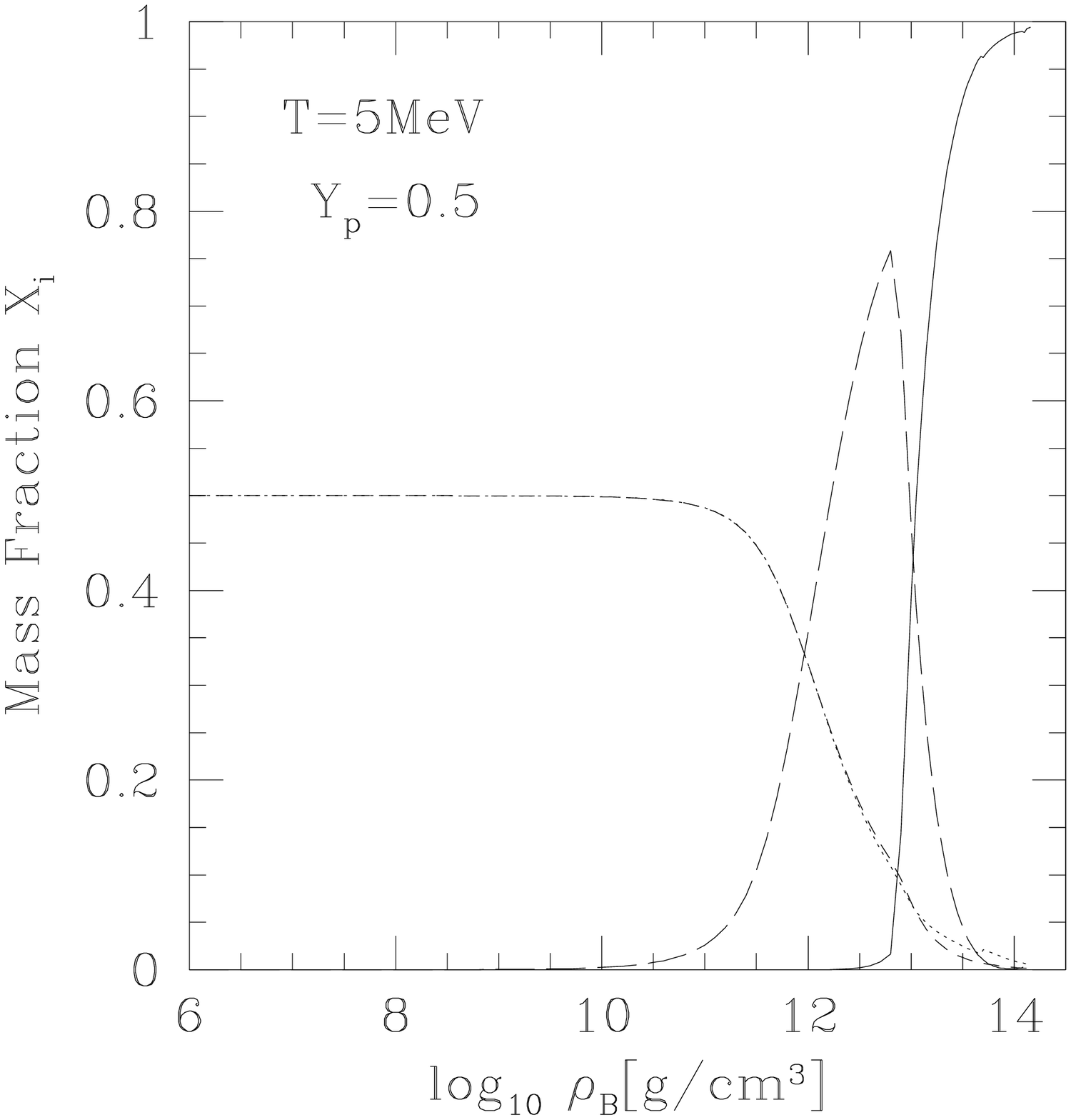}} \\
               \resizebox{63mm}{!}{\plotone{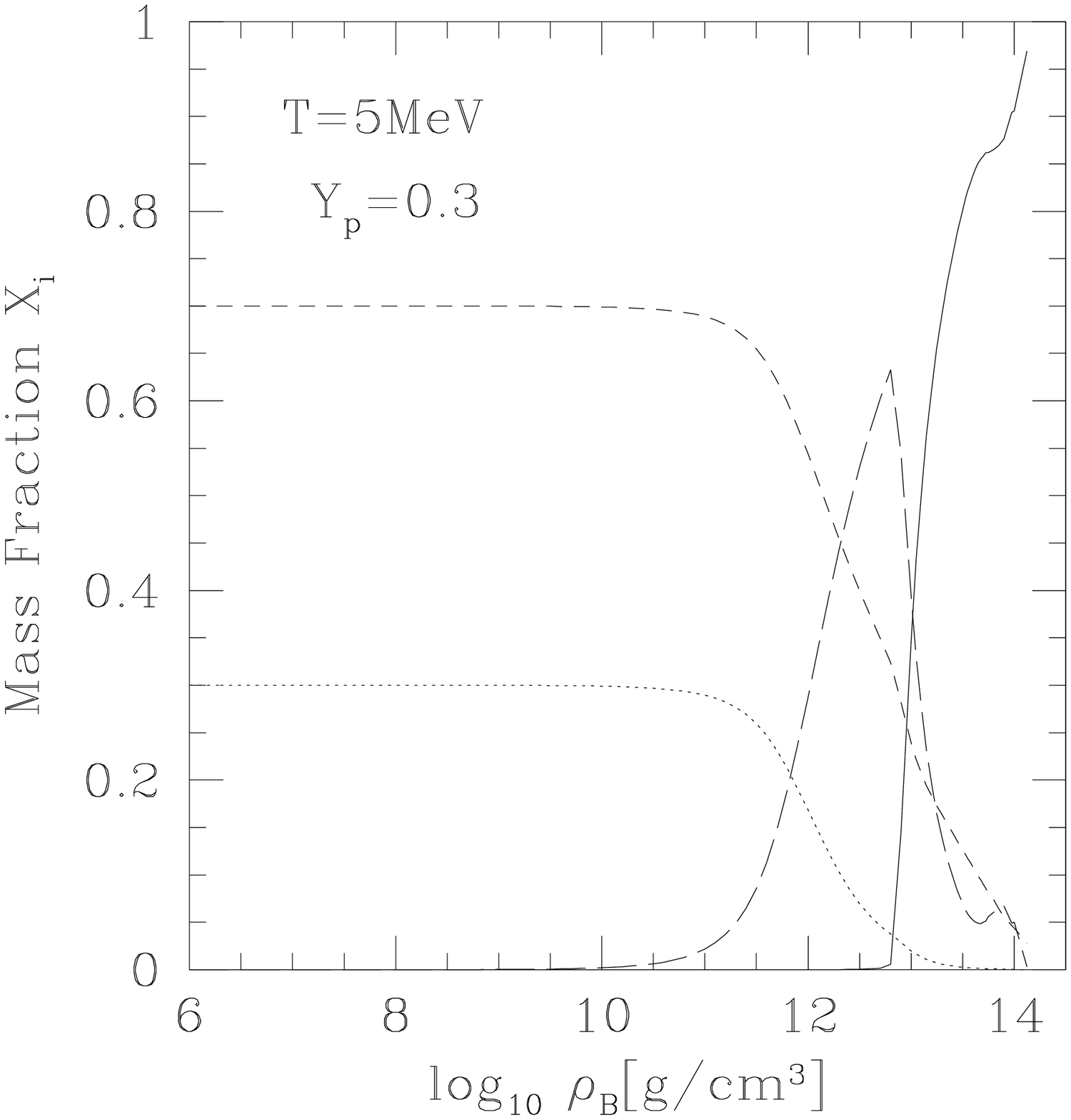}} \\
               \resizebox{63mm}{!}{\plotone{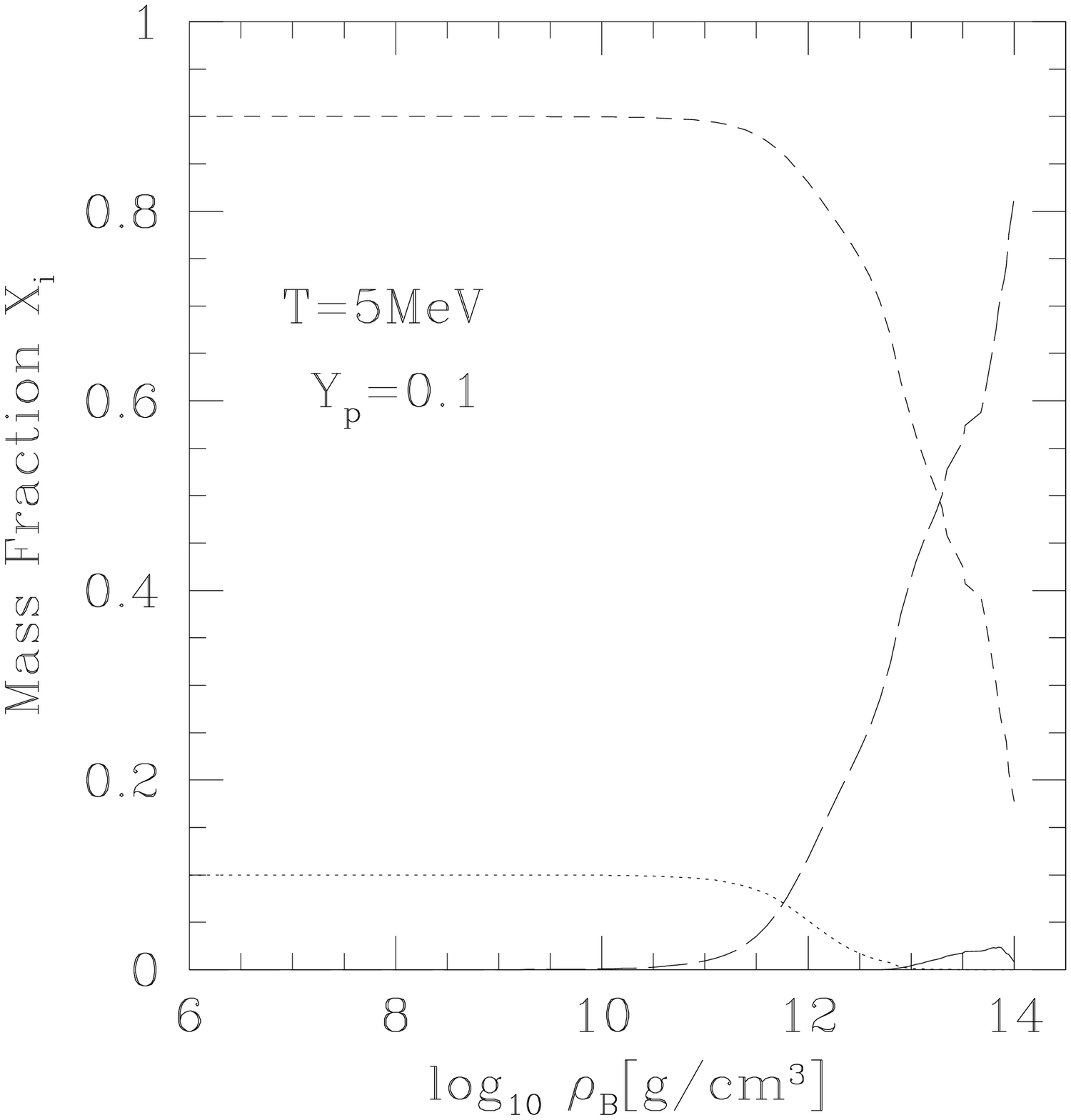}} \\
    \end{tabular}
\caption{The mass fractions of free protons (dotted lines), free neutrons (dashed lines), light nuclei ($ Z\leq 5$, long dashed lines) 
and heavy nuclei ($Z \geq 6$, solid lines) as a function of density for $T=5$MeV and $Y_p=0.5, 0.3, 0.1$ from top to bottom.}
    \label{cmp5}		
   \end{center}
\end{figure}

In order to expedite comparison with the existing EOS's with SNA, we show in Fig.~\ref{cmp1} the mass fractions of light ($Z_i\leq 5$) 
and heavy ($Z_i\geq 6$) nuclei together with free nucleons in the vapor for $T=1$MeV. Irrespective of the values of $Y_p$ selected
here ($Y_p=0.5, 0.3, 0.1$), the free nucleons dominate other constituents up to $\rho_B \sim 10^{7}$g/cm$^3$ at this temperature. 
At larger densities ($10^7$g/cm$^3 \lesssim \rho_B \lesssim 10^{10}$g/cm$^3$), some of free nucleons are taken into symmetric light 
nuclei such as deuterons and alpha particles. For densities larger  than $\rho_B \sim 10^{10}$g/cm$^3$, light nuclei disappear and 
heavy nuclei are populated dominantly. Near the nuclear saturation density, the matter is mostly made of nuclei in the bubble or 
uniform states. In the case of $Y_p=0.5$, the free protons coexist with symmetric light nuclei and neutron-rich heavy nuclei at 
$\rho_B \sim 10^{10}$g/cm$^3$ and only symmetric heavy nuclei survive at larger densities. For $Y_p=0.3, 0.1$, on the other hand, 
most of free protons are contained in nuclei and free neutrons coexist with heavy nuclei. In the case of $Y_p=0.1$, in particular, 
there is a density regime near the nuclear saturation density, where very neutron-rich light nuclei such as the one with $(Z, A) = (1, 10)$ 
exist in the pasta or uniform states as we mention later again. Fig.~\ref{cmp5} shows the matter compositions at a bit higher 
temperature, $T=5$MeV. It is evident for all the three values of $Y_p$ that free nucleons are dominant up to higher densities, 
$\rho_B \sim 10^{12}$g/cm$^3$. It is mentioned incidentally that the mass fractions of light nuclei in our EOS are larger in general 
than those of alpha particles in the H. Shen's EOS. This is partly because the H. Shen's EOS has only alpha particles as light nuclei 
whereas ours takes into account other nuclei such as deuterons. The difference somewhat affects the thermodynamical properties. 
For example, the entropy per baryon in our EOS is larger than that in the H. Shen's EOS at $\rho_B \sim 10^{12}$g/cm$^3$ and 
$T \gtrsim 5$MeV owing to larger degrees of freedom in the light nuclei in the former as shown in \ref{secreth}. 

\begin{figure}
\begin{center}
\begin{tabular}{ll}
   \plottwo{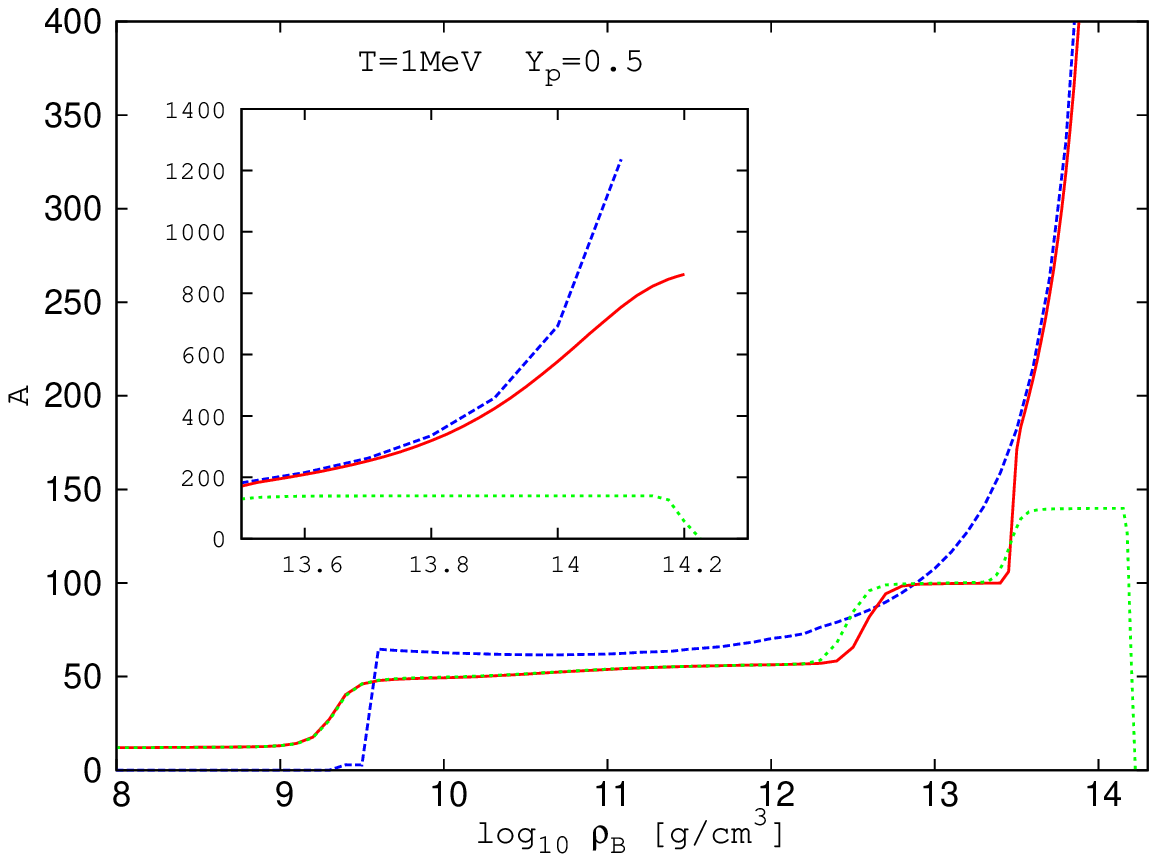}{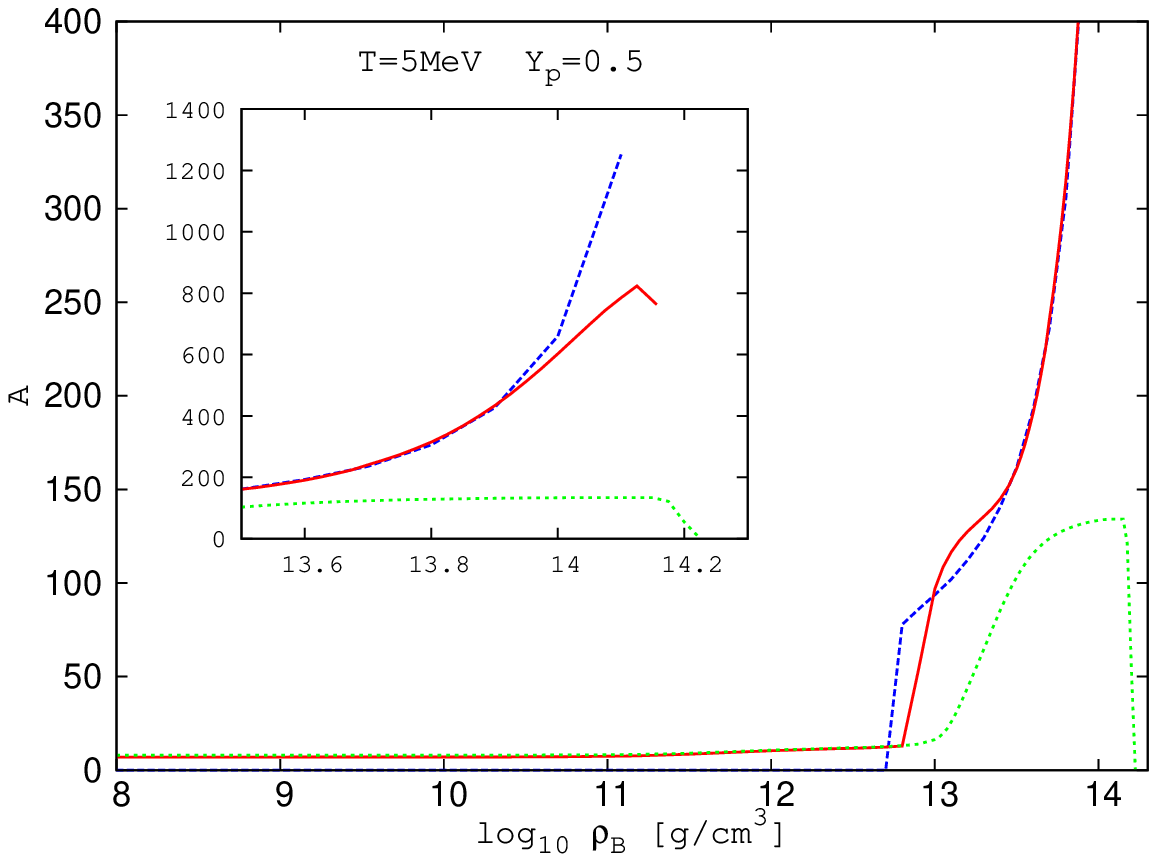} \\
   \plottwo{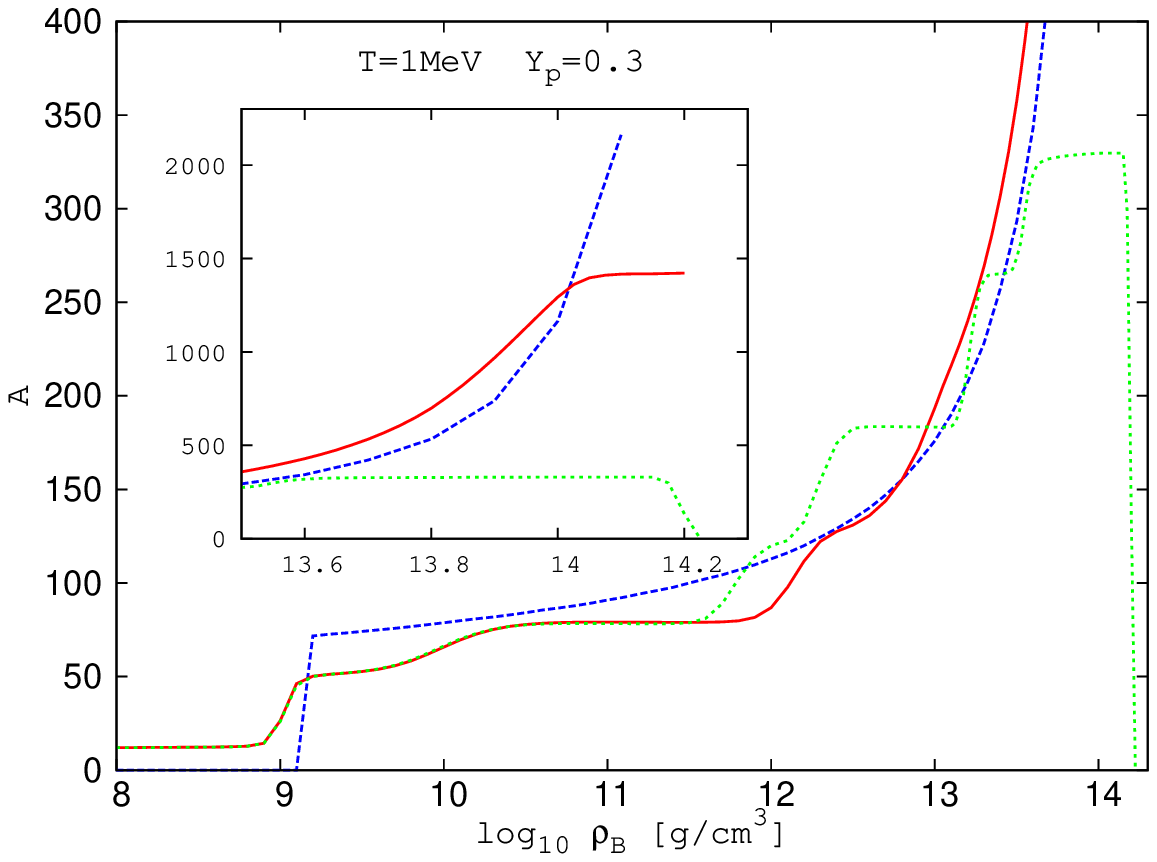}{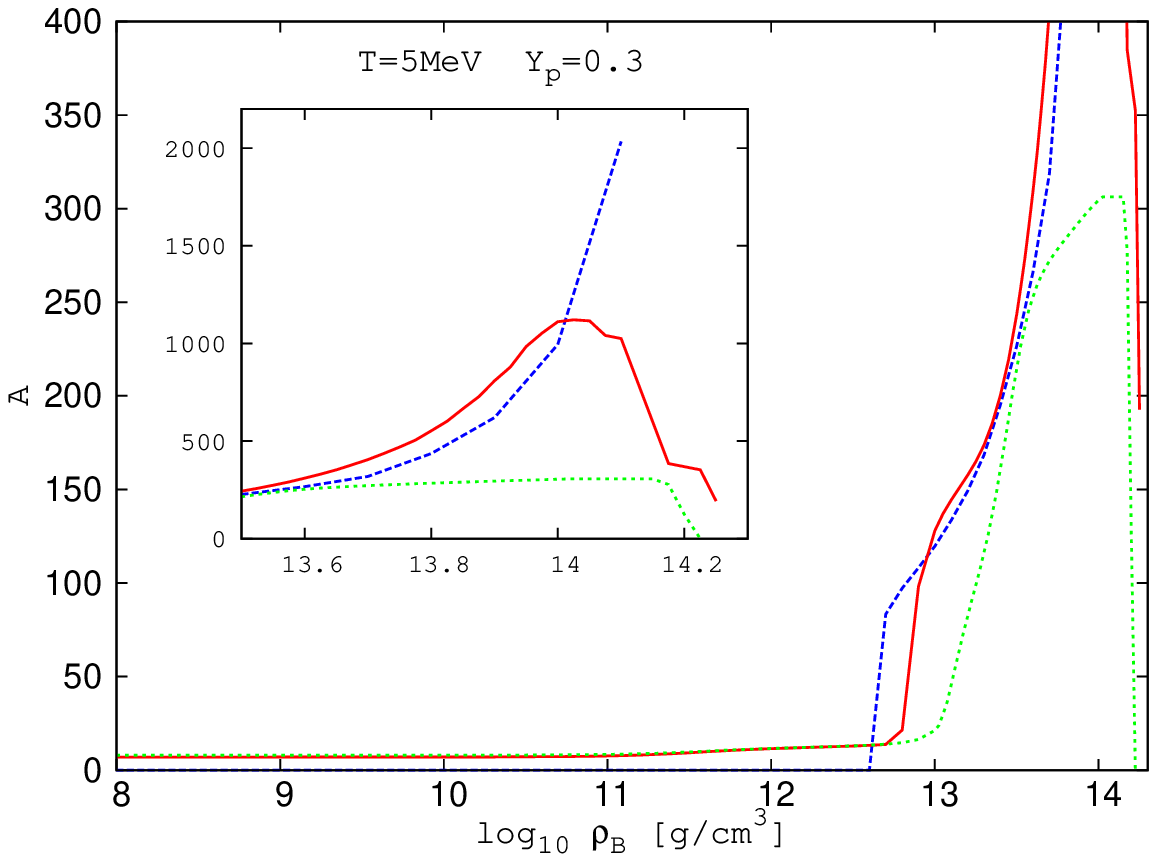} \\
   \plottwo{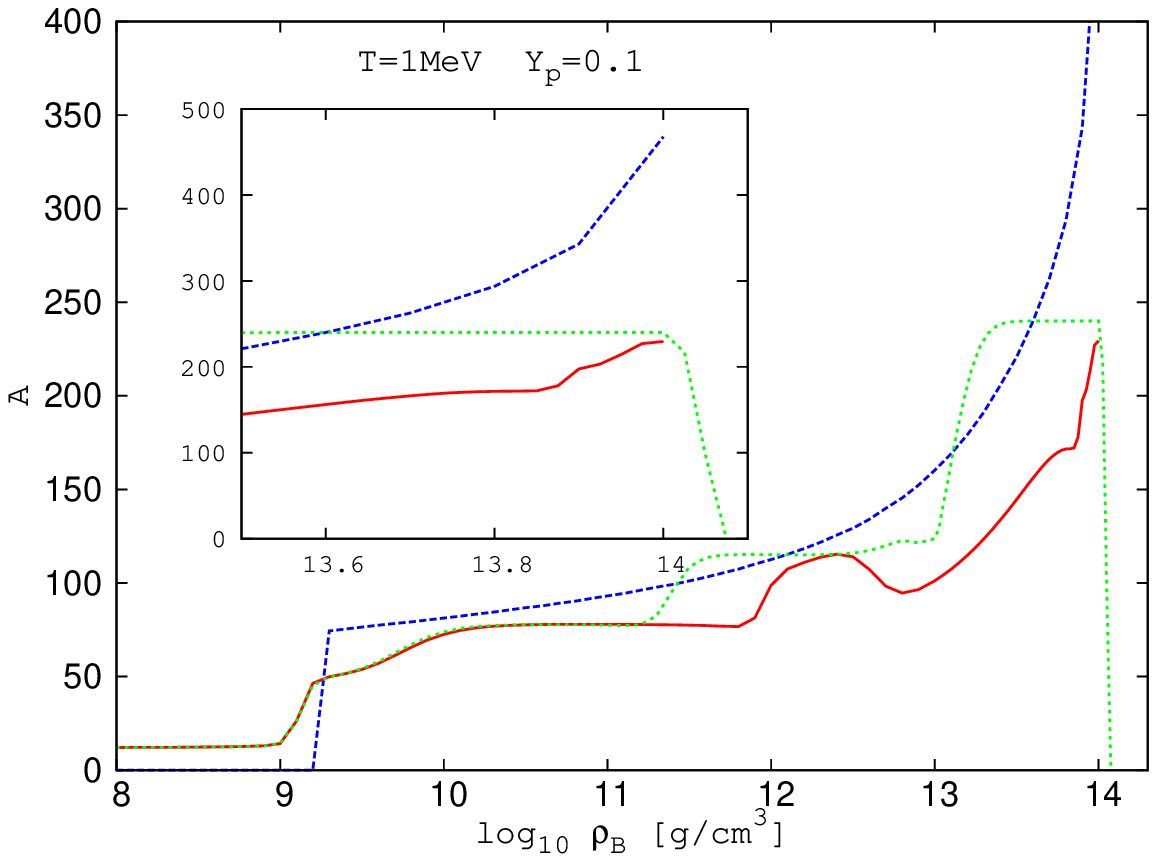}{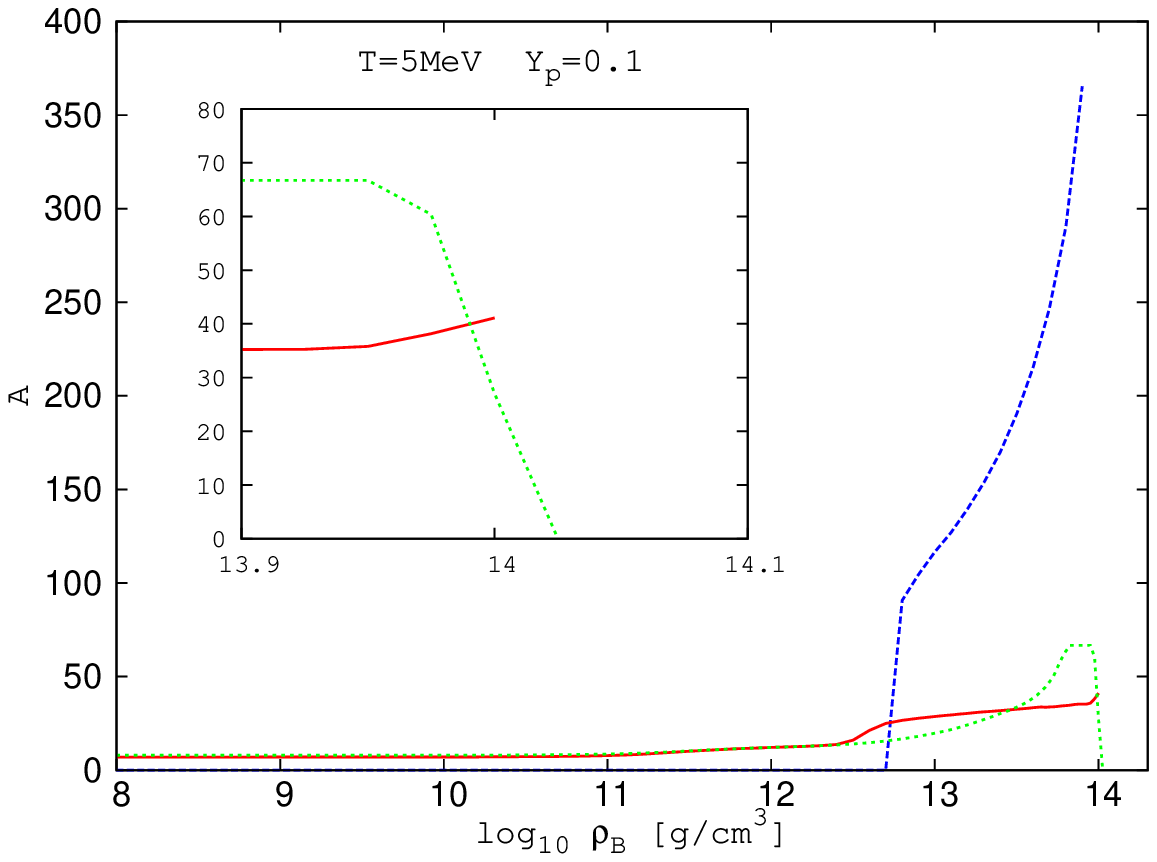} \\
    \end{tabular}
\end{center}
\caption{The average mass number, $\bar{A}$, of heavy nuclei with $Z \geq 6$ for our EOS (solid red lines) and 
the Hempel's EOS (dotted green lines) together with 
the mass number of representative nucleus for the H. Shen's EOS (dashed blue liens) as a function of density for $T=1$MeV and 
$Y_p=0.1, 0.3, 0.5$. The insets are the close-ups of the high density regimes.}
\label{ma1}
\end{figure}

The average mass number, $\bar{A}$, of heavy ($Z_i\geq 6$) nuclei is displayed in Fig.~\ref{ma1}. For comparison, not only our own 
results but those of Hempel's and the mass number of the representative nuclei, $A_{rep}$, for the H. Shen's EOS are also shown. 
It is found that for $T=1$MeV (the left column of the figure), the former two give the average mass numbers that are systematically 
smaller than the H. Shen's EOS at $\rho_B \lesssim 10^{12}$g/cm$^3$, around which the supernova core becomes optically thick and 
neutrinos are trapped and equilibrated thermally and chemically with the matter. It is also clear that the average mass number grows 
step-wise for our and Hempel's EOS's whereas it is monotonic for the H. Shen's EOS. 
This is simply due to the shell effect: nuclei are abundant in the vicinity of the neutron magic numbers ($N=28, 50, 80$). At higher 
densities, $\rho_B \gtrsim 10^{12}$g/cm$^3$, the effect is suppressed in our EOS, as mentioned already, because the nuclear bulk energy is 
interpolated from the experimentally derived value and the one obtained theoretically by RMF, which does not include the shell effect. 
Note also that at these densities there appear very heavy and/or neutron-rich nuclei, for which no experimental mass data are available 
and we employ RMF for the estimation of their bulk energies (see section~\ref{subsecb}). 
Heavy nuclei continue to grow up until the nuclear saturation density in our EOS, whereas the growth stops at some density and 
nuclei disappear before the saturation density in the Hempel's EOS (see the insets of the figure), which is at odds with the 
conventional idea. This point will be mentioned shortly again. 

The discernible difference between our results and the Hempel's at $\rho_B \sim 10^{12}$g/cm$^3$ for $Y_p=0.1, 0.3$ is attributed to the 
difference in the masses of neutron-rich nuclei, which we calculate by RMF whereas \citet{Hempel2010} obtained from the 
mass table by \citet{Geng2005}. 
The dip in the mass number at $\rho_B \sim 10^{13} \rm{g/cm^3}$ for $Y_p=0.1$ in our EOS is artificial, being produced by the neglect of 
the shell effect in our mass formula. As nuclei become very neutron-rich, there happens a rather rapid shift of abundant nuclei 
from those with experimental data to those without, resulting in the observed change of the mass number.
The mass number does not increase until higher densities are reached. As shown in the right column of Fig.~\ref{ma1} for $T=5$MeV, 
for example, the mass number starts to increase at $\sim 10^{13}$g/cm$^3$. This is because the increase of entropy by the translational 
degrees of freedom of free nucleons is effective to minimize the free energy at this high temperature, whereas the decrease of internal 
energy by the binding energy of nuclei is more advantageous at lower temperatures. Another interesting feature at this temperature is the 
decrease of the mass number near the nuclear saturation density for $Y_p=0.3, 0.5$ even in our EOS. There are two reasons for this: one is 
the formation of bubble nuclei and the reduction of surface energy as a result; the other reason is that the less heavy nuclei become, 
the greater entropy they have thanks to their larger translational degrees of freedom and they lower the free energy. It should be noted 
that unlike in the Hempel's EOS, nuclei are not dissociated completely in our EOS. 

\begin{figure}
\begin{center}
\begin{tabular}{ll}
  \resizebox{52mm}{!}{\plotone{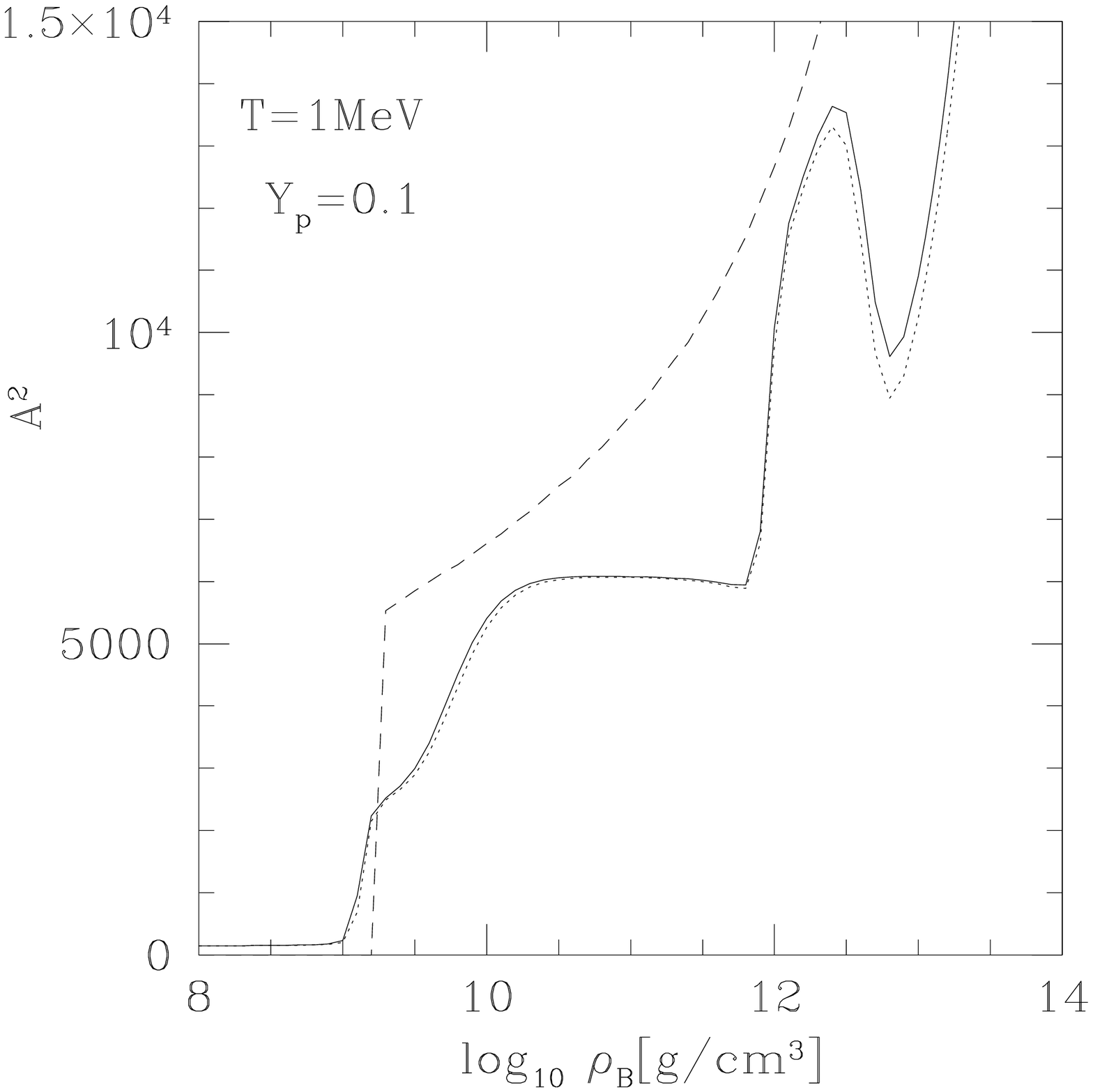}}   \resizebox{52mm}{!}{\plotone{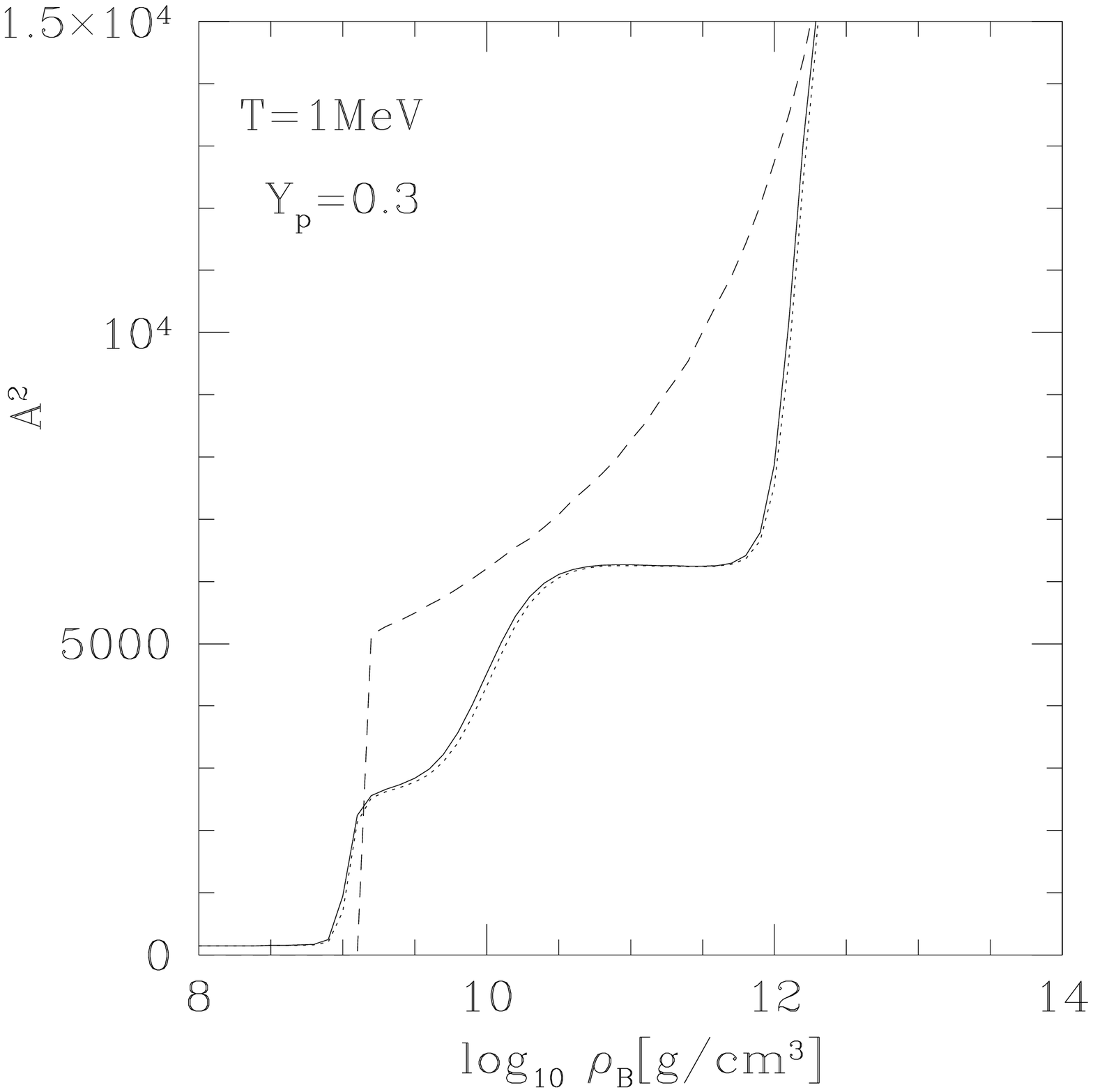}}   \resizebox{52mm}{!}{\plotone{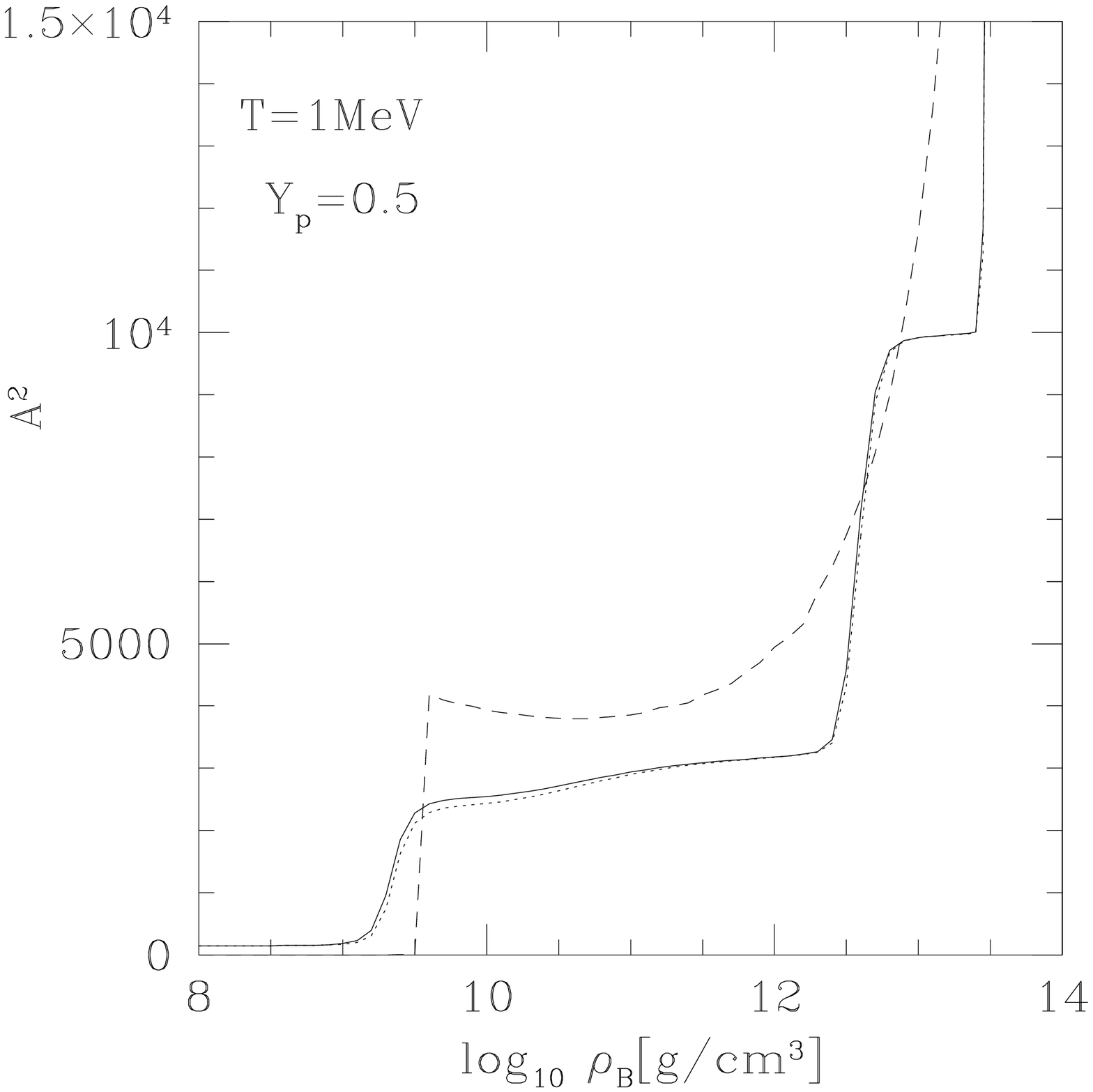}} \\
  \resizebox{52mm}{!}{\plotone{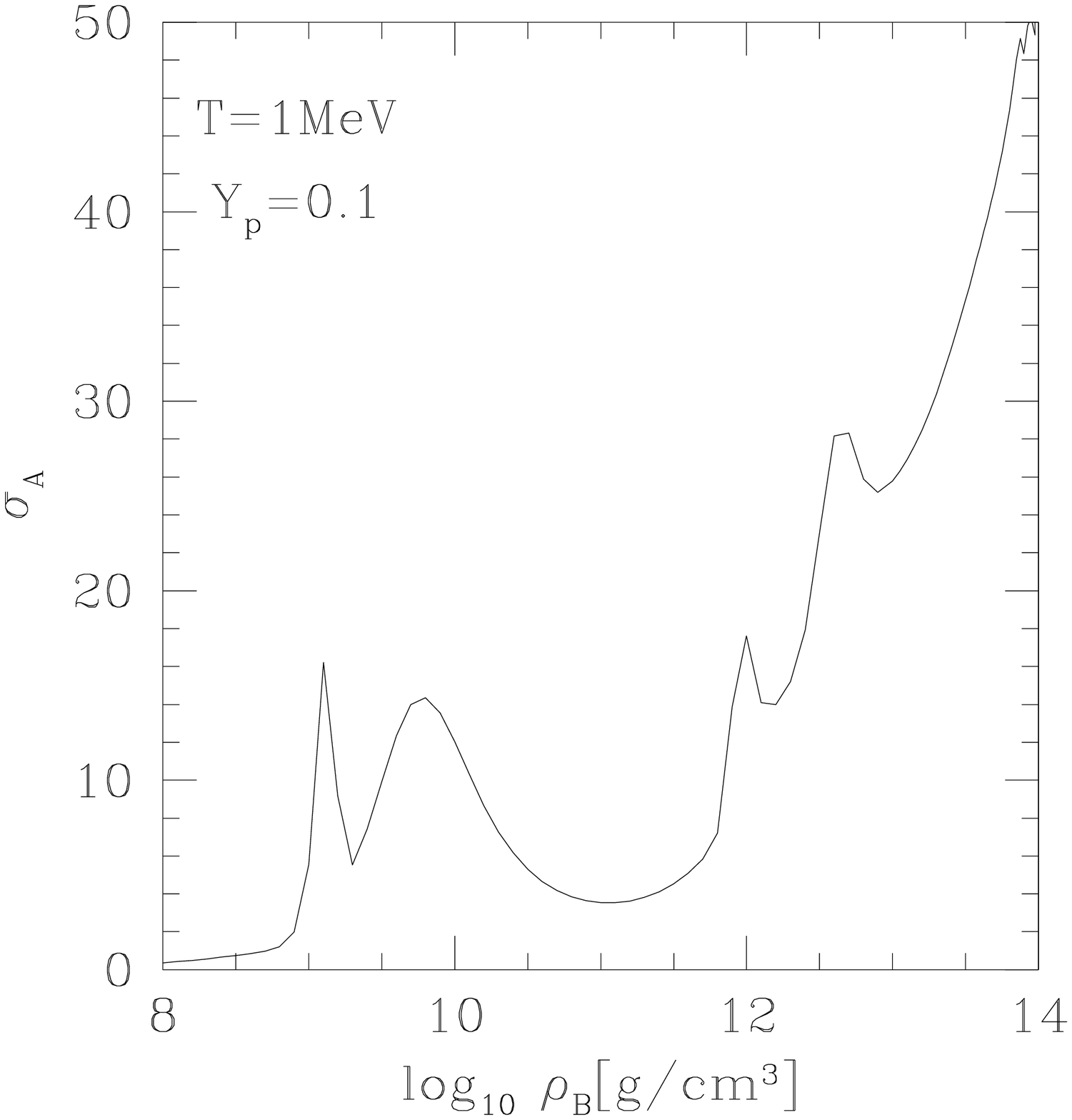}}   \resizebox{52mm}{!}{\plotone{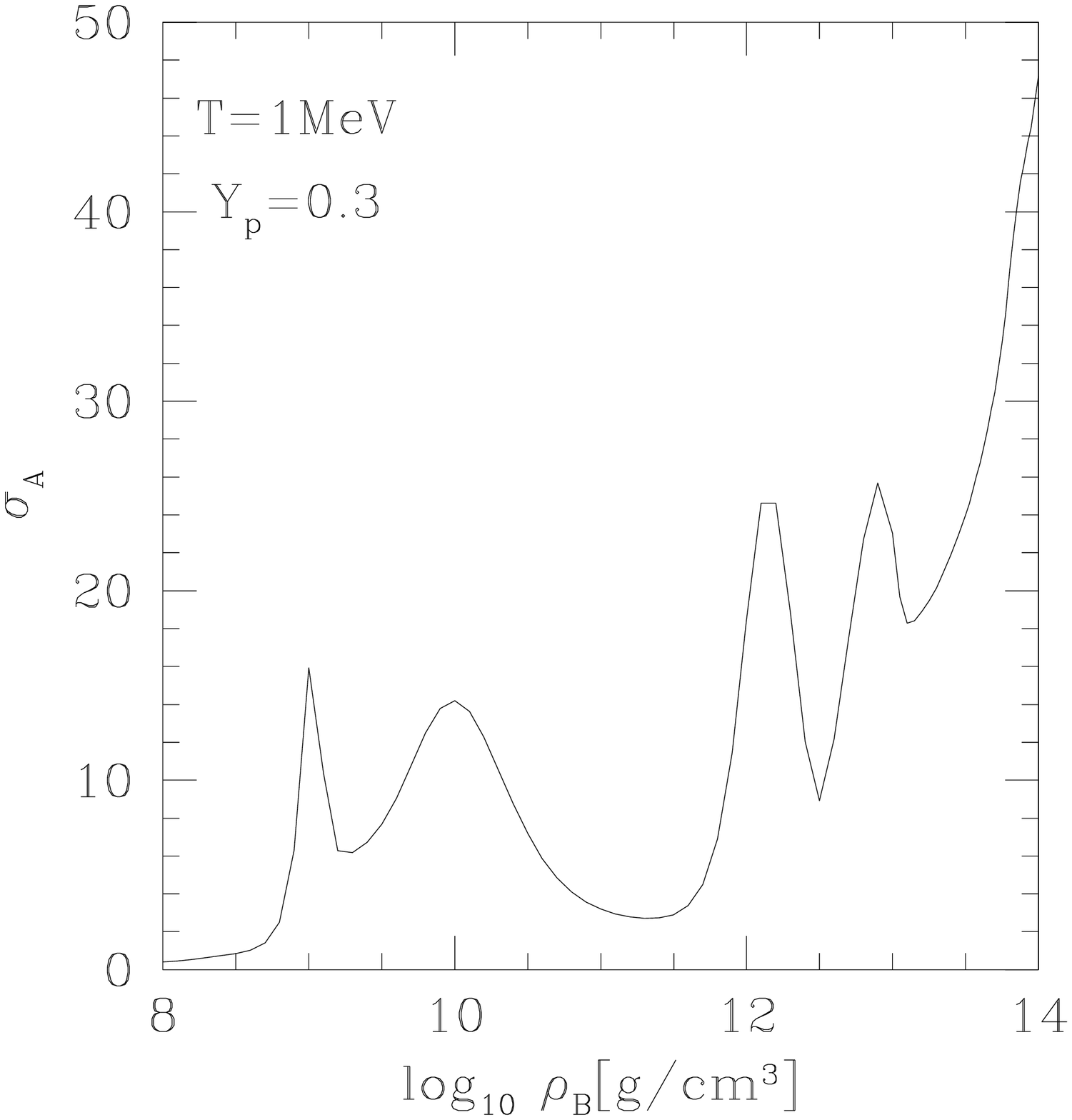}}   \resizebox{52mm}{!}{\plotone{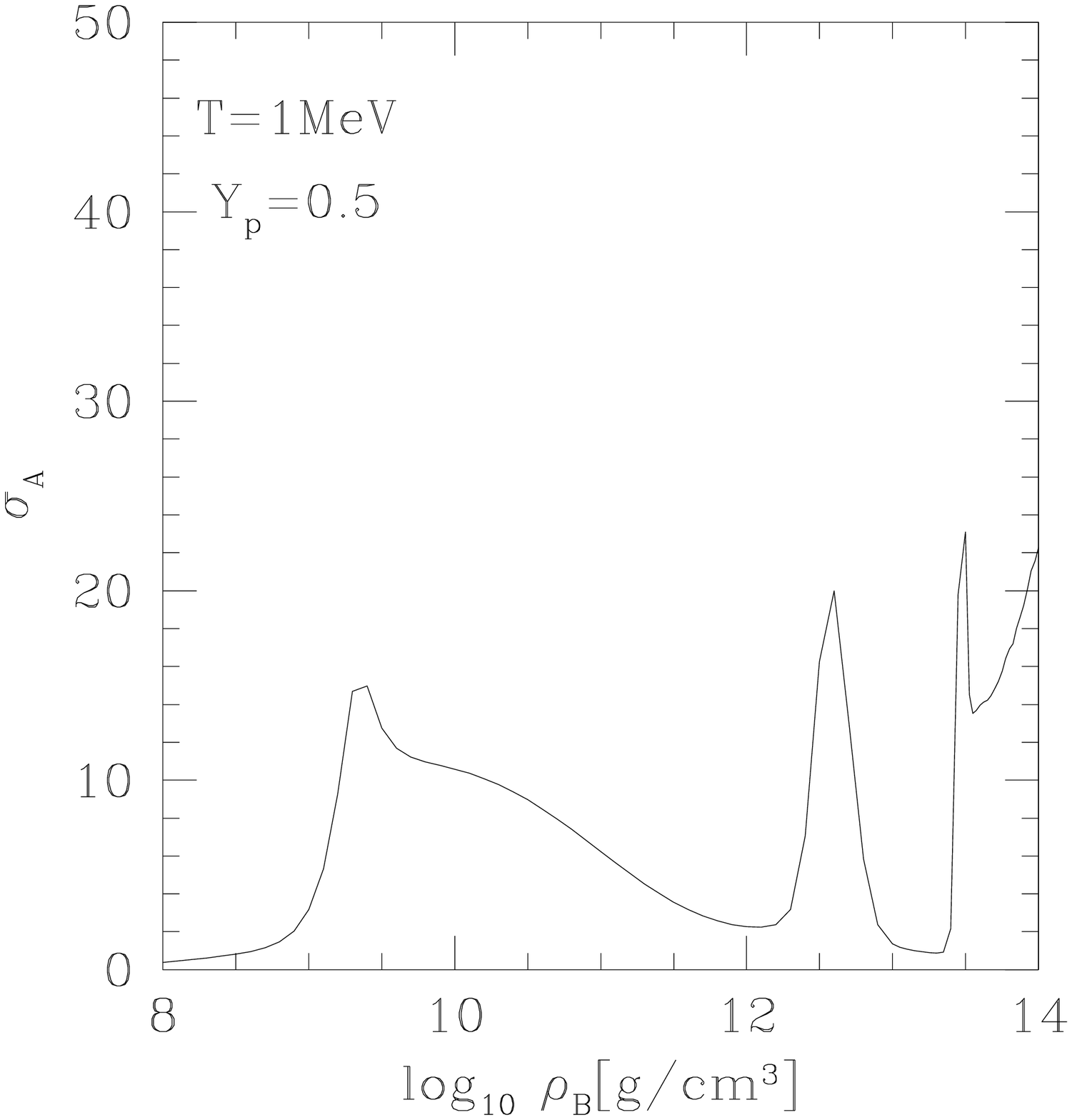}} \\
  \resizebox{52mm}{!}{\plotone{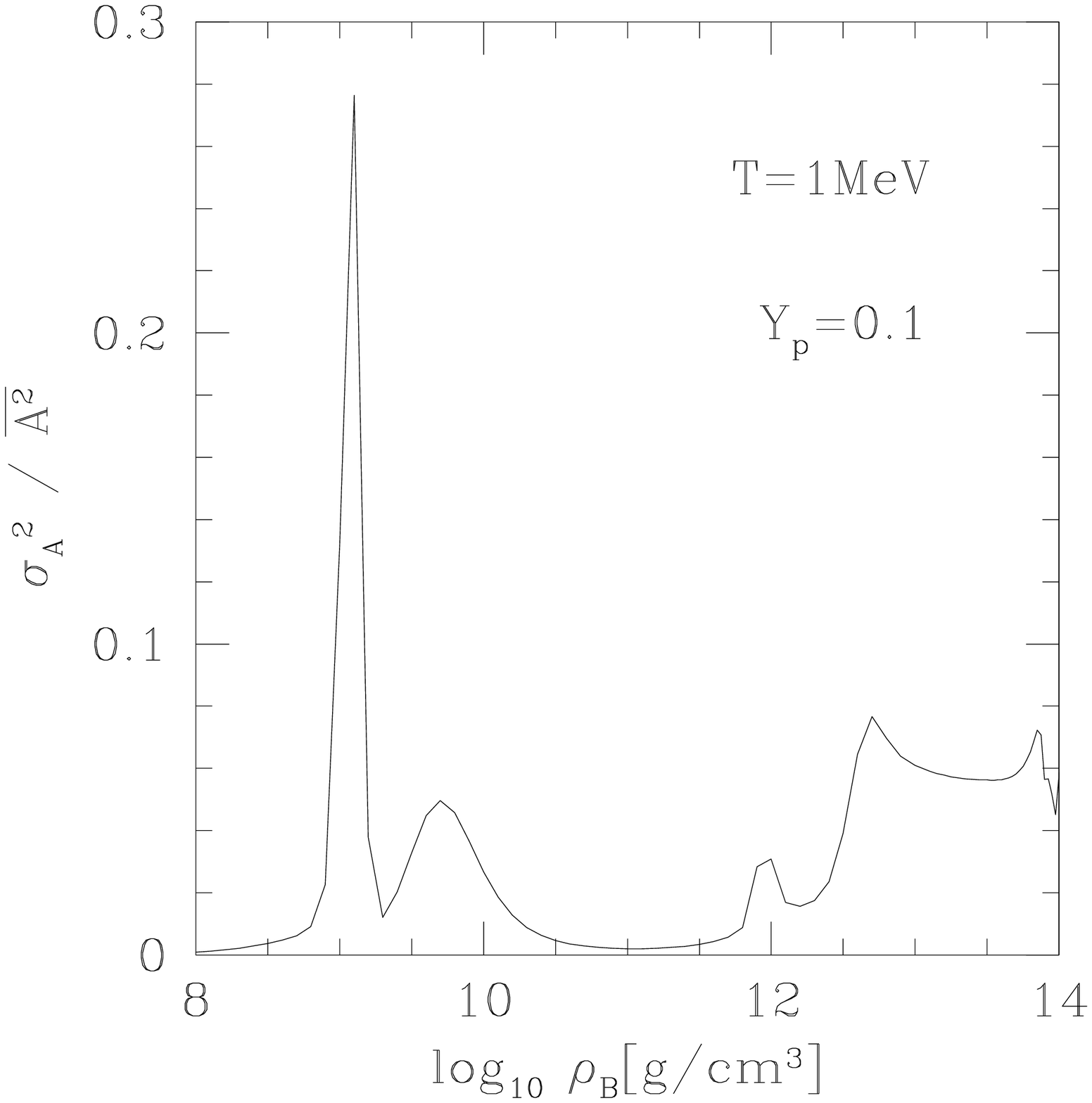}}   \resizebox{52mm}{!}{\plotone{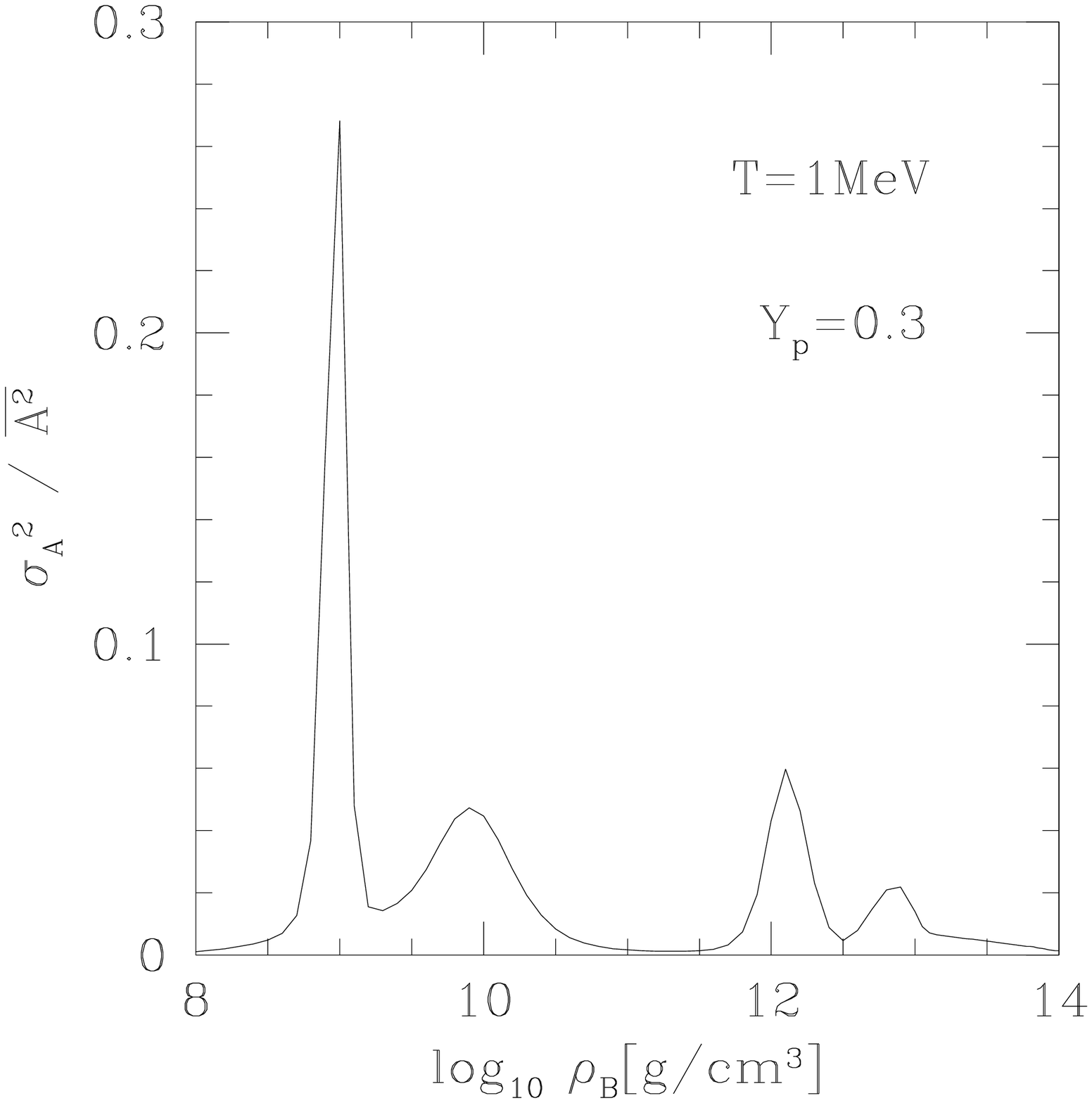}}   \resizebox{52mm}{!}{\plotone{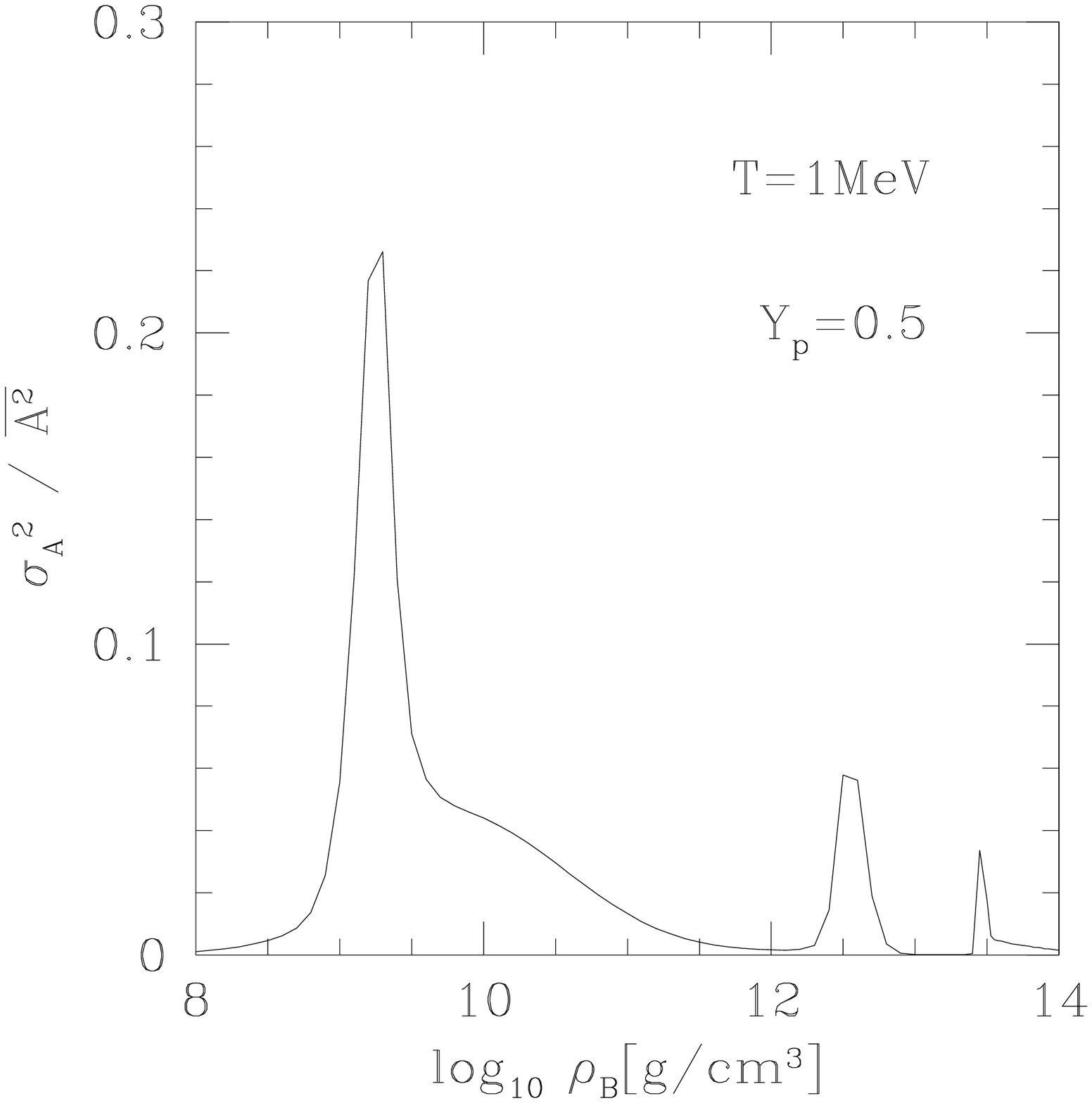}} \\
    \end{tabular}
\end{center}
\caption{The square of mass numbers (top), the standard deviation of mass number, $\sigma_A=\sqrt{\overline{A^2}-\bar{A}^2}$, (middle) 
and the dispersion normalized  by the average mass number squared, $\sigma_A^2 / \overline{A^2}$, (bottom) of heavy nuclei with $Z \geq 6$ for $T=1$MeV and
$Y_p=0.1\ ({\rm left}), 0.3\ ({\rm middle})\ {\rm and}\ 0.5\ ({\rm right})$. In the top panels, the solid and dotted lines show 
the average mass number squared, $\overline{A^2}$, and the square of average mass number, $\bar{A}^2$, in our EOS, respectively, whereas the 
dashed lines display the mass number squared of the representative nucleus for the H. Shen's EOS.}
\label{aadis}
\end{figure}

The rate of coherent scatterings of neutrinos on nuclei is proportional not to the mass number of nuclei but to its square as long as
the wave length of neutrino is much larger than the nuclear size \citep{Burrows2003}. As mentioned already, in SNA the rate is
evaluated by the mass number squared for the representative nucleus, the validity of which needs confirmation. In the top panels of 
Fig.~\ref{aadis} we compare the average mass number squared, $\overline{A^2 }$, with the square of the average mass number, $\bar{A}^2$,
and the mass number squared of the representative nuclei in the H. Shen's EOS. In the middle and bottom panels, the standard deviations, $\sigma_A=\sqrt{\overline{A^2}-\bar{A}^2}$, and the dispersion normalized by the average mass number squared, $\sigma_A^2/\overline{A^2 }$, are also shown. It is intriguing to find that the average mass number squared is almost indistinguishable from 
the square of the average mass number. The reason for this small dispersion is that heavy nuclei are rather concentrated in the 
vicinities of the neutron magic numbers. In this sense, SNA is not so bad an approximation in evaluating the rate of coherent 
scatterings of neutrinos on nuclei. It should be noted, however, that the mass numbers of the representative nuclei in the H. Shen's EOS
are systematically larger than the average mass numbers obtained in our EOS. Incidentally, at very high densities 
($\rho_B \gtrsim 10^{14}$g/cm$^3$), most of nuclei are in the pasta phase.

\begin{figure}
\begin{center}
\begin{tabular}{ll}
               \resizebox{61mm}{!}{\plotone{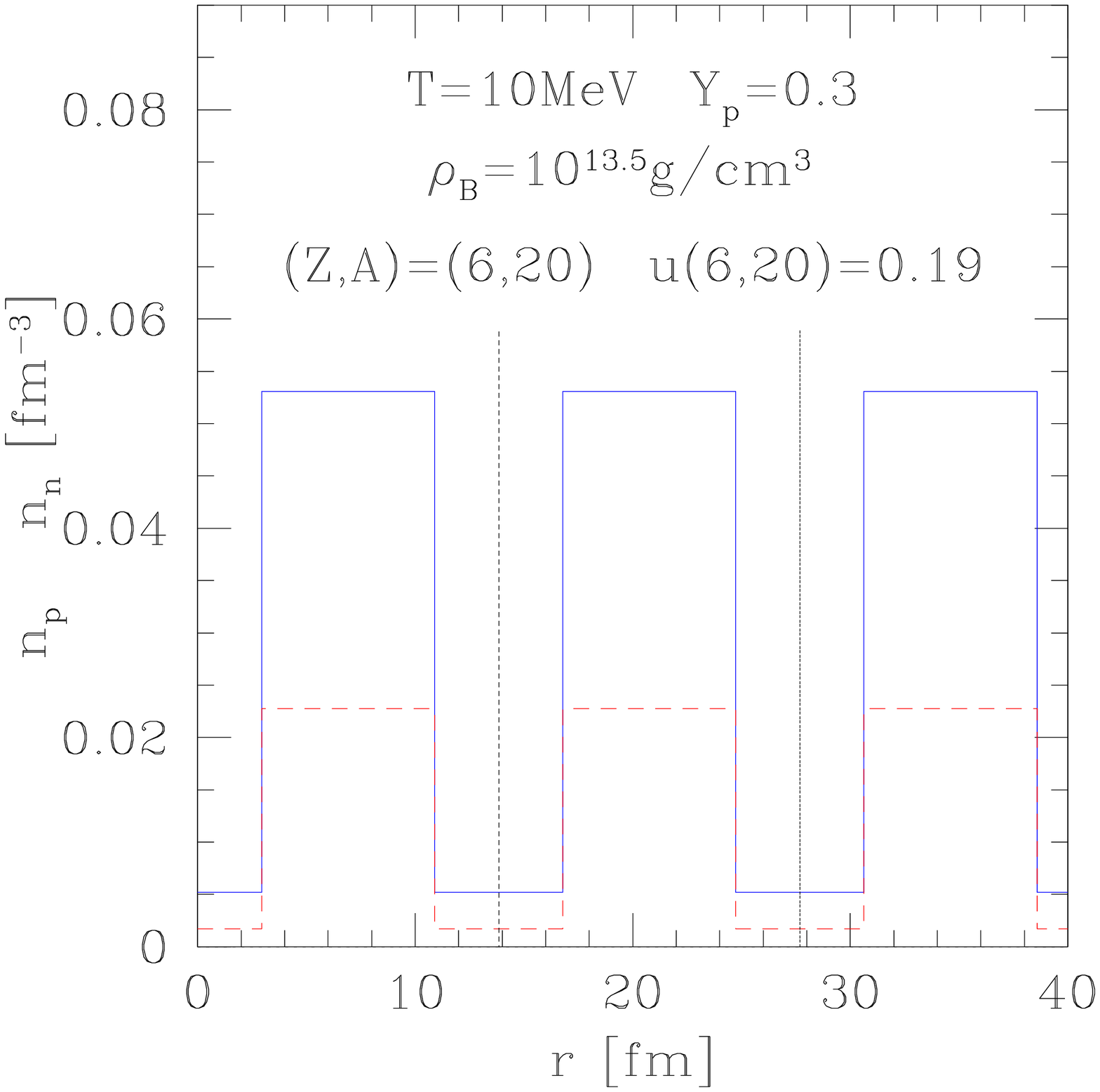}} \\
               \resizebox{61mm}{!}{\plotone{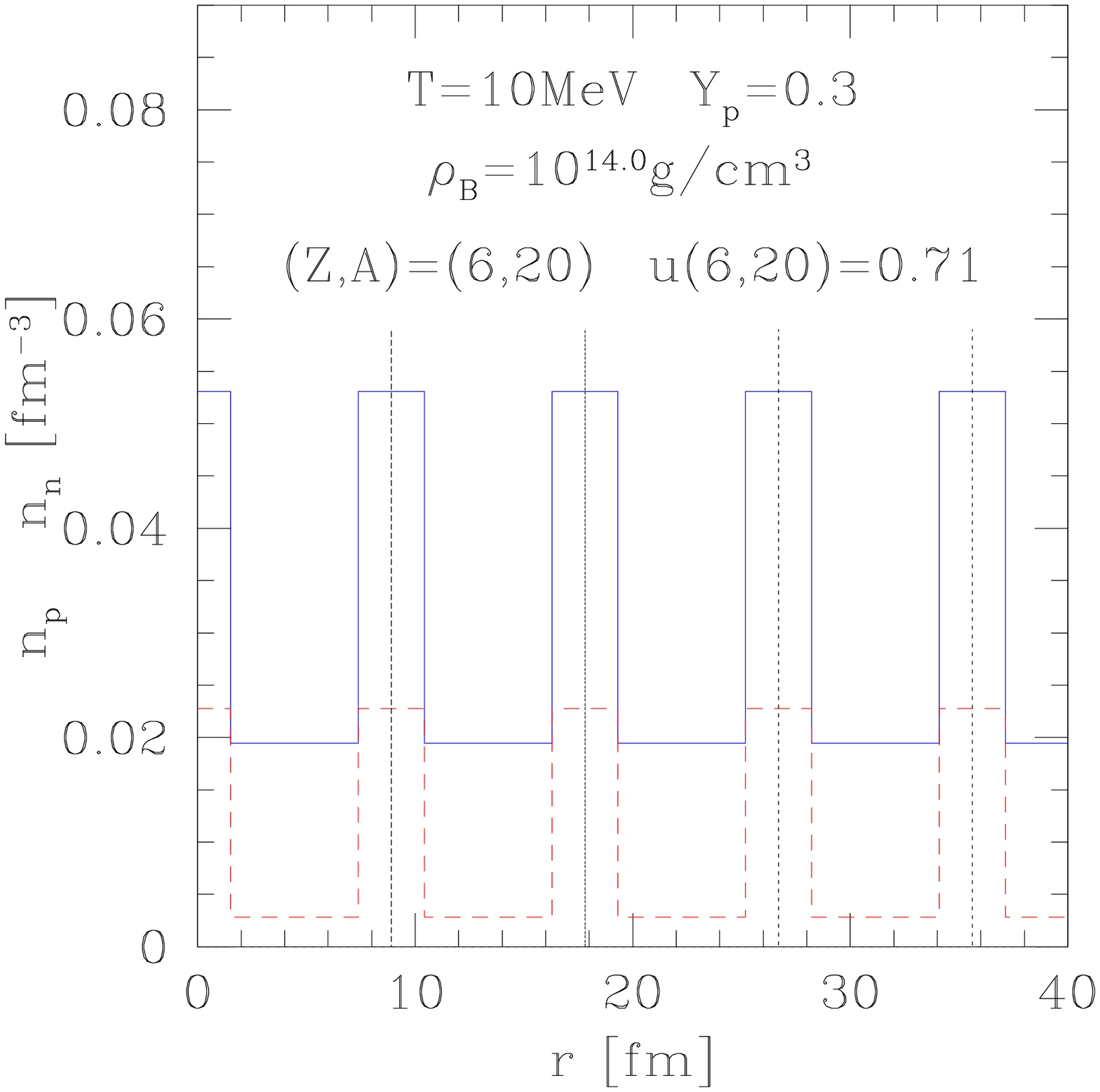}} \\
               \resizebox{61mm}{!}{\plotone{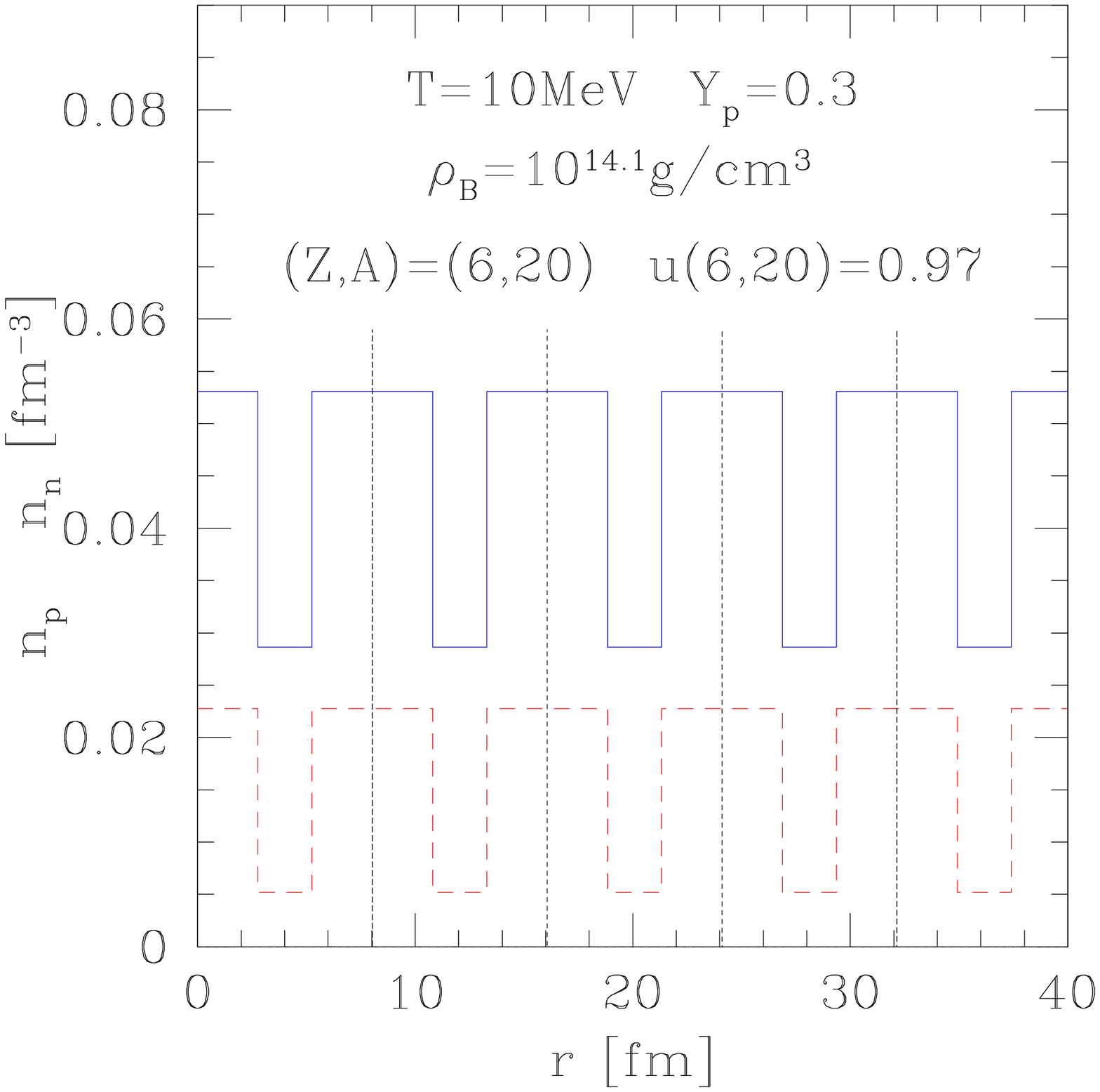}} \\
    \end{tabular}
\end{center}
\caption{Schematic pictures of the number densities of proton (dashed red lines) and neutron (solid blue lines) in the W-S cells for 
the nucleus with $(Z, A) = (6, 20)$ at $T=10$MeV, $Y_p=0.3$ and $\rho_B = 13.5, 14.0, 14.1$g/cm$^3$ from top to bottom. 
The dotted black lines indicate the cell boundaries. }
\label{pc1}
\end{figure}

We demonstrate here that the transition to uniform nuclear matter occurs in one of the following two ways depending on the proton 
fraction of the entire system, $Y_p$.
The first case corresponds to relatively large proton fractions, $Y_p \gtrsim 0.3$. At $\rho_B  \gtrsim 10^{13}$g/cm$^3$,
most of the nuclei in abundance have a similar proton fraction $Z_i/A_i$, which is close to $Y_p$ as shown in Fig.~\ref{dis13}.
As the density rises up to the nuclear saturation density, their W-S cells get smaller and these abundant nuclei enter the pasta phase almost 
simultaneously. In the bubble phase, the volumes occupied by the nuclei in their W-S cells become larger and 
finally the nuclei coalesce with each other. The sequence is schematically illustrated for the nucleus with $(Z, A)=(6,20)$ 
in Fig.~\ref{pc1}. Note that the density in the vapor, or the bubble, remains lower than that in the nuclei up to the saturation 
density in this case. 

\begin{figure}
\begin{center}
\begin{tabular}{ll} 
\epsscale{.82}
     \plottwo{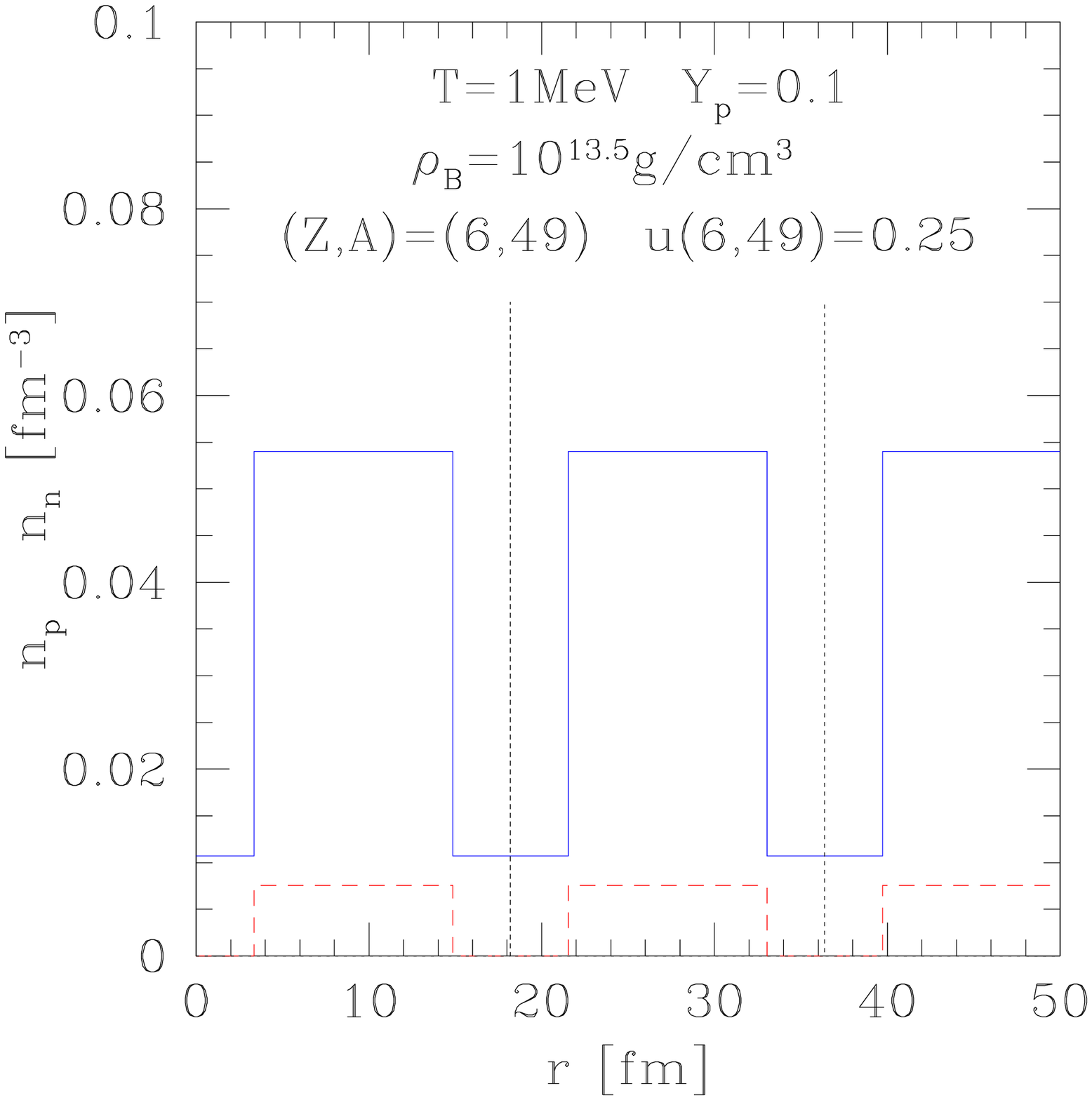}{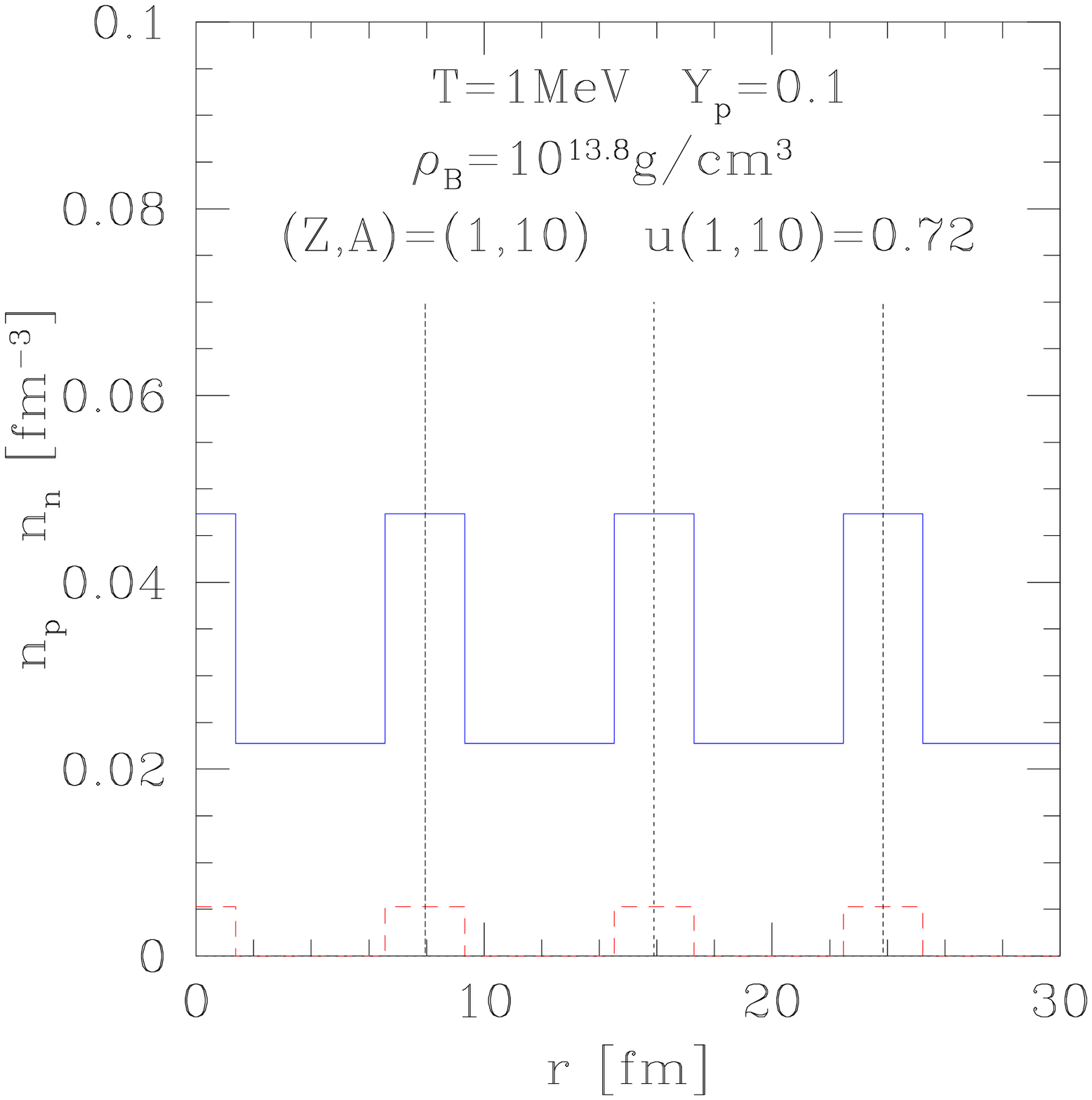} \\
      \plottwo{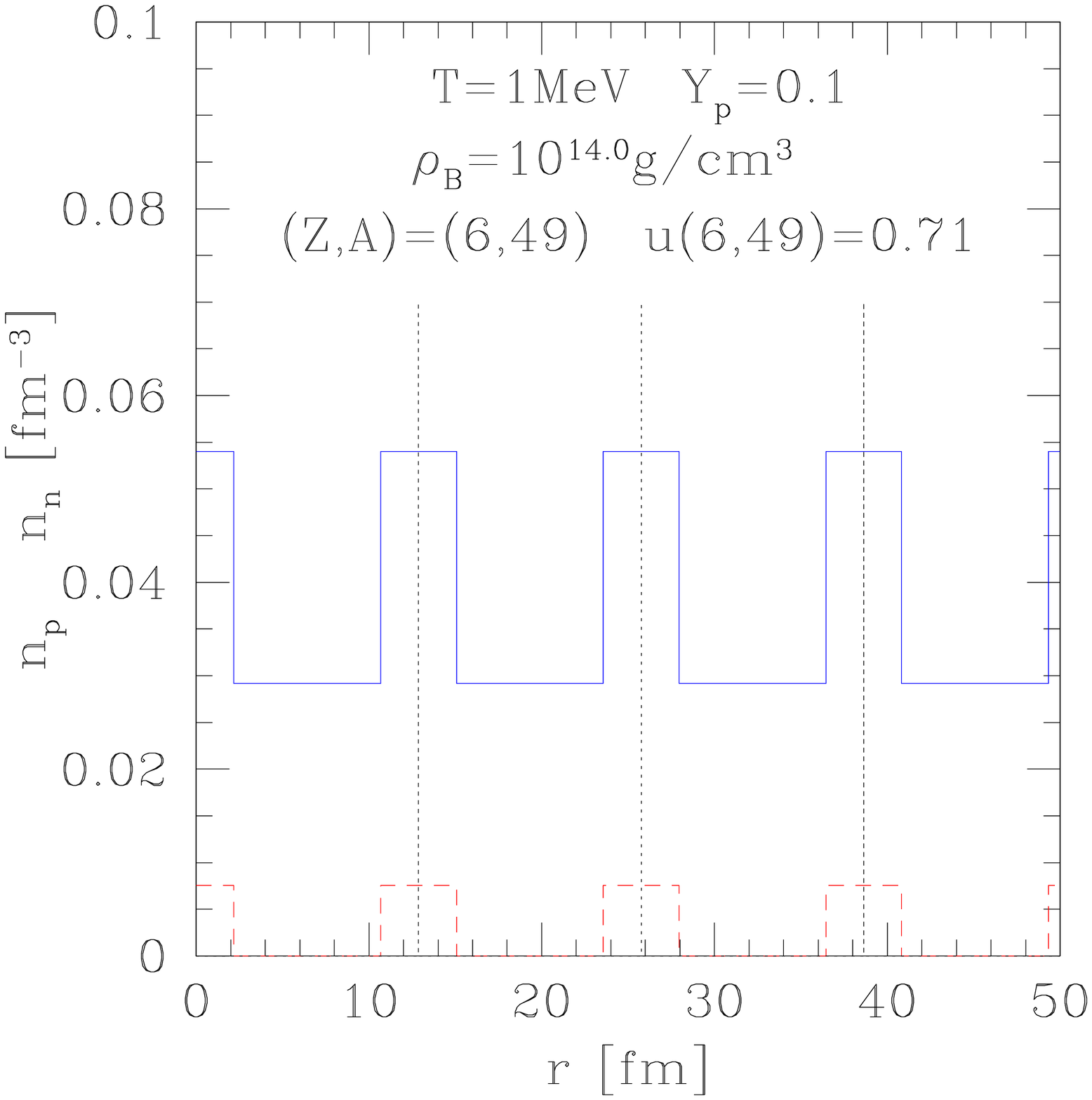}{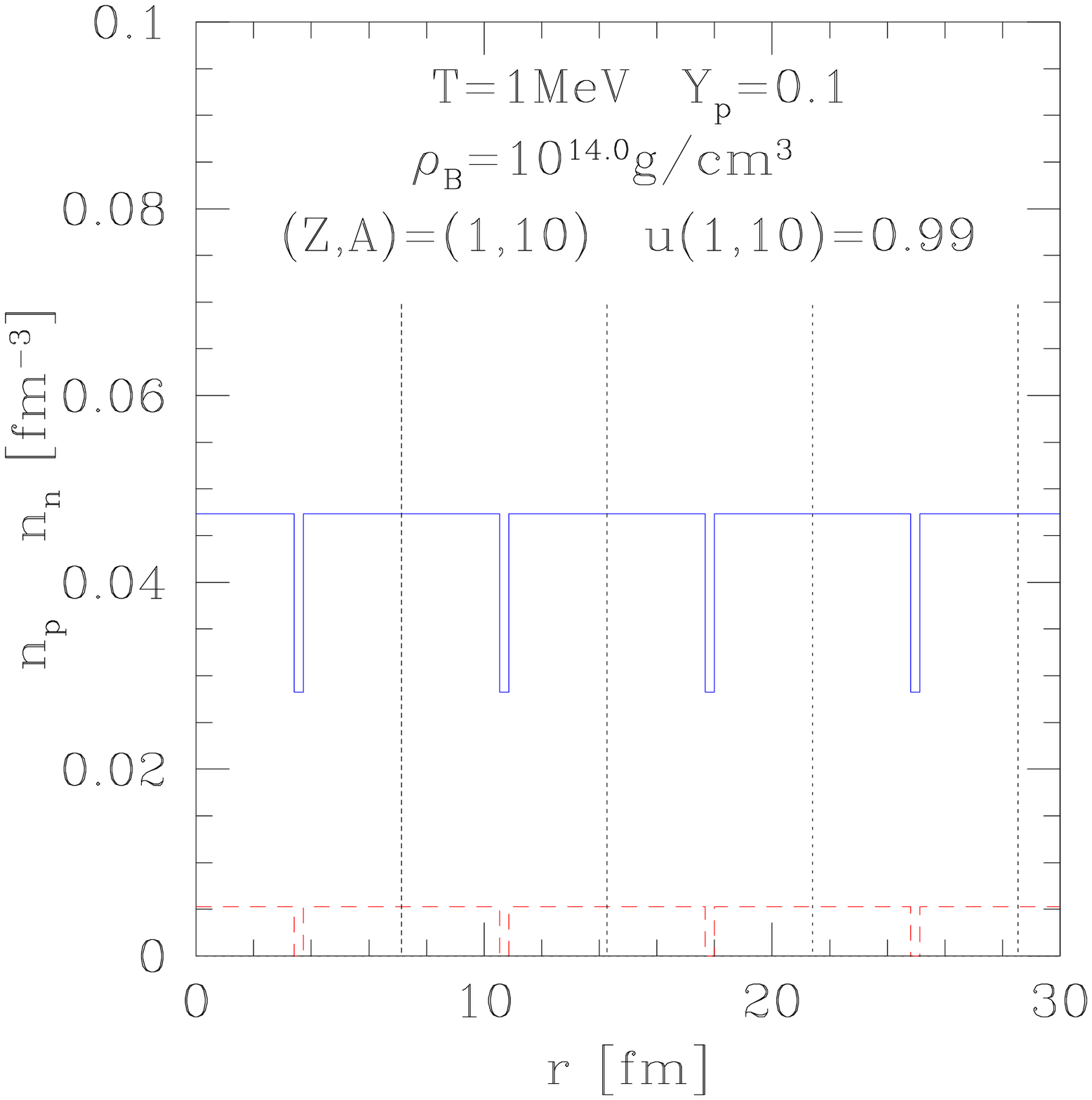} \\
      \plottwo{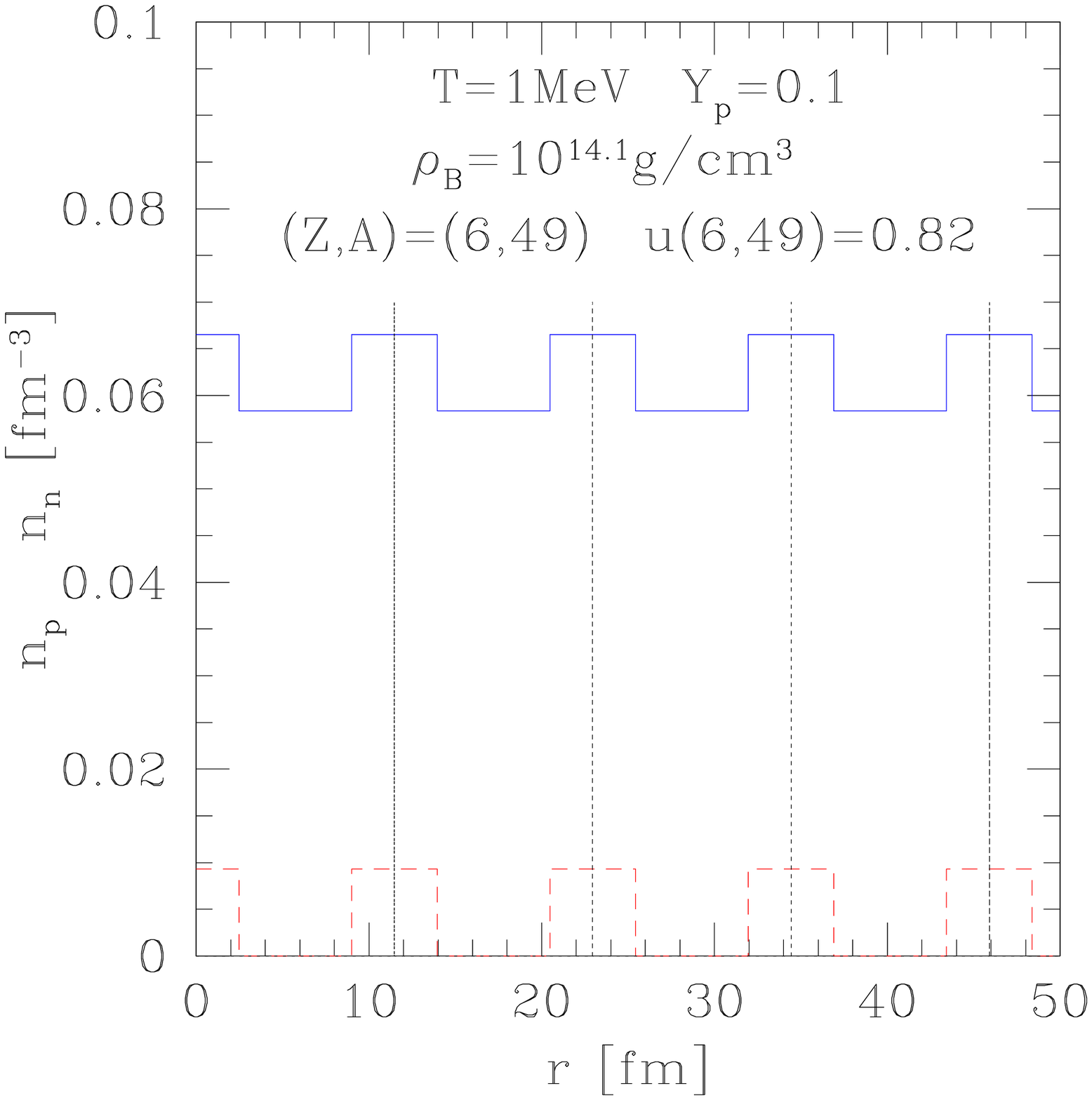}{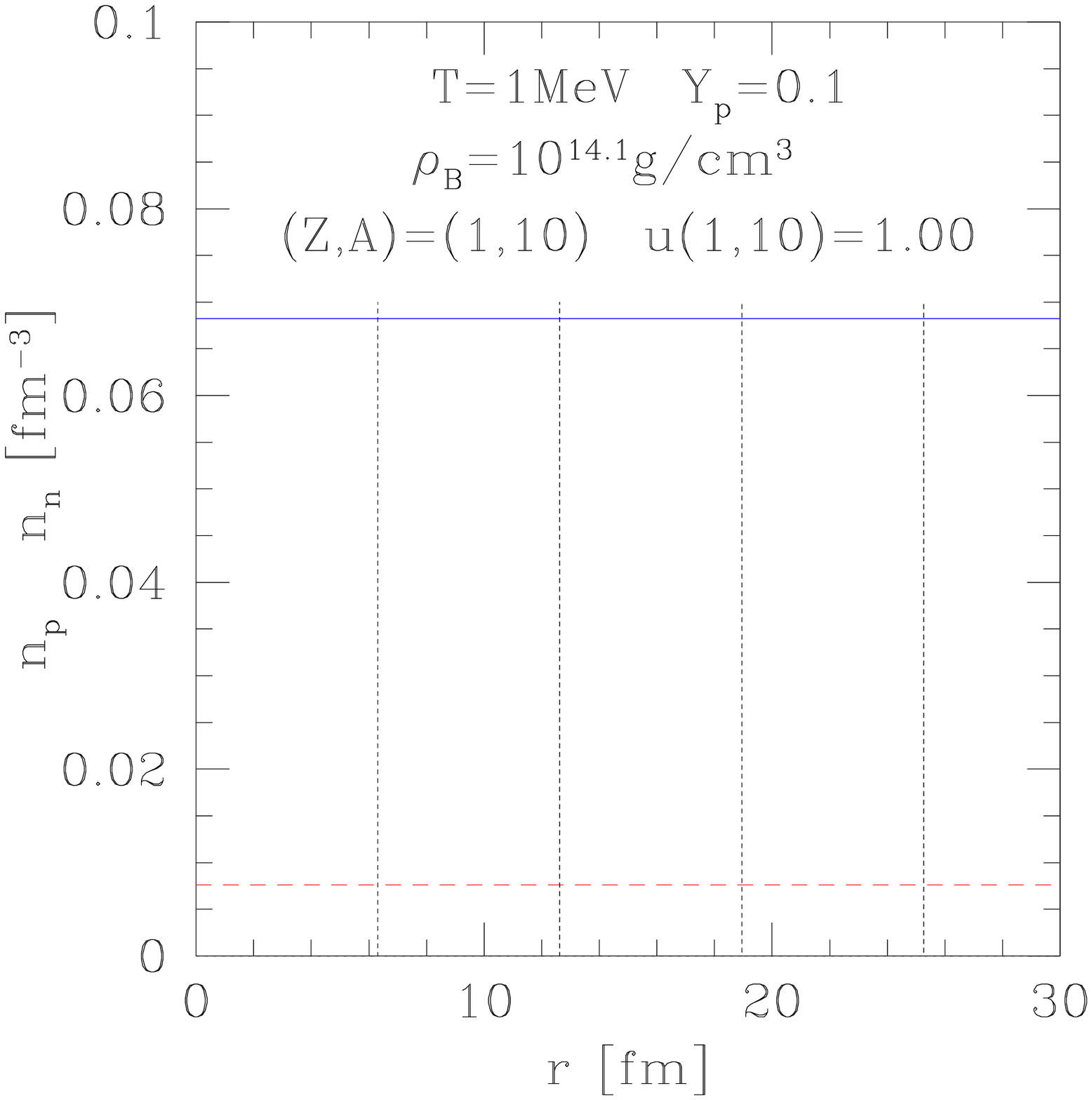} \\
    \end{tabular}
\end{center}
\caption{Schematic pictures of the number densities of proton (dashed red lines) and neutron (solid blue lines) in the W-S cells for 
the nuclei with $(Z, A) = (6, 49)$ (left) and $(Z, A) = (1, 10)$ (right) at $T=1$MeV, $Y_p=0.1$ and $\rho_B = 13.5({\rm left}), 
13.8({\rm right}), 14.0, 14.1$g/cm$^3$ 
from top to bottom. The dotted black lines indicate the cell boundaries. }
\label{pc2}
\end{figure}

The second case occurs for more neutron-rich systems, $Y_p \lesssim 0.3$. The proton fractions, $Z_i/A_i$, of nuclei 
having large mass fractions are not necessarily close to $Y_p$ of the entire system and take various values at $\rho_B \gtrsim 10^{13}$g/cm$^3$. 
This is due to the fact that more symmetric nuclei have greater binding energies. Hence the nuclei are less neutron-rich than the vapor 
in general. As a result of this diversity of the proton fractions of the nuclei in abundance, these nuclei reach the pasta phase at 
different densities. Put another way, nuclei of different shapes coexist with substantial mass fractions at a particular density 
above $\rho_B \sim 10^{13}$g/cm$^3$. For example, at $Y_p=0.1$, $\rho_B = 10^{13.9}$g/cm$^3$ and $T=1$MeV, a very heavy nucleus 
with $(Z, A) = (48, 228)$ still remains in the normal phase whereas a very neutron-rich light nucleus with $(Z, A) = (1, 10)$ as well as
a neutron-rich, slightly heavy nucleus with $(Z, A)=(6,49)$ are already in the bubble phase. This light nucleus becomes uniform in the
W-S cell already at $\rho_B = 10^{14.1}$g/cm$^3$ whereas the others are still non-uniform. This actually corresponds to the formation 
of a big nucleus with no bubbles inside, which coexists with other bubble nuclei. As the density increases further beyond the nuclear
saturation density up to the point, where the entire system becomes uniform, the density of this big nucleus also increases, which 
we referred to as overcompression in section~\ref{secnuclei}. The reason for the occurrence of such overcompression is that the binding 
energies of nuclei that are less neutron-rich than the whole system are greater in general and the non-uniform matter is energetically 
favored still at the nuclear saturation density for the whole system.
As shown in Fig.~\ref{pc2}, the transition to uniform nuclear matter for the very neutron-rich 
light nuclei occurs by coalescence with each other just as in the first case mentioned above. On the other hand, the transition  
for the nuclei with $(Z, A) = (6, 49)$ proceeds not by coalescence but by the increase in the density of the vapor. Although
it is not shown in the figure, it is interesting that the very heavy nucleus with $(Z, A) = (48, 228)$ does not reach the bubble phase 
and is in the intermediate pasta phase just prior to the transition to uniform nuclear matter. 
\subsection{Thermodynamical Quantities \label{secreth}}
Once the minimization is completed, the baryonic contribution to the free energy, to which our discussions are limited in this paper, is
obtained as a function of $\rho_B, T, Y_p$. All the thermodynamical quantities are then derived from it through thermodynamical relations.
It turns out that most of them are not very different from the ones in the Shen's and Hempel's EOS's: for the temperatures, $T=1, 5, 10$MeV 
considered in this paper, the three EOS's are essentially identical at low densities, $\rho_B \lesssim 10^{8}$g/cm$^3$, since 
the matter is simply composed of free nucleons as an ideal Boltzmann gas in all the EOS's. At higher densities, 
$10^{8} \lesssim \rho_B \lesssim 10^{13}$g/cm$^3$, a small difference appears, since some of the free nucleons are taken into nuclei, 
the treatment of which is different among the EOS's. At still greater densities, $\rho_B \gtrsim 10^{13}$g/cm$^3$, the difference is 
more apparent. In the following we will look at the behavior more in detail.

\begin{figure}
   \begin{center}
    \begin{tabular}{c}
               \resizebox{63mm}{!}{\plotone{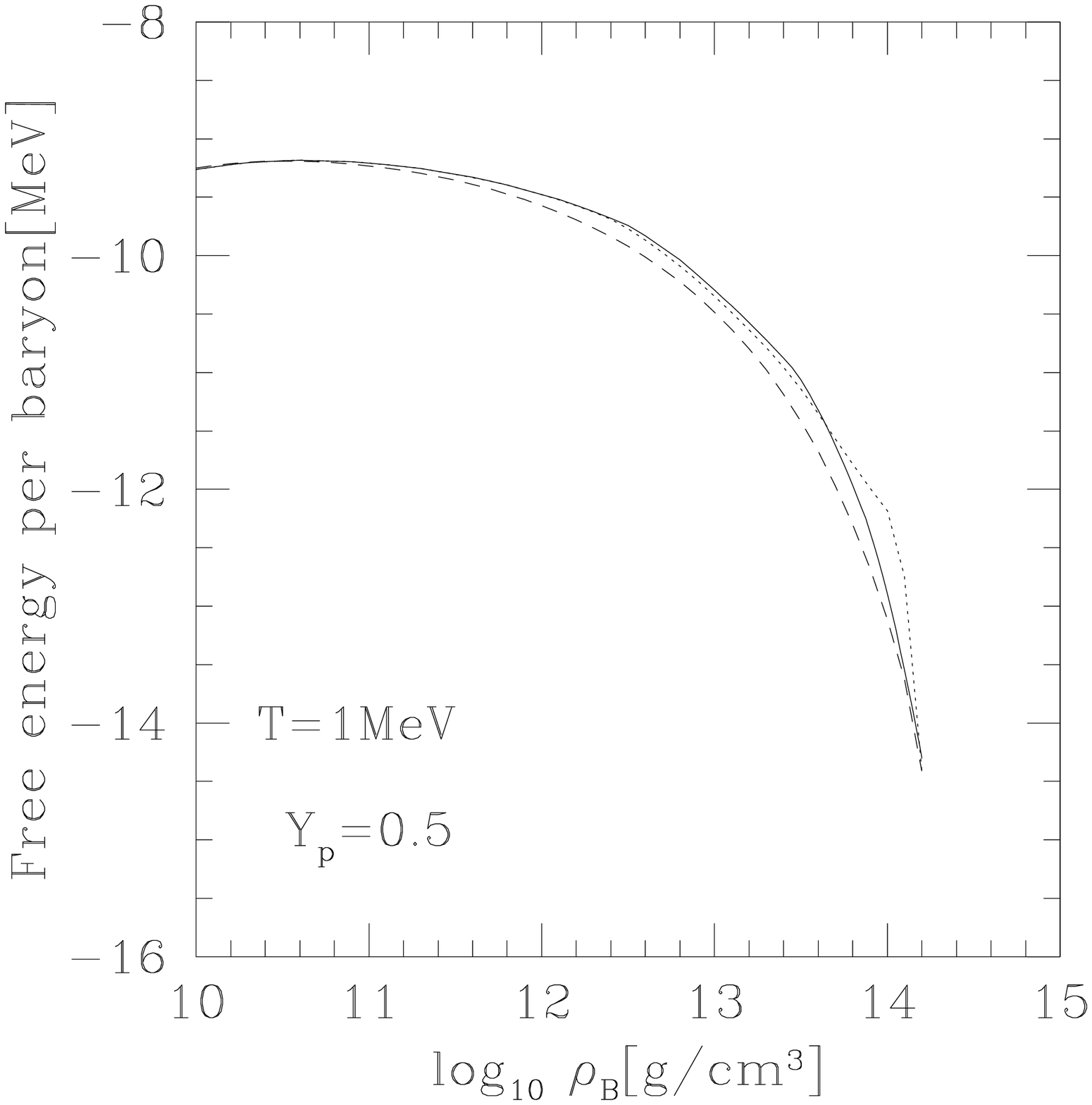}} \\
               \resizebox{63mm}{!}{\plotone{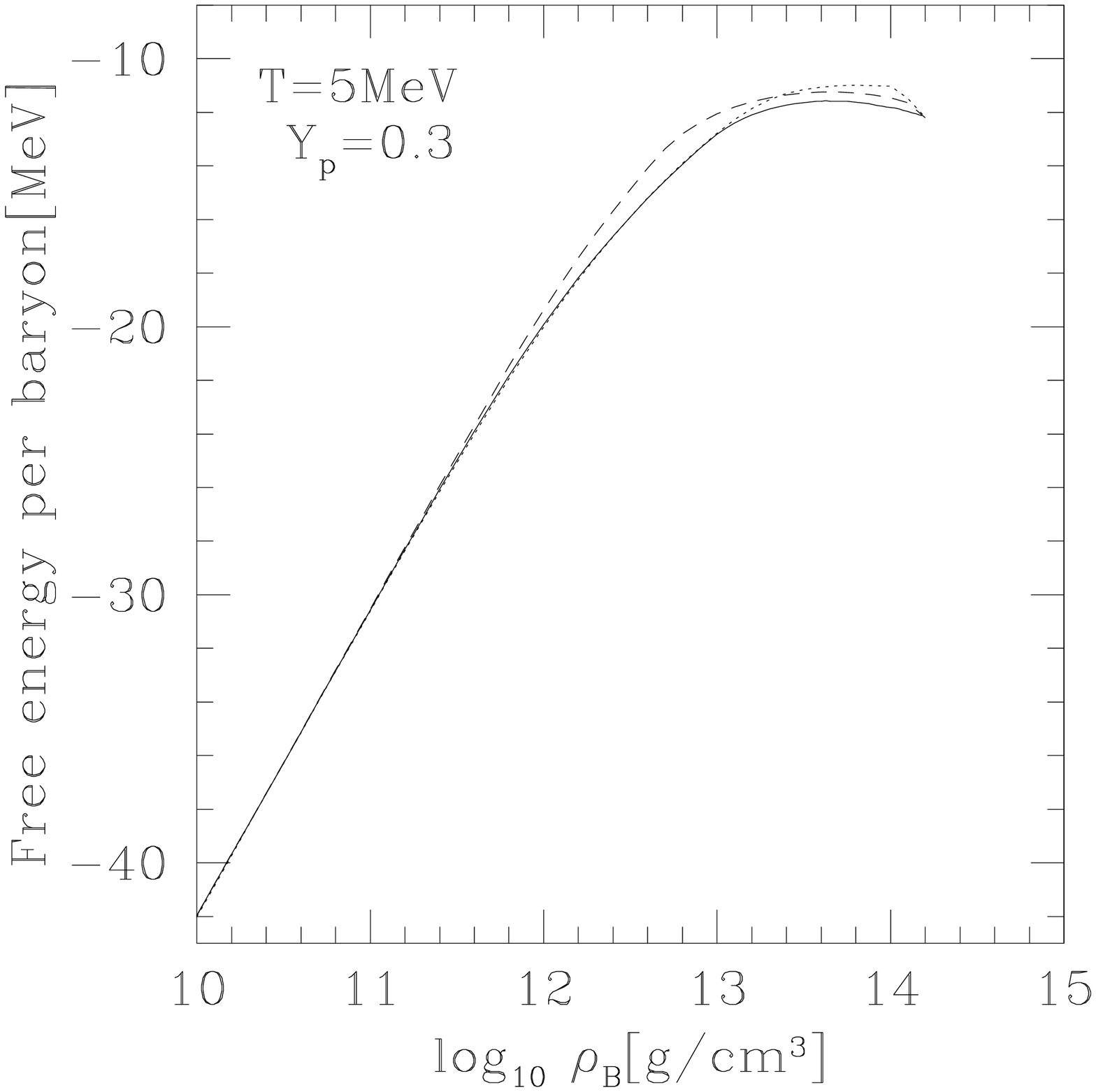}} \\
               \resizebox{63mm}{!}{\plotone{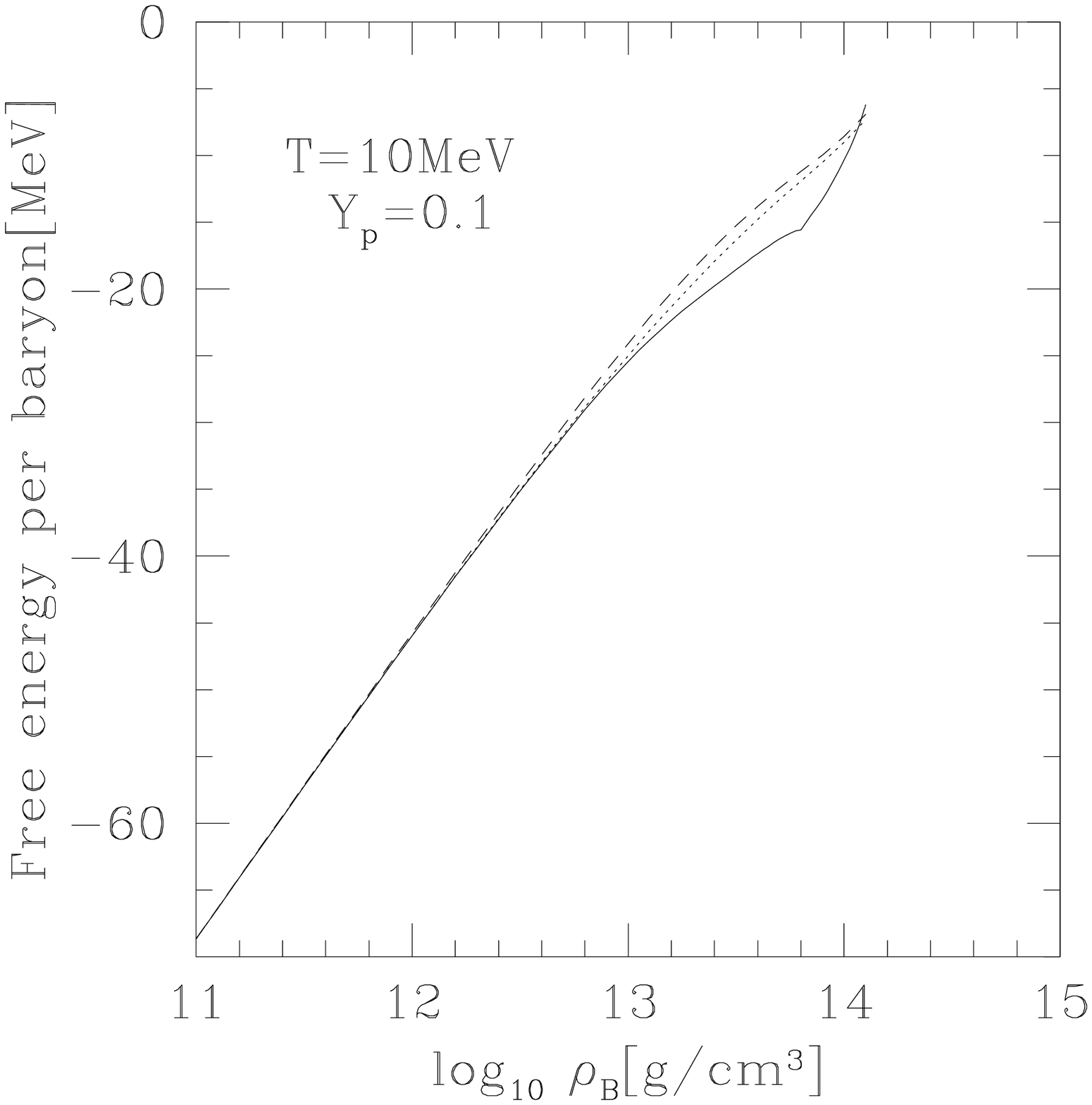}} \\
    \end{tabular}	
   \end{center}
\caption{The free energy per baryon for our (solid lines), H. Shen's (dashed liens) and Hempel's (dotted lines) EOS's as a function of 
density for ($T=1$MeV, $Y_p=0.5$), ($T=5$MeV, $Y_p=0.3$) and ($T=10$MeV, $Y_p=0.1$) from top to bottom.}
\label{fr}
\end{figure}

Fig.~\ref{fr}  shows the free energies per baryon as a function of density for the three combinations of temperature and proton fraction:
($T=1$MeV, $Y_p=0.5$), ($T=5$MeV, $Y_p=0.3$), ($T=10$MeV, $Y_p=0.1$). In all the cases, the free energies are close to one another 
among the three EOS's. Note that our EOS coincides with the H. Shen's EOS at the highest density as it should do. This is also true of 
the Hempel's EOS although how uniform nuclear matter is reached is different between the two EOS's as described in the previous section. 
Looking more closely, we find for $T=1$MeV and $Y_p=0.5$ that the Hempel's free energy is larger than those of our and H. Shen's EOS's at
$\rho_B \sim 10^{14} \rm{g/cm^3}$ and the hump is followed by a steep decline. The difference originates from the fact that 
the Hempel's EOS includes only nuclei with $Z \lesssim 100$ and that unlike our treatment only the Coulomb energy is manipulated by 
using the W-S cell to account for the existence of other nuclei. Their prescription lead to the total dissociation of heavy nuclei to free 
nucleons before reaching the nuclear saturation density. The difference between our EOS and the H. Shen's EOS, on the other hand, comes from 
the fact that the latter includes only single representative heavy nuclei. In the case of $T=5$MeV and $Y_p=0.3$, our and Hempel's 
free energies are smaller than that of H. Shen's EOS at $10^{12} \lesssim \rho_B \lesssim 10^{13}$g/cm$^3$. Under this condition light 
nuclei (deuterons in particular) are abundant as shown in Fig.~\ref{cmp5}. The H. Shen's EOS takes into account alpha particles alone 
as light nuclei. The greater degrees of freedom in the light nuclei in our EOS, especially the substantial population of deuteron, make 
the difference observed. The deviation of Hempel's free energy from ours at $\rho_B \gtrsim 10^{13}$g/cm$^3$ occurs for the same reason as 
in the case of $T=1$MeV and $Y_p=0.5$. For $T=10$MeV and $Y_p=0.1$ our free energy is smaller than those of the other 
two EOS's at $\rho_B \gtrsim 10^{13}$g/cm$^3$. The reason is that the existence of pasta phase is taken into account in our EOS. In fact,
under this condition our EOS predicts the co-existence of various neutron-rich nuclei in the pasta phase or in the uniform phase that 
corresponds to a big nucleus with no bubbles inside, and free neutrons (see section\ref{secrecmp} and Fig.~\ref{pc2}). 
This is in sharp contrast to the matter composed of normal nuclei and free neutrons considered in the other two EOS's. 

\begin{figure}
   \begin{center}
    \begin{tabular}{c}
               \resizebox{63mm}{!}{\plotone{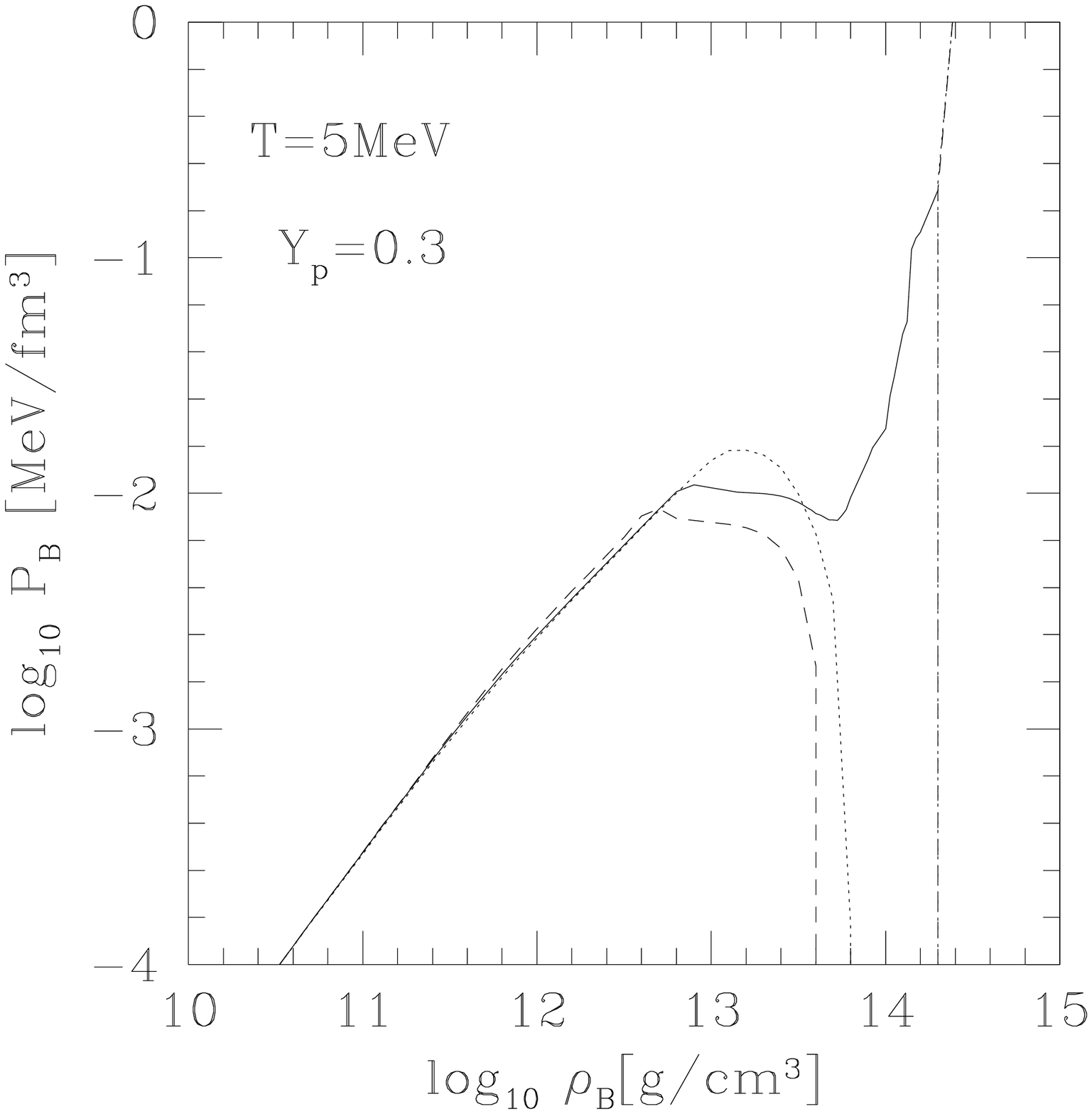}} \\
               \resizebox{63mm}{!}{\plotone{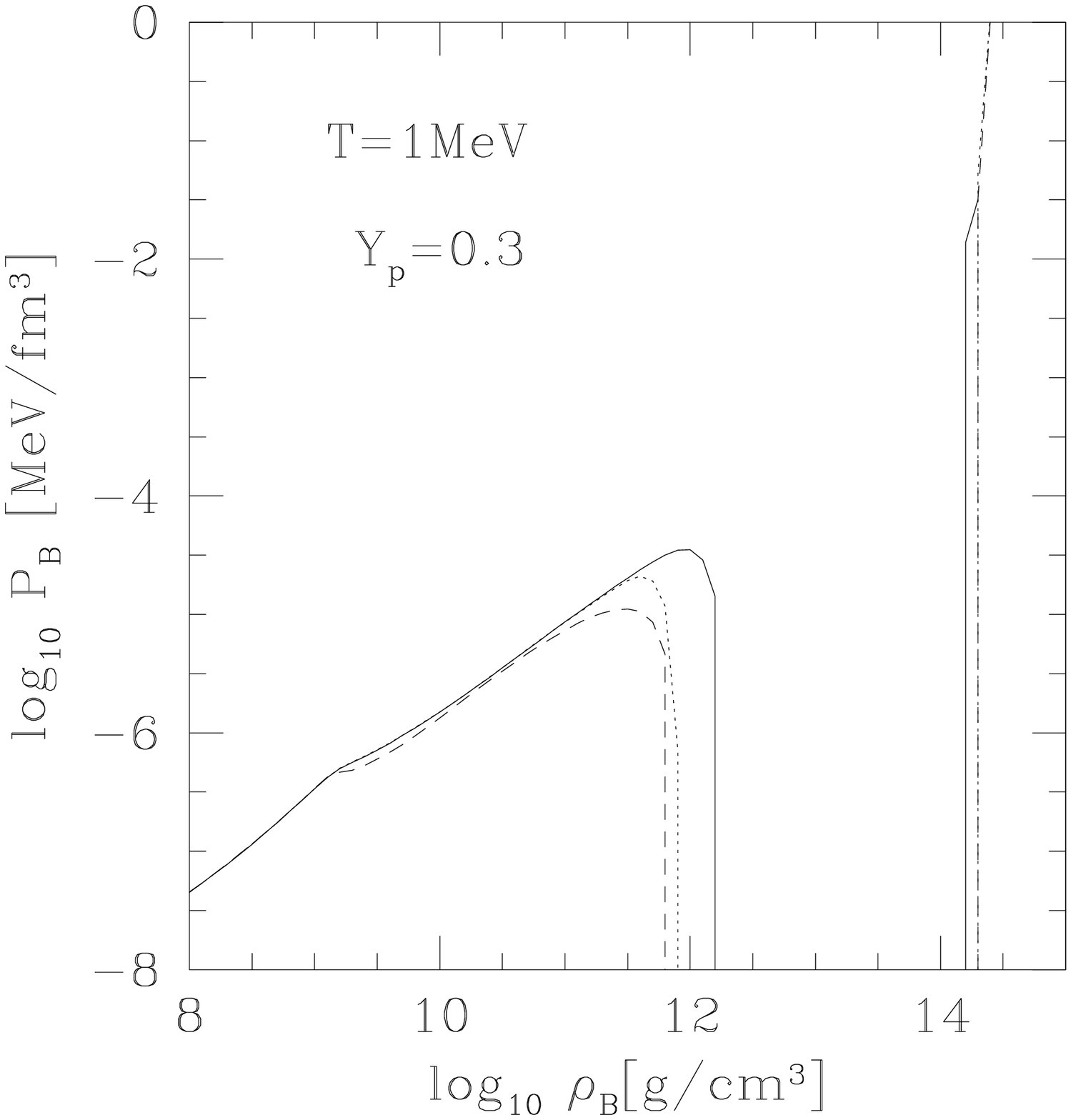}} \\
               \resizebox{63mm}{!}{\plotone{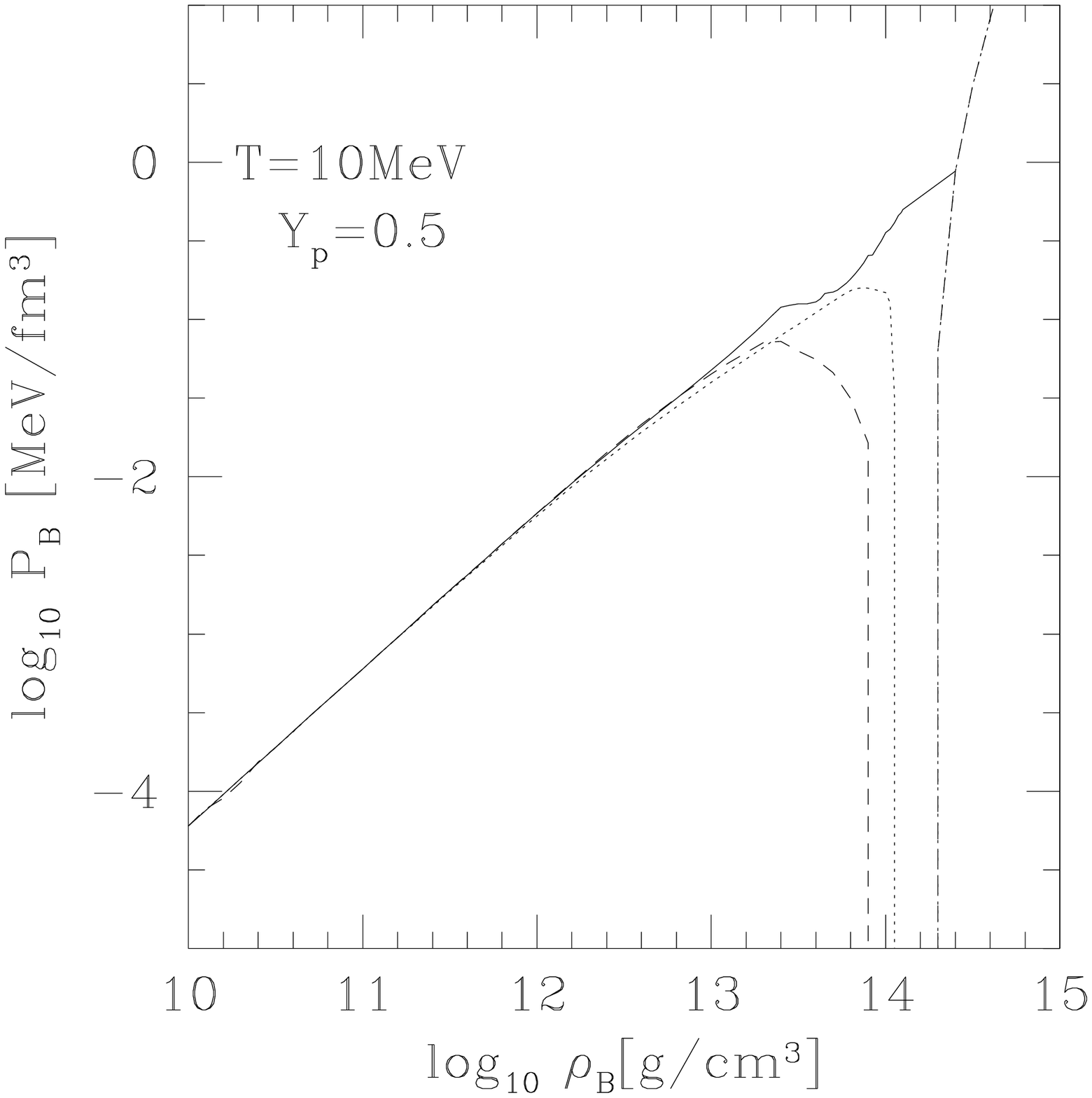}} \\
    \end{tabular}	
   \end{center}
\caption{The baryonic pressure for our (solid lines), H. Shen's (dashed liens) and Hempel's (dotted lines) EOS's as a function of 
density for ($T=5$MeV, $Y_p=0.3$), ($T=1$MeV, $Y_p=0.3$) and ($T=10$MeV, $Y_p=0.5$) from top to bottom.}
\label{pr}
\end{figure}

The pressure is shown as a function of density for three combinations of temperature and proton fraction in Fig.~\ref{pr}. The three 
EOS's agree with one another at low densities, $\rho_B \lesssim 10^{12}$g/cm$^3$. The difference appears, however, in the density regime 
where the pressure becomes negative for the H. Shen's and Hempel's EOS's. The baryonic pressure can be negative when the Coulomb-energy 
contribution, which is negative owing to the attractive nature of the Coulomb interactions, dominates over the other positive 
contributions. Note that the positive leptonic pressure is much larger than the baryonic one normally and the total pressure 
never becomes negative. For $T=5$MeV and $Y_p = 0.3$, our EOS gives positive pressures at $\rho_B \sim 10^{14}$g/cm$^3$, where 
the other EOS's predict negative values. This is due to the existence of the pasta phase, which our EOS takes into account but 
the others do not. In fact, the absolute values of the Coulomb-energy contribution are reduced when the change of the charge 
distribution in the W-S cell is considered in the pasta phase. As a result the pressure in our EOS remains positive. It is also noted 
that the contribution from the excluded-volume effect, $\sum_i {p_i^{ex}}$, gives a non-negligible positive contribution to the 
pressure. At the lower temperature, $T=1$MeV, even our EOS gives negative baryonic pressures and the main difference from the 
other EOS's lies in the lowest value, $\rho_B \sim 10^{12}$g/cm$^3$, of the densities that give negative pressures. It is found 
that our EOS predicts a slightly higher density than the others. This is because our EOS gives larger mass fractions of free neutrons,
which give a positive contribution to the pressure. The difference from the H. Shen's EOS is a consequence of the shell effect, 
which is naturally included in the experimental mass data but ignored in the Thomas-Fermi approximation adopted in the H. Shen's EOS. 
The difference from the Hempel's result, on the other hand, comes from the difference 
in the theoretical estimate of the masses of neutron-rich nuclei with no experimental data. In the case of $T=10$MeV and $Y_p = 0.5$, 
our and Hempel's pressures are larger than that of H. Shen's EOS at $\rho_B \gtrsim 10^{13.5}$g/cm$^3$. This is because the H. Shen's EOS 
neglects the contribution from the translational motions of nuclei, which is not negligible at this high temperature. 
It is also noted that at high densities, $\rho_B \sim 10^{13}$g/cm$^3$, the contribution from the excluded-volume effect is not negligible, 
either, the treatment of which is different between our and Hempel's EOS's (see section~\ref{subsect}). 

\begin{figure}
   \begin{center}
    \begin{tabular}{c}
               \resizebox{63mm}{!}{\plotone{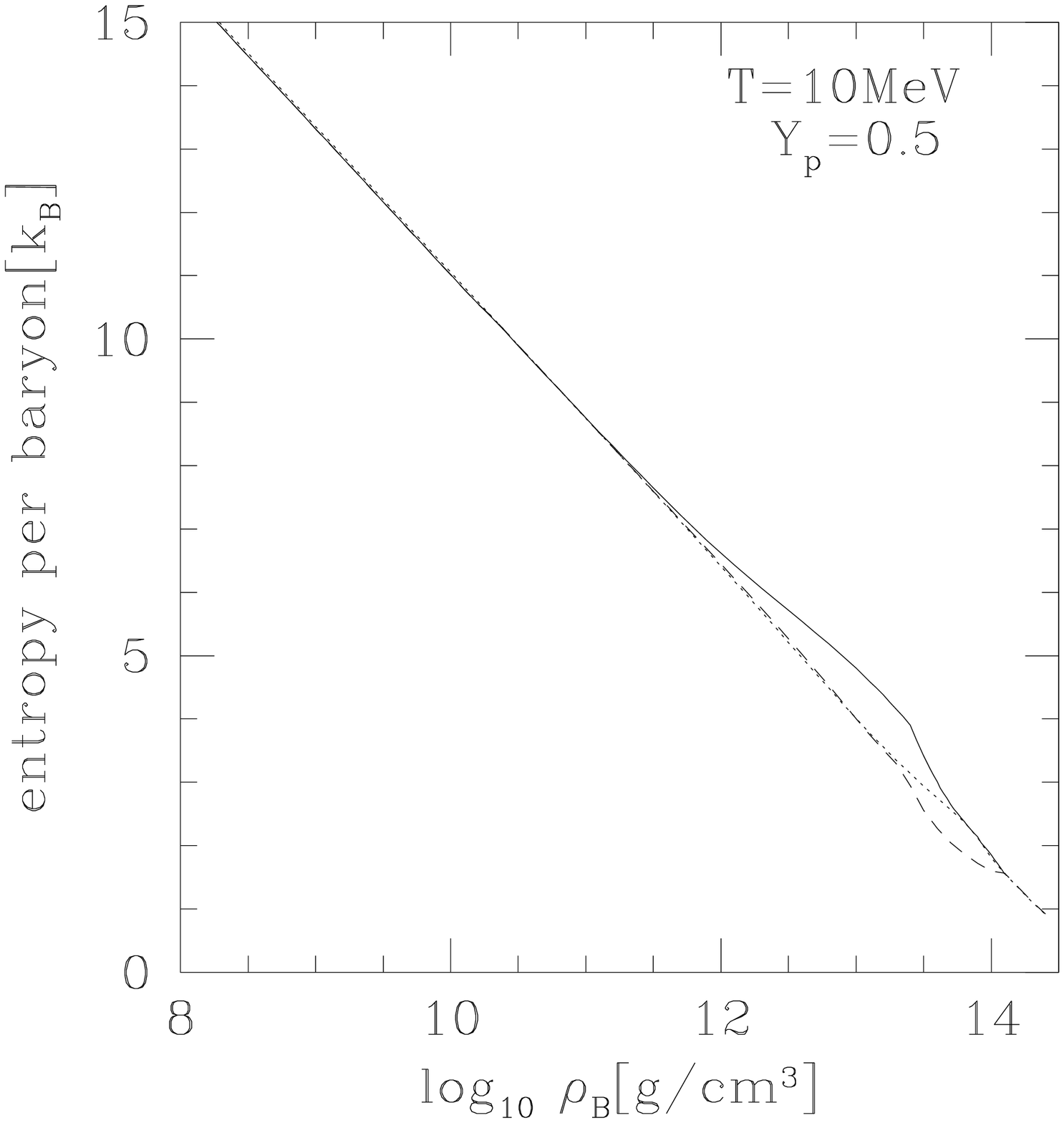}} \\
               \resizebox{63mm}{!}{\plotone{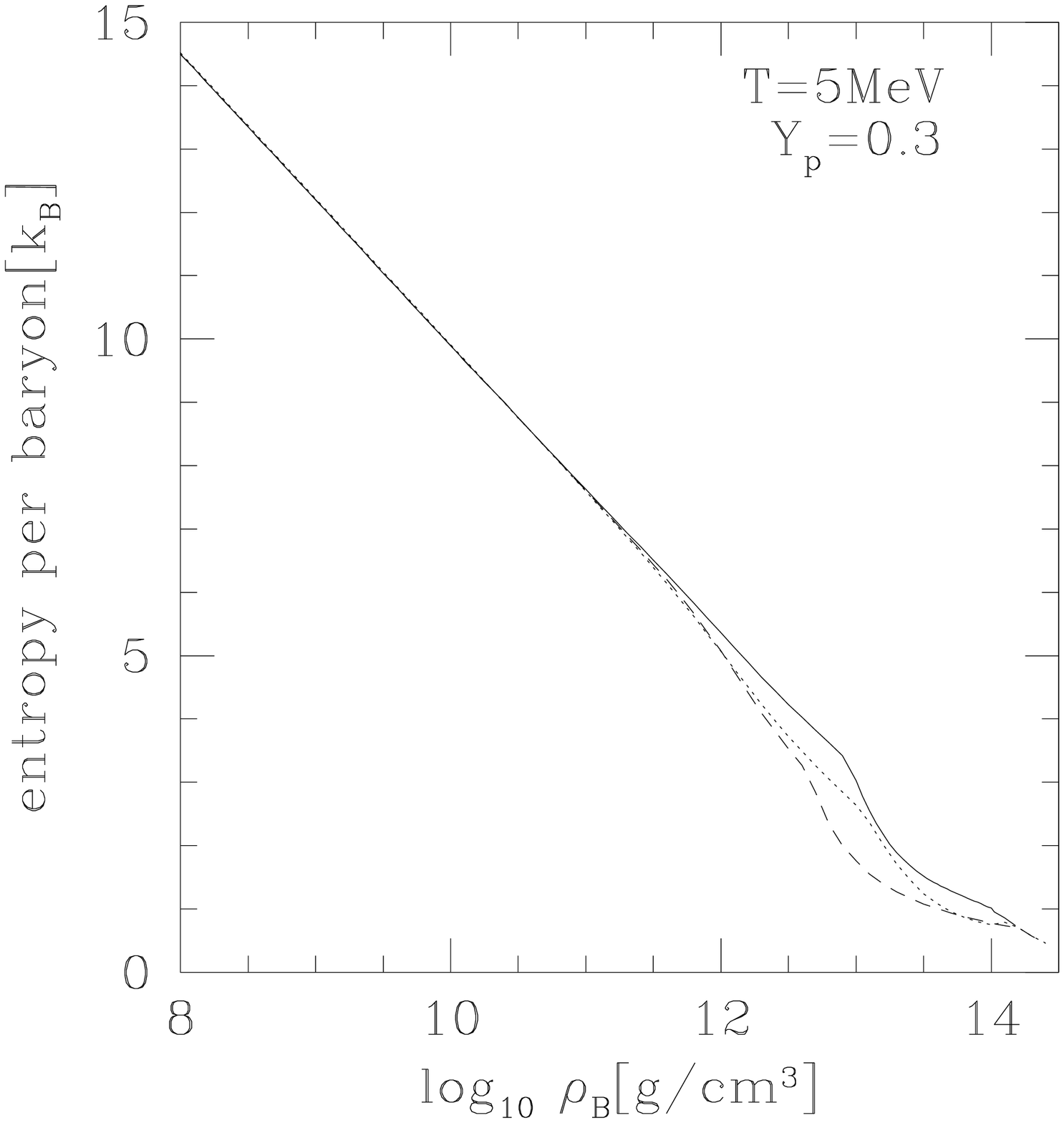}} \\
               \resizebox{63mm}{!}{\plotone{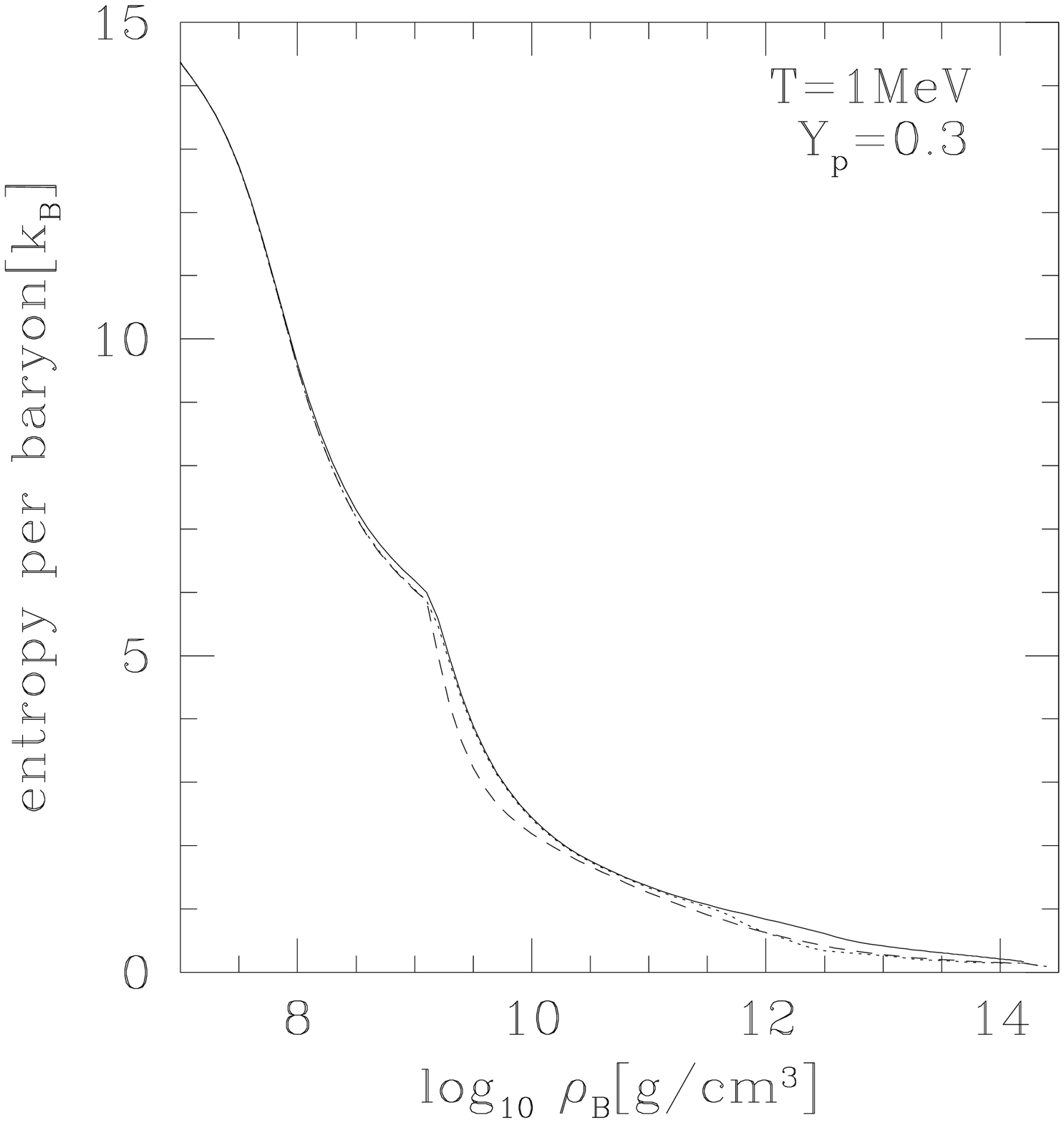}} \\
    \end{tabular}	
   \end{center}
\caption{The entropy per baryon for our (solid lines), H. Shen's (dashed liens) and Hempel's (dotted lines) EOS's as a function of 
density for ($T=10$MeV, $Y_p=0.5$), ($T=5$MeV, $Y_p=0.3$) and ($T=1$MeV, $Y_p=0.3$) from top to bottom.}
\label{en}
\end{figure}

The entropy per baryon is displayed as a function of density for three combinations of temperature and proton fraction in Fig~\ref{en}. 
For high temperatures, $T=5, 10$MeV, our results are larger than those of H. Shen's EOS at $\rho_B \sim 10^{12}$g/cm$^3$ owing to larger 
population of light nuclei such as deuteron as well as the inclusion of the entropy from excited states of nuclei, $g_i(T)$ in 
Eq.~(\ref{eq:ex}). The difference at $\rho_B \sim 10^{13}$g/cm$^3$ is originated from the existence of heavier nuclei in our EOS than 
in the others. Thanks to the larger degrees of freedom, the entropy is actually larger in our EOS. At lower temperatures, $T=1$MeV, 
the entropy per baryon in our EOS is almost equal to those in the other EOS's.

\section{Summary and Discussions}
We have calculated the baryonic equation of state at sub-nuclear densities, which is meant for the use in core-collapse simulations 
and provides the abundance of various nuclei up to the atomic number of $\sim 1000$. The formulation is based on the mass formula and 
the model free energy is constructed so that it could reproduce the ordinary NSE results at low densities and make a continuous transition 
to the EOS for supra-nuclear densities, for which we have adopted the H. Shen's EOS in this paper. The experimental mass data have been 
used, whenever available, to obtain the bulk energy of nuclei. For very heavy and/or very neutron-rich nuclei with no experimental mass
data available, we have utilized for the evaluation of the bulk energy the theory used in the EOS for supra-nuclear densities, that is, 
RMF in this paper. At high densities, where the nuclear structure is affected by the presence of other nuclei, nucleons and electrons, 
we have approximated the bulk energy of nuclei by interpolation of the value obtained experimentally and the one derived theoretically. 
Assuming the charge neutrality in the W-S cell, we have calculated the Coulomb energy of nuclei. Close to the nuclear saturation 
density, the existence of the pasta phase has been taken into account in calculating the surface and Coulomb energies. The free energy of
the nucleon vapor outside nuclei is calculated by RMF. These prescriptions not only ensure the continuous transition to the EOS for supra-nuclear  
densities when the matter becomes uniform but also has remedied the peculiar behavior of the EOS by \citet{Hempel2010}, in which all 
nuclei are dissociated completely to nucleons before reaching the nuclear saturation density. In our EOS, the transition to uniform 
nuclear matter proceeds either via the coalescence of nuclei or the rise of the density in the nucleon vapor, depending on the proton 
fraction. 

For some representative combinations of density, temperature and proton fraction, we have made a comparison of the abundance of nuclei 
as well as thermodynamical quantities obtained in this paper with those provided by the H. Shen's and Hempel's EOS's. It is found that 
the thermodynamical quantities such as the free energy, pressure and entropy per baryon, agree well among these EOS's. The matter 
compositions show noticeable differences, however. The average mass numbers, for example, are systematically smaller at 
$\rho_B \lesssim 10^{12} $g/cm$^3$ in our EOS than the mass numbers of the representative nuclei in the H. Shen's EOS. The difference is
 ascribed to the neglect of the shell effect in the latter EOS. We have also demonstrated that very heavy nuclei with $Z \gtrsim 100$, 
which were completely ignored in the Hempel's EOS, are indeed abundant at $\rho_B \gtrsim 10^{13}$g/cm$^3$ and lead to the difference in 
the entropy per baryon between the two EOS's. It is interesting that the mass fractions of light nuclei such as deuteron cannot be 
ignored at high temperatures and, in fact, are larger than that of alpha particles in the H. Shen's EOS. The average of mass number 
squared is important for the evaluation of the coherent scattering rate and has been frequently approximated by the mass number squared 
of the representative nuclei in supernova simulations that employ an EOS with SNA. We have shown that the root mean square of mass number 
is very close to the average mass number owing to the shell effect and SNA is justified in this sense. It should be noted again, however, 
that the average masses in our EOS are systematically different from the masses of the representative nuclei in the H. Shen's EOS and 
the coherent scattering rate will be reduced in core collapse simulations with our EOS.

Although we have not made a direct comparison, it is interesting to make some comments on other EOS's. As mentioned in Introduction, 
the Lattimer \& Swesty's EOS \citep{Lattimer1991}, which is based on the compressible liquid drop model with SNA and a free energy 
per baryon with the Skyrme-type parametrization, is another EOS that is commonly employed in core collapse simulations. 
Very recently, \citet{Shen2011} published a new EOS for use in astrophysical simulations, which employs different theories, 
the Virial expansion at low densities and SNA with the Hartree approximation at high densities. It is noted first that since these
two EOS's behave quite similarly to the H. Shen's EOS, the results obtained for the latter in this paper will be mostly applicable 
to the former two. For example, thermodynamic quantities are not very different although the entropy in our EOS tends to be a bit 
higher than those in others because of the inclusion of various light nuclei in our EOS. On the other hand, the step-wise growth 
of average mass number at low densities is common only to our and G. Shen's EOS, since the latter also employs experimental mass 
data. At high densities, the Lattimer \& Swesty's and G. Shen's EOS's employ SNA just as the H. Shen's EOS. Not including bubble nuclei, 
the H. Shen's and G. Shen's EOS's predict very big droplet nuclei where the Lattimer \& Swesty's and our EOS give bubble nuclei. 
It is interesting, however, to mention incidentally that \citet{Shen2011} also found a shell-like hollow nuclei, which others did not.

There is an ample room for improvement in our EOS. The interpolation of the bulk energy and the treatment of the pasta phase are entirely
phenomenological and need justification or sophistication somehow. Our results show that the shell effect is important for the abundance
of nuclei. In the present prescription, however, the effect is suppressed maybe too strongly at high densities. We are planning to adopt an
empirical formula for this effect from the literature. With these caveats in mind we are currently constructing an EOS table, which will 
be available for supernova simulations and put in public domain. In so doing, electron capture rates, which are consistent with the
abundance given by the EOS, should be also included. In our formulation, it is quite simple to change the EOS for supra-nuclear densities.
In fact, all we need is the free energy density of uniform nuclear matter predicted by the particular EOS. The combination with another
EOS is under progress at present. 
Different RMF parametrizations will be also discussed in our forthcoming paper.

\acknowledgments 
The authors gratefully acknowledge the contribution of A. Ohnishi and C. Ishizuka to the core idea. We would like to thank M. Takano for useful 
information on the mass formula. S.F. is grateful to K. Nakazato and  N. Yasutake for their useful comments. 
This work was partially supported by Grant-in-Aid for Scientific Research of the Ministry of Education, Culture, Sports, Science and Technology (19104006, 21540281, 22540296) and Scientific Research  on innovative Areas (20105004, 20105005)

\newpage


\begin{thebibliography}{}

\bibitem[Akmal et al.(1998)] {Akmal1998} Akmal, A., Pandharipande, V. R. \& Ravenhall, D. G. 1998 Phys. Rev. C 58,1804 
\bibitem[Audi et al.(2003)] {Audi2003} Audi, G., Wapstra, A.H. \& Thibault, C. 2003, Nucl. Phys.. A 729, 337 
\bibitem[Bethe et al.(1990)] {Bethe1990} Bethe, H. A. 1990, Rev. Mod. Phys., 62, 801
\bibitem[Blinnikov et al.(2009)] {Blinnikov2009} Blinnikov, S. I., Panov, I. V., Rudzsky, M. A. and Sumiyoshi, K. arXiv: 0904.3849 [astro-ph].
\bibitem[Burrows et al.(1984)] {Burrows1984} Burrows, A. and Lattimer,J. M. 1984 Astrophys. J. 285, 294 
\bibitem[Fai et al.(1982)] {Fai1982} Fai,G. \& Randrup, J. 1982, Nucl. Phys. A 381, 557 
\bibitem[Gandolfi et al.(2010)] {Gandolfi2010} Gandolfi, s., Illarionov, A. Yu., Fantoni, S., Miller, J. C., Pederiva, F. \& Schmidt, K. E. 2010 Mon. Not. R. Astron. Soc. 404, 35, 39 
\bibitem[Geng et al.(2005)] {Geng2005} Geng,L.,Toki, H. \& Meng, J. Prog. 2005 Theor. Phys. 113, 785
\bibitem[Janka et al.(2007)] {Janka2007}Janka, H.-T., Langanke, K., Marek, A., Martinez-Pinedo, G., \& Muller, B. 2007, Phys. Rep.,442, 38
\bibitem[Hashimoto et al.(1984)]{Hashimoto1984} Hashimoto, M., Seki, H. and Yamada, M. 1984 Prog. Theor. Phys. 71, 320 
\bibitem[Hebeler et al.(2010)]{Hebeler2010} Hebeler, K. \& Schwenk, A.  2010 Phys. Rev. C 82, 014314 
\bibitem[Hempel et al.(2010)]{Hempel2010}   Hempel, M. \& Schaffner-Bielich, J. 2010  Nucl. Phys. A 837, 210
\bibitem[Hix et al.(2003)]{Hix2003} Hix, W. R., Messer, O. E. B., Mezzacappa, A., Liebend?rfer, M., Sampaio, J., Langanke, K., Dean, D. J., \& Mart?nez-Pinedo, G. 2003, Phys. Rev. Lett., 91, 201102
\bibitem[Langanke et al.(2003)]{Langanke2003} Langanke, K. \& Martinez-Pinedo, G. 2003, Rev. Mod. Phys., 75, 819
\bibitem[Nakazato et al.(2009)]{Nakazato2009} Nakazato, K., Oyamatsu, K. \& Yamada, S., Phys.Rev.Lett.,103, 132501 
\bibitem[Lattimer et al.(1991)]{Lattimer1991} Lattimer, J. M. \& Swesty, F. D. 1991, Nucl. Phys., A535, 331
\bibitem[Ravenhall et al.(1983)]{Ravenhall1983} Ravenhall,D. G., Pethick,C. J. \& Wilson, J. R. 1983 Phys. Rev. Lett. 50, 2066 
\bibitem[H. Shen et al.(1998)]{Shen1998} Shen, H., Toki, H., Oyamatsu, K. \& Sumiyoshi, K. 1998a, Nucl. Phys., A637, 435
\bibitem[H. Shen et al.(1998b)]{Shen1998b} Shen, H., Toki, H., Oyamatsu, K. \& Sumiyoshi, K. 1998b, Prog. Theor. Phys., 100, 1013
\bibitem[G. Shen et al.(2011)]{Shen2011} Shen G., Horowitz C. J. \& Teige S., 2011, Phys. Rev. C, 83, 035802  
\bibitem[Slattery et al.(1998)]{Slattery1980} Slattery, W.L., Doolen, G.D. \& DeWitt, H.E. 1980, Phys. Rev., A21, 2087
\bibitem[Steiner et al.(2005)]{Steiner2005} Steiner, A. W., Prakash, M. \& Lattimer, J. M.  2005   Phys. Rev., 411, 325
\bibitem[Sugahara et al.(1994)]{Sugahara1994} Sugahara,Y. and Toki, H. 1994 Nucl. Phys. A579, 557
\bibitem[Sumiyoshi et al.(1995)]{Sumiyoshi1995} Sumiyoshi, K., Toki, H. \& Oyamatsu, K. 1995 Nucl. Phys. A595, 327
\bibitem[Toki et al.(1995)]{Toki1995} Toki, H., Hirata, D., Sugahara, Y., Sumiyoshi, K. \& Tanihara, I. 1995 Nucl. Phys. A588, 357
\bibitem[Thompson et al.(2003)] {Burrows2003}Thompson, T. A., Burrows, A. \& Pinto, P. A. 2003, Astrophys. J.  592, 434
\bibitem[Vautherin (1994)]{Vautherin1994} Vautherin D. {\it supernova}  Bludman S A, Mochkovith, R. and Zinn-Justin, J. (eds) 354, 91 
\bibitem[Watanabe et al.(2005)] {watanabe2005} Watanabe, G., Maruyama, T., Sato, K., Yasuoka, K. \& Ebisuzaki, T. 2005Phys. Rev. Lett. 94, 031101 
%
\end{thebibliography}
\end{document}